\documentclass[sn-mathphys-num]{sn-jnl}% Math and Physical Sciences Numbered Reference Style

\usepackage{multirow}
\usepackage{graphicx}%
\usepackage{amsmath,amssymb,amsfonts}%
\usepackage{amsthm}%
\usepackage[utf8]{inputenc} %unicode support

\usepackage[title]{appendix}%
\usepackage{xcolor}%

\usepackage{lmodern}
\usepackage{hyperref}
\usepackage[caption=false]{subfig}
\usepackage{float}
\usepackage{svg}
\usepackage{comment}
\usepackage{afterpage}

\usepackage{booktabs}

\usepackage{multirow}
\usepackage[running]{lineno}
\usepackage{acronym}
\usepackage{bm}
\usepackage{natbib}

\title{Network-based diversification of stock and cryptocurrency portfolios}

\author[1]{\fnm{Dimitar} \sur{Kitanovski}}\email{dimitar.kitanovski@finki.ukim.mk}
\author[1]{\fnm{Igor} \sur{Mishkovski}}\email{igor.mishkovski@finki.ukim.mk}
\author[2,3]{\fnm{Viktor} \sur{Stojkoski}}\email{vstojkoski@eccf.ukim.edu.mk}
\author*[4,1]{\fnm{Miroslav} \sur{Mirchev}}\email{mirchev@csh.ac.at}

\affil[1]{\orgdiv{Faculty of Computer Science and Engineering}, \orgname{Ss. Cyril and Methodius University in Skopje}, 
\orgaddress{
\country{North Macedonia}}}
\affil[2]{\orgdiv{Faculty of Economics}, \orgname{Ss. Cyril and Methodius University in Skopje}, \orgaddress{
\country{North Macedonia}}}
\affil[3]{\orgdiv{Center For Collective Learning, ANITI}, \orgname{University of Toulouse}, \orgaddress{
\country{France}}}
\affil[4]{\orgname{Complexity Science Hub}, 
\orgaddress{
\city{Vienna}, 
\country{Austria}}}

\begin{document}
%\linenumbers
\abstract{Maintaining a balance between returns and volatility is a common strategy for portfolio diversification, whether investing in traditional equities or digital assets like cryptocurrencies. One approach for diversification is the application of community detection or clustering, using a network representing the relationships between assets. We examine two network representations, one based on a standard distance matrix based on correlation, and another based on mutual information. The Louvain and Affinity propagation algorithms were employed for finding the network communities (clusters) based on annual data. Furthermore, we examine building assets' co-occurrence networks, where communities are detected for each month throughout a whole year, and then the links represent how often assets belong to the same community. Portfolios are then constructed by selecting several assets from each community based on local properties (degree centrality), global properties (closeness centrality), or explained variance (Principal component analysis), with three value ranges (max, med, min), calculated on a minimal spanning tree or a fully connected community sub-graph. We explored these various strategies on data from the S\&P 500 and the Top 203 cryptocurrencies with a market cap above 2M USD in the period from Jan 2019 to Sep 2022. Moreover, we study in more detail the periods of the beginning of the COVID-19 outbreak and the start of the war in Ukraine. The results confirm some of the previous findings already known for traditional stock markets and provide some further insights, while they reveal an opposing trend in the crypto-assets market.}



\maketitle

\section{Introduction}

In the domain of investment theory, portfolio diversification stands as a pillar, encapsulated by the timeless adage, ``Don't put all your eggs in one basket.'' Rooted in Markowitz's seminal work~\cite{markowitz1952utility}, this principle advocates for risk mitigation by dispersing investments across various financial instruments, industries, or categories. The ensuing challenge for investors has always been navigating the ever-shifting sands of global finance to enact this diversification effectively.

From the underpinnings of the Modern Portfolio Theory to the sophisticated constructs of the Capital Asset Pricing Model~\cite{sharpe1964capital,lintner1965security,mossin1966equilibrium,fama1993common}, our comprehension of investment strategies has seen substantial refinement over the last century. However, as the complexities of financial markets continued to unfold, it was discovered that traditional models can sometimes fall short in capturing intricate inter-dependencies and non-linear relationships among asset returns. For instance, while the Pearson correlation coefficient has been a cornerstone in understanding asset relationships, it is increasingly clear that it falls short in capturing the full spectrum of market dynamics, especially in the face of non-linear interactions~\cite{hinich1985evidence,franses1996forecasting,caginalp2011nonlinearity,hartman2018nonlinearity,inglada2020comprehensive}. This limitation calls for more advanced methodologies that can adequately map out both the linear and non-linear dependencies between various financial instruments. 

Addressing these limitations, the application of network science in finance has emerged as a transformative approach over the last couple of decades \cite{allen2009networks,huang2009network,chi2010network,BridaRisso2010,silva2016structure,battiston2016price,kim2017predicting, iori2018empirical}. By mapping assets as a network of interconnected nodes, network science offers a sophisticated lens for examining financial markets, capturing the intricate and often non-linear dynamics that drive market behavior. But despite the recent advancements in network science, challenges in applying it in portfolio management still persist.  

In this paper, we use a network-based approach for portfolio diversification, addressing key issues in traditional investment strategies. The approach encompasses a five-step process: First, we analyze both linear and non-linear relationships between financial assets using Pearson correlation and mutual information. Second, we transform these relational metrics into distance matrices, setting the stage for network creation. Third, we employ these matrices to construct network representations of financial assets. The fourth step involves applying community detection algorithms to these networks, segmenting the market into distinct communities. Last, we select financial assets from each community, using principal component analysis or network metrics such as degree and closeness centrality. This comprehensive framework allows for a more nuanced and effective portfolio construction, adapting to the complex dynamics of modern financial markets.

We apply our method to data spanning both conventional assets (stocks) and digital assets (cryptocurrencies), each offering unique insights under varied market conditions. Our analysis is methodically divided into three distinct periods: the entire dataset range for a comprehensive understanding, the COVID-19 pandemic period for examining market responses to global health crises, and the beginning of the Russian invasion of Ukraine to observe the impact of geopolitical tensions. 

We find that the construction and analysis of investment portfolios, whether they consist of stocks or cryptocurrencies, reveal intricate patterns and trends that are critical to understanding the dynamics of financial markets. Our comprehensive examination across various periods, including the beginning of the COVID-19 pandemic and the start of the Russian invasion of Ukraine, underscores the influence of external crises on portfolio performance. Notably, in stock portfolios, the network-based approach often outperformed baseline portfolios. This method was effective not only in maximizing returns but also in managing volatility, thus offering a balanced approach to portfolio construction. However, the trends observed in cryptocurrency portfolios were more complex and less predictable, highlighting the unique and volatile nature of the digital asset markets. We should note that the period of analysis during the Russian invasion of Ukraine also includes the large fall in the value of cryptocurrencies, which happened before the invasion in the Autumn of 2021.

We also find a pronounced divergence in the behavior of investment portfolios during periods of global crises, which significantly disrupted traditional patterns observed in calmer periods. In the face of overlapping crises such as COVID-19 and the Russian invasion of Ukraine, even well-constructed portfolios encountered challenges, often registering low or negative returns and high volatility. This was particularly evident in the realm of cryptocurrencies, where the market's response to global events was markedly different from that of stocks. Despite these challenges, our analysis revealed that certain network-based approaches consistently contributed to forming more resilient portfolios. These findings underscore the necessity of adaptive and robust portfolio construction strategies to navigate the complex and rapidly changing landscape of financial markets.

The rest of the paper is organized as follows. In Section \ref{sec:rw}, we explore existing literature and previous studies, establishing the context and background for our research. Then, in Section \ref{sec:md}, we detail our analytical approach and describe the data sets used. Section \ref{sec:res} presents our findings, where we subdivide the analysis into three distinct periods: the entire period, the beginning of the COVID-19 pandemic, and the beginning of the Russian invasion of Ukraine, which allows us to examine the impact of different global events on portfolio diversification strategies for both stocks and cryptocurrencies. Finally, in Section \ref{sec:con}, we summarize our key findings and discuss the limitations of this approach and the potential avenues for future research, highlighting the implications of our study in the broader context of financial analysis and portfolio management.

\section{Related work}\label{sec:rw}

Correlations have a long history of application in the analysis of financial systems \cite{Mantegna_Stanley_1999, Bouchaud_Potters_2003}. Correlation matrices can contain a large number of random values, which can be analyzed and cleaned using tools from random matrix theory \cite{Bun_2017}. Many studies have explored the application of various correlation-based clustering techniques \cite{dose2005clustering, tola2008cluster, Nanda2010ClusteringIS,Duarte2020}. On the other hand, various network approaches, often based on correlations, have also been widely applied in the study of financial markets \cite{allen2009networks,silva2016structure,battiston2016price,BridaRisso2010}, including stocks, currencies \cite{Mizuno_2006,basnarkov2019correlation,basnarkov2020lead} and cryptocurrencies \cite{Stosic2018CollectiveBO,papadimitriou2020evolution}, which can all be represented as complex networks \cite{newman2003structure}. Network analysis of financial markets can provide valuable insights into their underlying structure and hidden patterns, which can be particularly valuable for portfolio construction, optimization, and management \cite{battiston2016complexity,niu2021implicit}.

The pioneering work that introduced the representation of the relationships between assets with networks was done in \cite{mantegna1999hierarchical}. In this work, a \ac{MST} was created by considering correlations among stock prices in the \ac{DJIA} and \ac{SP500} indices. The authors subsequently explored the use of MSTs in financial networks in several other works where they investigated different stock markets and indices, as well as correlations calculated over different time spans, which were summarized in \cite{bonanno2004networks}. In \cite{onnela2002dynamic,onnela2003dynamics}, the MST representation was studied with a dynamical dimension using S\&P 500 data, and it was noticed that the network has a scale-free characteristic. It was also observed that in times of crisis, the network structure becomes even more centralized. The distance to the highest-degree node decreases, and the average path length decreases. Moreover, the authors discovered that the portfolios with minimal risk tend to contain assets located at the periphery of the MST.

A denser correlation-based network representation of financial markets was explored in \cite{tumminello2005tool,Tumminello_2006}, where a \ac{PMFG} was created by filtering the full correlation matrix. The PMFG provides a richer representation while still preserving the hierarchical structure of the MST. In both approaches using MST and PMFG, portfolios composed of assets situated at the periphery usually showed lower risk and better returns \cite{pozzi2013spread}. Additionally, the increased distance between assets further improves the portfolio performance. Similar conclusions were brought in \cite{peralta2016network}, using a strategy for portfolio management that balances between a specifically developed assets centrality measure for financial networks and individual assets properties. Another approach of a \ac{DBHT} was explored in \cite{Musmeci2014ClusteringAH} and compared in detail with MST and PMFG. A study of the Chinese stock market \cite{ren2017dynamic}, using an MST representation and various network centrality measures and distances, argued that the portfolio performance depends on the market conditions. Portfolios composed of more central assets perform better when there is a draw-up in the near future, while more peripheral portfolios perform better when there is a draw-down in the horizon. Another group of studies examined a hierarchical risk parity model \cite{dePrado59,raffinot2017hierarchical}. Other studies have explored portfolio optimization based on correlation networks, like global-motion correlation in \cite{li2019portfolio}, correlation and assortativity in \cite{ricca2024portfolio}, and correlation, centrality, and random matrix theory filtering in \cite{giudici2022network}. Some authors have also explored some measures that capture non-linear asymmetric relationships, such as transfer entropy \cite{ioannidis2023portfolio}.

The cryptocurrency market has also been recently studied from a network perspective. In \cite{Stosic2018CollectiveBO, chaudhari2020cross, gavin2021community}, its MST representation and community structure were examined in detail, but its application to portfolio diversification was just slightly discussed, and the network characteristics are not employed in the asset selection process. The price synchronization of cryptocurrencies was studied in \cite{papadimitriou2020evolution}. In a recent paper, \cite{kitanovski2022cryptocurrency}, we have presented some very preliminary results of our study, which we have extensively expanded in this work. Another recent paper, \cite{das2023portfolio}, addresses the problem of portfolio optimization using only standard clustering techniques. Some authors have also studied network-based portfolio diversification of a diverse set of assets, such as stocks, bonds, cryptocurrencies, commodities, and forex \cite{vyrost2019network, wang2024portfolio}.

This section provides an overview of some of the most important works in this area, while a more comprehensive review can be found in \cite{marti2021review}, which collects numerous works in the study of financial markets represented as networks, in addition to the classical correlation and clustering analyses. Another popular direction of improving portfolio management nowadays is the application of artificial intelligence, recently reviewed in \cite{sutiene2024enhancing}, which also includes network-based approaches. One such study \cite{pacreau2021graph} has employed a supervised multi-relational graph neural network that includes sector and supply-chain information besides price correlations for portfolio management, while another study \cite{soleymani2021deep} has explored the application of reinforcement learning. Nevertheless, the application of artificial intelligence for portfolio construction should be made with precaution and measures to preserve its sustainability, accuracy, fairness, and explainability \cite{babaei2022explainable,giudici2024safe}.

\section{Methods and data}\label{sec:md}

\subsection{Portfolio diversification preliminaries} \label{portfolio}

Standard methods for portfolio diversification use the concepts of expected return \(E(R)\) and expected variance \(\sigma^2\) to find the optimal share $w_i$ of investment in asset $i$ \cite{markowitz1952utility}. 
Consider the case of $N$ assets with $E(R_i)$ and $\sigma^2_i$ being, respectively, the expected individual return and variance of each asset $i$. Then, the expected return is
\begin{equation}
E(R) = \sum_{i=1}^{N} w_i E(R_i).
\end{equation}
Analogously, the expected portfolio variance is
\begin{equation}
\sigma^2 = \sum_{i=1}^{N} \sum_{j=1}^{N} w_i w_j \mathit{C}_{ij} \sigma_i \sigma_j.
\end{equation}
There are two caveats that underpin this approach. First, a central role in portfolio diversification is played by the Pearson correlation coefficient $\mathit{C}_{ij}$, which quantifies the linear relationships between different assets $i$ and $j$. To this end, we are often interested in constructing portfolios of assets that are negatively related in order to reduce volatility. In recent years, however, there has been a growing understanding that linear correlations might not fully capture the underlying linkages, particularly when the relationships are more intricate and non-linear~\cite{dionisio2004mutual,fiedor2014networks,guo2018development}. Indeed, in complex financial markets, assets often exhibit non-linear dependencies, which the Pearson correlation coefficient fails to capture. This motivates the need to adopt advanced measures that can holistically capture the network of linear and non-linear dependencies between assets. 
Second, it is evident that we need to adequately choose the assets and their corresponding weights in order to generate an optimal portfolio. Indeed, the optimal portfolio is most often found as a constrained optimization problem that seeks to maximize $E(R)$ while minimizing $\sigma^2$ given the weights and a set of assets that can be part of the portfolio. But before we can apply this optimization procedure, we need to select the potential assets that could be part of our portfolio.

Before we describe the network-based portfolio diversification, let us define some quantities that will be used throughout our study for the portfolio selection and evaluation. If the portfolio value at day $d$ is $P_d$, the portfolio \textbf{annual returns} are \begin{equation}
    R_\mathrm{a} = \frac{P_\mathrm{D} - P_\mathrm{0}}{P_\mathrm{0}},
\end{equation}
where $D=365$. If the \textbf{daily log returns} are given as 
\begin{equation} \label{eq:log_return}
    r_d = \ln\left(\frac{P_d}{P_{d-1}}\right),
\end{equation}the daily log return volatility is $\sigma_d = \mathrm{std}(r_d)$ and the \textbf{annualized log return volatility} is
\begin{equation}
    \sigma_\mathrm{a} = \sigma_\mathrm{d} \times \sqrt{D}.
\end{equation}
For consistency, we assume $365$ trading days for both asset types, and for the missing prices, we assign the previous day's price. Finally, the \textbf{Sharpe ratio} \cite{sharpe1994sharpe} is defined as 
\begin{equation}
    \mathrm{SR}= \frac{R_\mathrm{a} - R_f}{\sigma_\mathrm{a}},
\end{equation} 
where $R_f$ is a risk-free rate, i.e. the annualized 13-week U.S. Treasury bill rate \footnote{The daily treasury bill rates are taken from the U.S. Department of Treasury (\href{https://home.treasury.gov/resource-center/data-chart-center/interest-rates/TextView?type=daily_treasury_bill_rates}{home.treasury.gov})}.

\subsection{Network-based portfolio diversification} 
\label{network_portfolio}
\begin{figure}
  \includegraphics[width=0.8\linewidth]{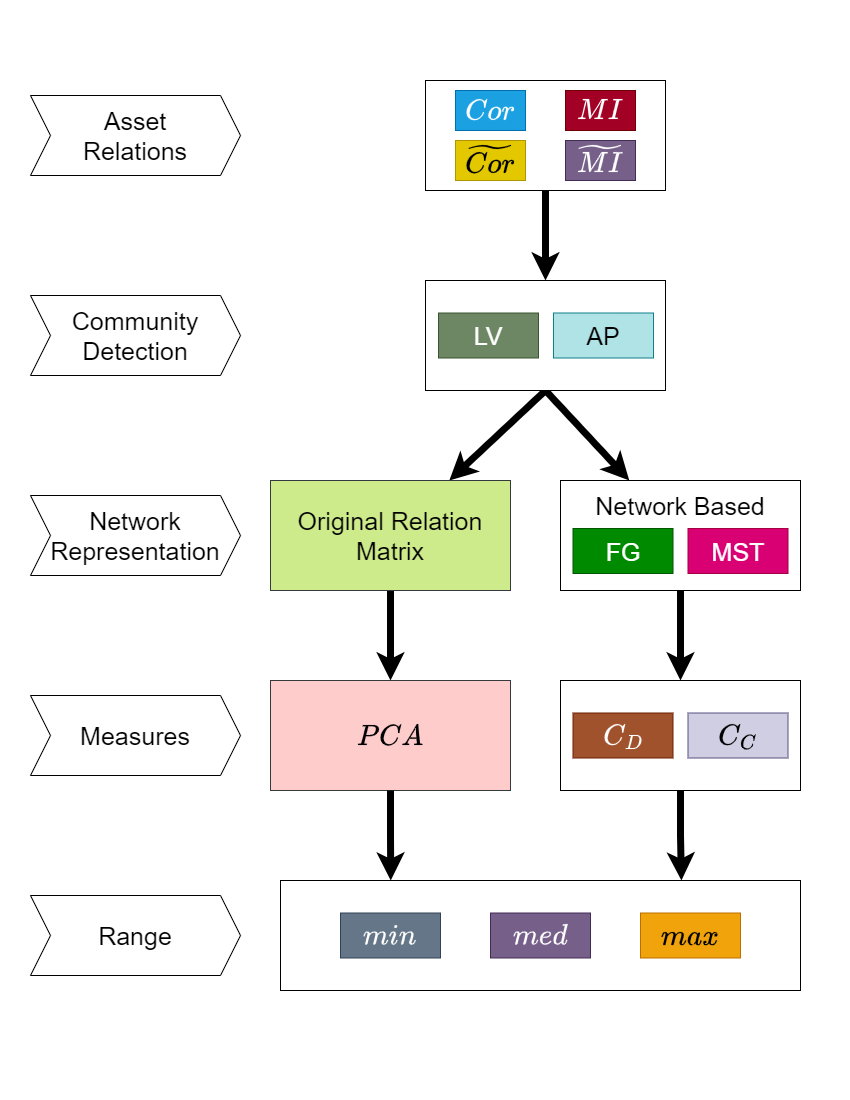}
  \caption{Methodological framework for constructing investment portfolios: a comprehensive overview of strategies employed. The symbols represent: Cor – Correlation, MI – Mutual information, ($\sim$) – co-occurrence, $C_C$ – Closeness centrality, $C_D$ – Degree centrality, PCA – Principal component analysis, FG – Full graph, MST – Minimum spanning tree, LV – Louvain, and AP – Affinity propagation.}
  \label{fig:port_meth}
\end{figure}

%To address these issues, we employ a network-based approach for portfolio diversification. Namely, to address the first problem, we use multiple measures for the relationships between assets. To tackle the second problem, we create networks between assets and exploit network concepts, such as centrality measures and community detection algorithms, to reveal the assets that could potentially be part of our portfolio. 
To address the portfolio diversification problem, we employ a network-based approach. First, we use two different measures to express the relationships between assets. Then, we create network representations and exploit network concepts, such as centrality measures and community detection algorithms, to identify the assets that could potentially be part of our portfolio. 
In short summary, our network-based portfolio diversification approach consists of the following five steps:

\begin{description}
     \item[1. Representing the relationships between financial assets] - First, we calculate the relationships between the financial assets using two types of measures i) \ac{Cor} that captures linear dependencies and ii) \ac{MI} that also captures non-linear dependencies. To provide a comprehensive analysis, we also introduce co-occurrence relationships, namely \ac{cCor} and \ac{cMI}.
      %$\widetilde{\mathit{Cor}}$ and $\widetilde{\mathit{MI}}$, derived from correlation and mutual information respectively.
      
     \item[2. Transformation of the relational matrix into a distance matrix] - During this step, the relational matrix is modified and a distance matrix $D$ is derived from it. The measure of similarity is transformed into distance.
      
     \item[3. Creating network representation] - This step includes creating a network representation from the financial assets using the distance matrix obtained in the previous step. We consider two types of network representations, a \ac{FG} and an extracted \acf{MST}.
      
     \item[4. Finding communities (clusters)] - In this step, our objective is to segment the market, so we employ two community detection (clustering) algorithms. Namely, the \ac{LV} community detection algorithm was applied to the MST, while the \ac{AP} clustering algorithm was applied directly to the correlation and mutual information matrices.
      
     \item[5. Choosing financial assets from communities] - Finally, we select representative assets from the previously detected communities. Each stocks' and cryptocurrencies' portfolio is equally-weighted and is composed of $P=25$ and $P=20$ assets, respectively. We strive to select an equal number of assets from each community, but if the $P$ assets can not be equally drawn from the obtained $Q$ communities, the remaining $R \equiv P \mod Q$ assets are drawn from the largest $R$ communities. The selection of assets is based on various statistical and network metrics. First, we used \ac{PCA} applied directly to the data (the correlation and mutual information matrices), where the first three components that best describe the data were taken. We also employ two network centrality measures: degree centrality ($C_D$), which is a local metric, and closeness centrality ($C_C$), which is a global metric. The closeness centrality was applied to the FG and the MST of the whole graph. On the other hand, degree centrality was applied to the FG and MST of each community sub-graph. In distance networks, nodes with a larger degree "centrality" are more distant from others, so the centrality notion is inverted. In the assets selection process, three ranges are considered from each metric, namely: minimal (\textit{min}), medial (\textit{med}), and maximal (\textit{max}). The minimal range contains assets with the lowest values for the corresponding metric, the medial range consists of the assets that lie around the median, and the maximal range includes assets with the largest metric values.

\end{description}
Figure \ref{fig:port_meth} depicts the process of constructing portfolios, and by moving through the steps, one can derive all portfolio construction strategies investigated in this study.

\subsection{Relational representation} \label{relationships}

We adopted two relational representations, namely Pearson correlation coefficient and mutual information, calculated using daily log returns $r_{di}$ for asset $i$ at day $d$, defined similarly as Eq.~\ref{eq:log_return}, but further normalized and denoted briefly as $r_i$.

%\textcolor{red}{To compute the normalized logarithmic return, we first obtained the daily logarithmic return (calculated using the annual daily standard returns) of the financial asset using Equation~\ref{eq:log_return}, resulting in the daily logarithmic return $r_d$. Furthermore, the obtained logarithmic return was used to calculate the normalized log return: 
%\begin{equation}
%    r_i = \frac{r_d - \mu}{\sigma},
%\end{equation}
%where $\mu$ and $\sigma$ represent the mean and standard deviation of $r_d$, respectively.}

First, we quantified the linear relationship between asset returns using the \acf{Cor}, which is used to produce a correlation matrix $\mathbf{C}_{N \times N}$ with elements calculated as
\begin{equation}
C_{ij}=\frac{\sum_t (r_i - \overline{r_i}) (r_j - \overline{r_j}))}{\sqrt{\sum_t (r_i - \overline{r_i})^{2} \sum_t (r_j - \overline{r_j})^{2}}},
\end{equation}
representing the correlation between two assets. The values can vary between -1 and 1, with -1 representing a completely negative correlation, 1 indicating a completely positive correlation, and 0 signifying the absence of any correlation.

Second, we used the \acf{MI} to capture the potential nonlinear relationships between asset returns, which is a measure originating from information theory that quantifies the amount of information shared between two random variables. Specifically, it gauges the reduction in uncertainty of one variable given knowledge of the other. We denote the \textit{MI} between all asset pairs using a matrix $\mathbf{M}_{N \times N}$ where its elements are calculated as 
\begin{equation}
    M_{ij} = H(r_i) - H(r_i|r_j),
\end{equation}
where $H(r_i)$ represents the entropy for the variable $r_i$, which is the average level of information inherent in the outcome of that variable, while $H(r_i|r_j)$ represents the conditional entropy, which is the amount of information that is sufficient to describe the outcome of $r_i$ if the value of $r_j$ is known. The resulting mutual information matrix $\mathbf{M}$ has values ranging from 0, which denotes no mutual information between two financial assets, to $\infty$, which denotes strong dependency. We believe mutual information fits well within our framework of portfolio diversification because it provides a symmetric measure of the relationship between the assets, just as correlation, but also captures the nonlinear dependencies.

An additional analysis involved the monthly determination of communities and a calculation of a yearly overlapping coefficient among communities between months, yielding values ranging from 20\% to 35\%. This outcome underscores the unpredictable nature of the stock and cryptocurrency markets. The aforementioned methodologies generate highly fluctuating communities, with interrelationships among assets undergoing significant changes every month. To address this volatility, a co-occurrence matrix of size $N \times N$ was constructed, which reflects the frequency of two assets belonging to the same community. The purpose of this matrix is to establish an asset relationships measure that mitigates the pronounced instability of the communities.

\subsection{Network representation}\label{network_representation}

The matrix values of the relational representation should meet specific requirements to be utilized as a network adjacency matrix for performing network analysis, and employing the correlation matrix directly to build a network representation is not appropriate \cite{mantegna1999hierarchical,gavin2021community}. Namely, the authors in \cite{mantegna1999hierarchical} demonstrate that the correlation coefficient cannot be used as a distance between two financial assets in the creation of a network structure because they do not fulfill the axioms that define the Euclidean metric. Instead, a comparable distance matrix is often derived using a distance measure and then used as an adjacency matrix for the network among assets. To create distance matrices from the correlation and mutual information matrices, an elementary transformation of the original matrices is needed.

From the given correlation matrix $\mathbf{C}$, we can construct the distance matrix $\mathbf{D}^C$ through a specific transformation as it was done in prior works \cite{mantegna1999hierarchical}

\begin{equation}
D^C_{ij} = \sqrt{2(1 - C_{ij})}.
\end{equation}
A value of \(D^C_{ij} = 0\) conveys that assets \(i\) and \(j\) have a perfect correlation, \(D^C_{ij}= 2\) indicates a total negative correlation, and a mid-value, \(D^C_{ij} = \sqrt{2}\) denotes the absence of any discernible correlation between the respective assets.

Alternatively, considering the mutual information matrix $\mathbf{M}$, we can derive the distance matrix $\mathbf{D}^M$ through the subsequent transformation

\begin{equation}
D^M_{ij} = |M_{ij} - \max{(M_{ij})}|.
\end{equation}
In this context, the interpretation of the matrix values is altered. Specifically, larger values of \(D^M_{ij}\) suggest a weaker relationship between assets \(i\) and \(j\), whereas smaller values of \(D^M_{ij}\) indicate a pronounced dependence between the two assets.

\subsection{Clustering and community detection}\label{community_detection}

Clustering and community detection algorithms can be used to examine a set of assets (objects) and group them into clusters and communities based on their properties or connectivity. Clustering is used in statistics and machine learning, where objects are grouped based on their properties, while community detection is typically used in networks where objects (nodes) are grouped based on their connectivity. The application of clustering and community detection in financial market analysis can provide market participants with a better understanding of the relationships between various entities in the financial market, which can be useful for a variety of purposes, including risk management, portfolio construction, and market monitoring. In this work, we apply two well-known algorithms, one community detection algorithm, namely the Louvain method \cite{khan2017network}, and one clustering algorithm, namely Affinity propagation \cite{dueck2009affinity}. Both methods do not require specification of the numbers of communities (clusters) in advance, while Affinity propagation naturally supports negative weights.

\acf{LV} is a well-known method for community detection in networks, based on the idea of network modularity, in which there are stronger internal connections within each community than the outside connections between communities. It is a bottom-up approach that starts by assigning each node in the network to a unique community, and then it iteratively integrates these communities by optimizing the network modularity score until no additional gains in modularity can be made. The result is a network partition where nodes are clustered together into communities that have few connections between them and high levels of internal cohesiveness.
In this study, we applied the Louvain method to the MST extracted from the network representing the financial markets, where nodes denote the financial assets and links denote their mutual relationships (distance). However, because the Louvain method considers links as a measure of similarity between node pairs, we have inverted the weights in the MST, which are provided to the algorithm. This MST with inverted weights, representing similarity, is utilized only as an input to the Louvain method, while everywhere else in the study the weights represent distance. The weight inversion was not performed in some previous studies, but it is very important for correct community detection.

\acf{AP} is a clustering algorithm that belongs to the category of exemplar-based methods. When identifying clusters, \textit{AP} takes into account "responsibility" and "availability". Responsibility, represented by the symbol $r(i, k)$, quantifies how well data point $i$ should serve as the benchmark for data point $k$. How well data point $k$ may serve as an example for data point $i$ is indicated by the availability symbol, represented as $a(i, k)$. Until the algorithm converges to a set of exemplars that represent the clusters in the data, these values are iteratively updated. Furthermore, the method looks for a set of exemplars that maximizes net similarity while using the fewest possible exemplars. In essence, it selects the dataset's most representative points as exemplars and then assigns further data points to these exemplars depending on how similar or unlike they are to other points in the dataset. 
When working with data where unfavorable associations or differences are present, as is the case here, this flexibility is especially helpful. Therefore, without any modifications, the algorithm can be called using the correlation and mutual information matrices that represent precomputed distances.

\subsection{Data}\label{dataset}
In this study, we utilized publicly available daily historical market price data for stocks and cryptocurrencies. Namely, we collected historical prices for stocks that make up the S\&P 500 index using Finnhub \footnote{Finnhub is an API that allows access to stock data (\href{www.finnhub.io}{www.finnhub.io})}, and historical cryptocurrency coins prices from CoinMarketCap \footnote{CoinMarketCap is a web page that provides data about cryptocurrencies prices, market capitalization and other information (\href{www.coinmarketcap.com}{www.coinmarketcap.com})} using a Python scraper called \emph{cryptocmd}. The large amount of available historical data for both stocks and cryptocurrencies presents an extensive array of research opportunities. Researchers can delve into diverse aspects such as market trends, volatility patterns, and the impact of external factors on financial instruments. This abundance of data allows for in-depth analyses that can contribute to a deeper understanding of market dynamics, risk management strategies, and the development of innovative investment approaches. 

\begin{figure}[h!tbp]
    \centering
    \subfloat[Pearson correlation coefficient]{\label{fig:correlation_distribution_stock_and_crypto}
        \includegraphics[width=0.49\linewidth,height=5cm,clip]{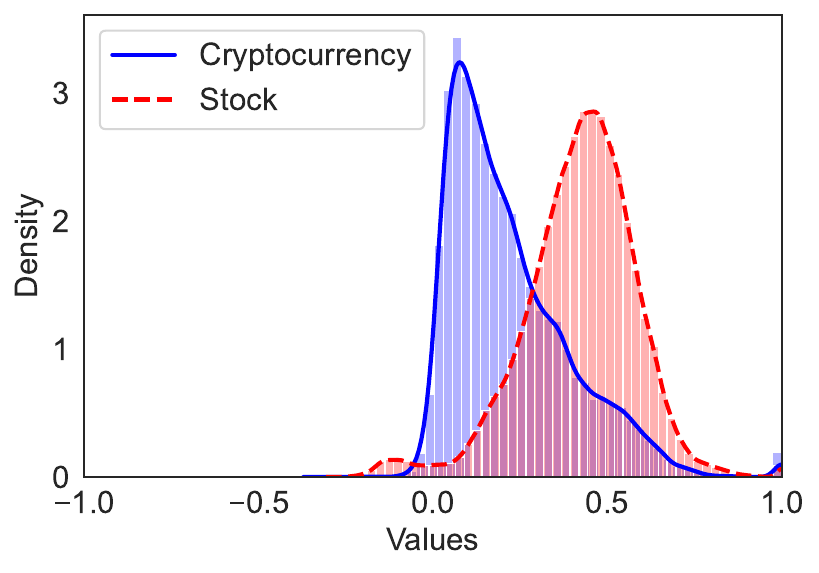}}
    \subfloat[Mutual information]{\label{fig:mi_distribution_stock_and_crypto}
        \includegraphics[width=0.49\linewidth,height=5cm,clip]{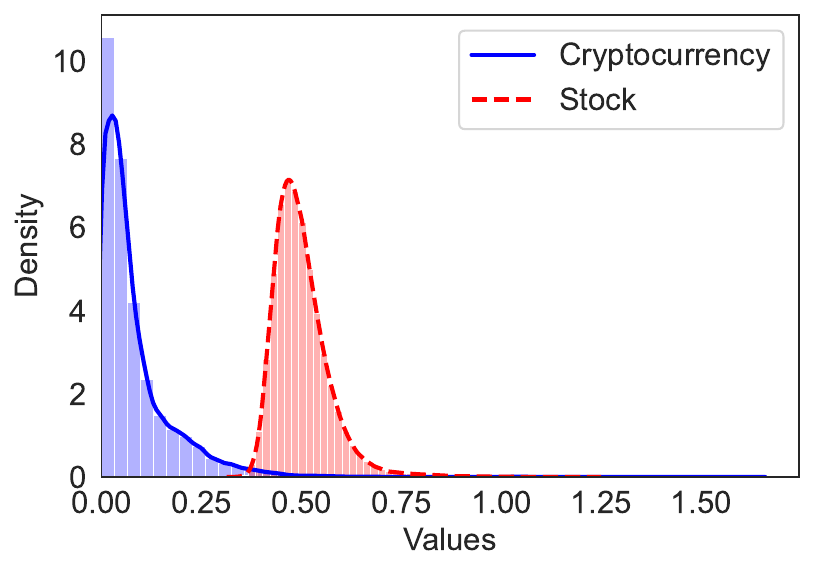}}
    \caption{Distribution of the Pearson correlation coefficient and mutual information in the cryptocurrency and stock market data from January 2019 to September 2022.}
    \label{fig:distribution_stock_and_crypto}
\end{figure}

The data collection for this study was performed for a period of more than five years, but the analyses presented in this paper are conducted on data from January 2019 to September 2022 in order to have a larger set of available crypto assets. We have included all 453 companies that continuously appear in the \acs{SP500} in this period, and for cryptocurrencies, we have selected the \ac{TOP 203} by market capitalization, excluding stable coins (a type of cryptocurrency fixed in price to another asset) and any cryptocurrencies with a market cap below 2M USD. It is noteworthy that the historical information gathered for stocks and cryptocurrencies spans two important periods that had a significant impact on the global financial markets: i) the COVID-19 pandemic and ii) the start of the armed confrontation between Russia and Ukraine. Figure \ref{fig:correlation_distribution_stock_and_crypto} shows the distribution of the Pearson correlation coefficient in the cryptocurrencies and stock market data in the observed period, while Figure \ref{fig:mi_distribution_stock_and_crypto} represents the distribution of the mutual information. It can be noticed that the distributions calculated for the stocks are different from those for cryptocurrencies as they contain larger values and have modes further away from zero.

%\afterpage{\clearpage}

\section{Results}\label{sec:res}

The goal of our research is to construct, analyze, and assess various diversified investment portfolios consisting either of stocks or cryptocurrencies. First, we provide several visualizations from the clustering and community detection methods applied to the entire period of observation. Then, we present the results from the portfolio diversification study, which is segmented by asset type: i) conventional assets (stocks) and ii) digital assets (cryptocurrencies). This distinction allows for a nuanced understanding of how different asset classes perform under varying market conditions. The examination of these assets is further refined by dividing the analysis into three distinct periods, each characterized by unique market dynamics and global events:

\begin{enumerate}
\item \textbf{Entire period} - The entire range of our dataset from January 2019 to September 2022. It provides a complete view of asset performance across various market cycles and conditions.

\item \textbf{Beginning of the COVID-19 pandemic} - The first year of the COVID-19 pandemic, i.e. from January 2020 to December 2020. This period is critical for understanding the impact of a global health crisis on asset behavior, as it presents unprecedented challenges and volatility in financial markets.

\item \textbf{Russian invasion of Ukraine} - The start of the Russian invasion of Ukraine, from August 2021 to July 2022. This time frame is significant for analyzing the effects of geopolitical tensions and uncertainties on asset performance, offering insights into how such events can influence market dynamics.
\end{enumerate}
Each segment of our analysis provides a detailed exploration of how various strategies for portfolio selection from the two asset types in different periods perform in terms of volatility and returns.

All portfolios were constructed on a window of one-year in-sample (training) data and evaluated on the following window of one-year out-of-sample (test) data. Moreover, for the analysis considering the entire observation period we use averaging by considering multiple windows slid by one month. Hence, the in-sample data is created from one-year periods (windows) beginning in January 2019 and ending in September 2021, while the out-of-sample data is created from one-year periods (windows) beginning in January 2020 and ending in September 2022.

Furthermore, we provide a comparison with the performance of portfolios that were constructed using randomly selected financial assets, the S\&P500 index (for stocks), and Top 203 assets (for cryptocurrencies). The portfolios comprised of randomly selected financial assets are constructed in the same manner as the other portfolios, namely by employing a time window that shifts by one month in the same time frame as the other portfolios, and their performance was averaged by doing 100 independent repetitions. We should also note that all portfolios constructed within our study are equally weighted, except the S\&P500 index and the Top 203 cryptocurrencies that take into account the market capitalization. Nevertheless, our findings provide a deep insight into the structure of financial markets but leave an open way for future studies, including weight optimization of the portfolio assets.

\subsection{Clustering and community detection}

\begin{figure}[hbt!]
    \centering
    \subfloat[Cryptocurrencies correlation]{\label{fig:cc_louvain}\includegraphics[width=0.49\linewidth]{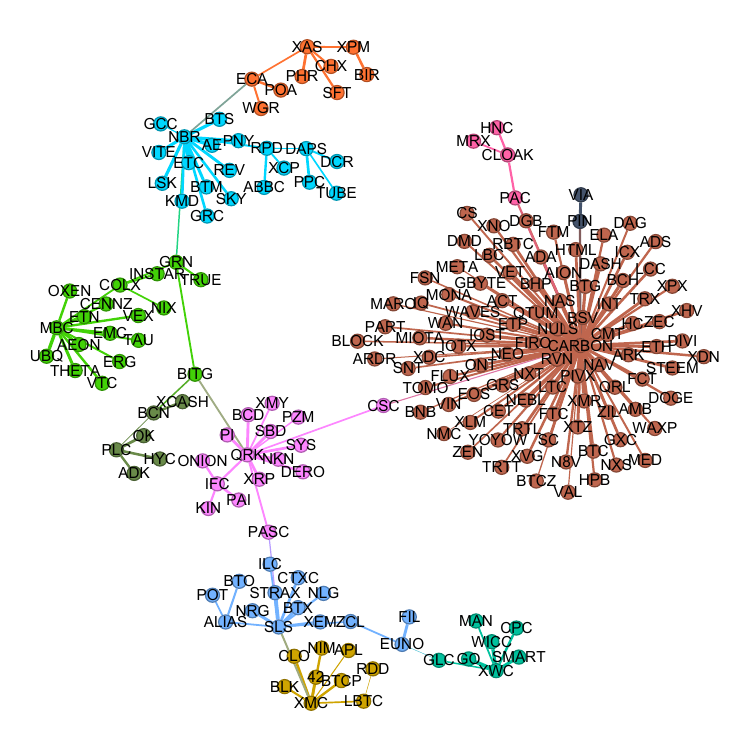}}
    \subfloat[Stock correlation]{\label{fig:cs_louvain}\includegraphics[width=0.49\linewidth]{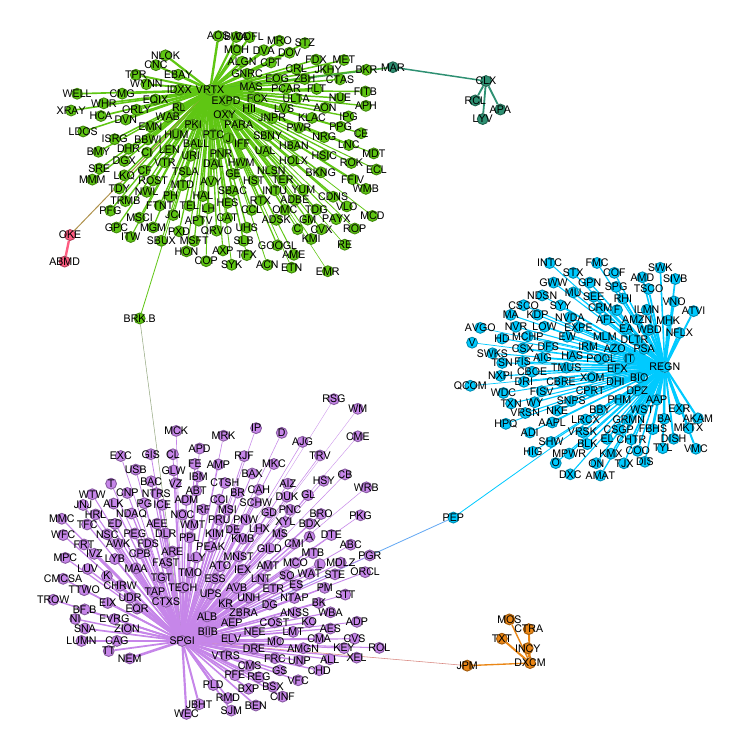}}\\
    \subfloat[Cryptocurrencies mutual information]{\label{fig:mc_louvain}\includegraphics[width=0.49\linewidth]{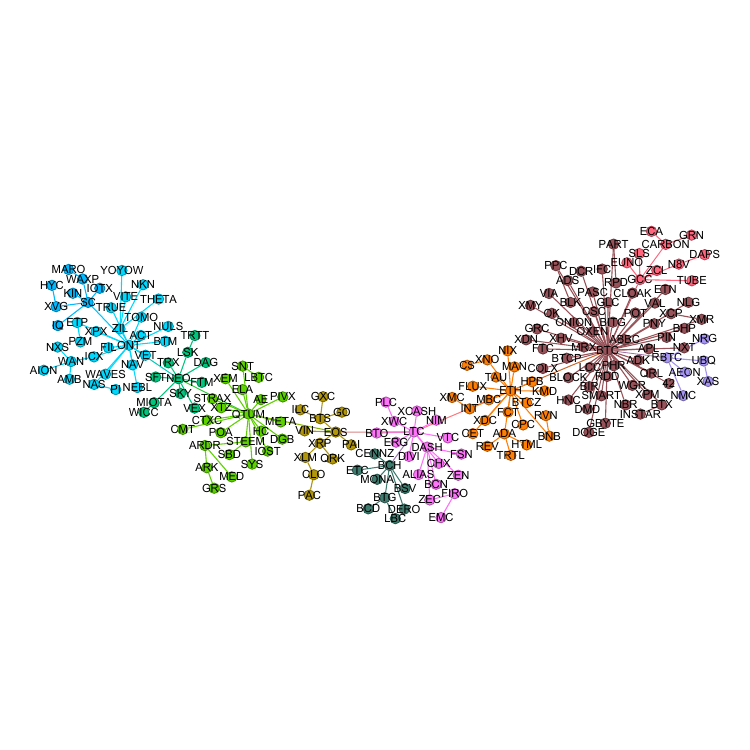}}
     \subfloat[Stock mutual information]{\label{fig:ms_louvain}\includegraphics[width=0.49\linewidth]{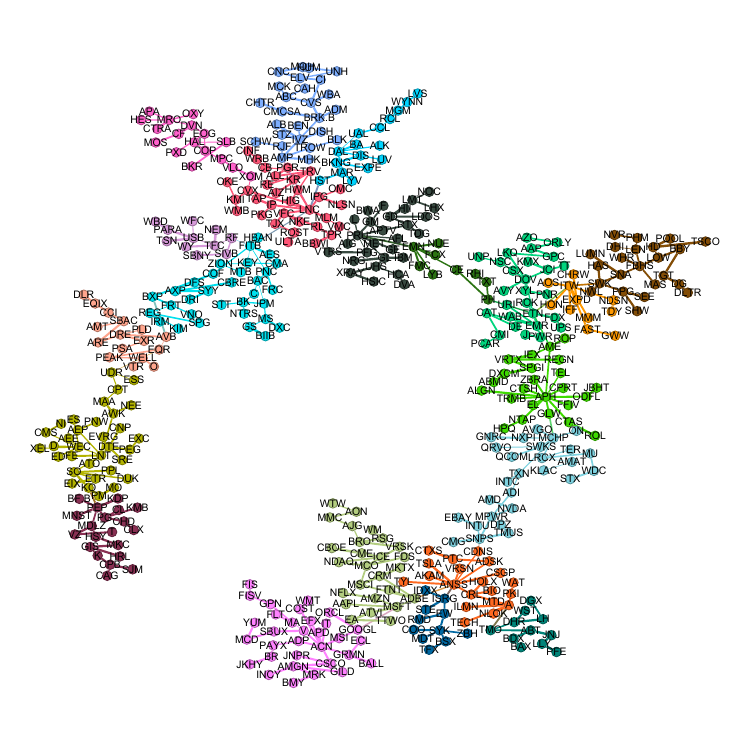}}
    \caption{Network communities of cryptocurrencies and stocks derived from historical prices from January 2019 to September 2022, using the Louvain algorithm. The networks are minimal spanning trees derived from the Pearson correlation coefficient and mutual information calculated from the normalized log-returns of asset prices.}
    \label{fig:louvain_communities}
\end{figure}

\begin{figure}[hbt!]
    \centering
    \subfloat[Cryptocurrencies correlation]{\label{fig:cc_aff}\includegraphics[width=0.49\linewidth,trim={9cm 5cm 10cm 5cm},clip]{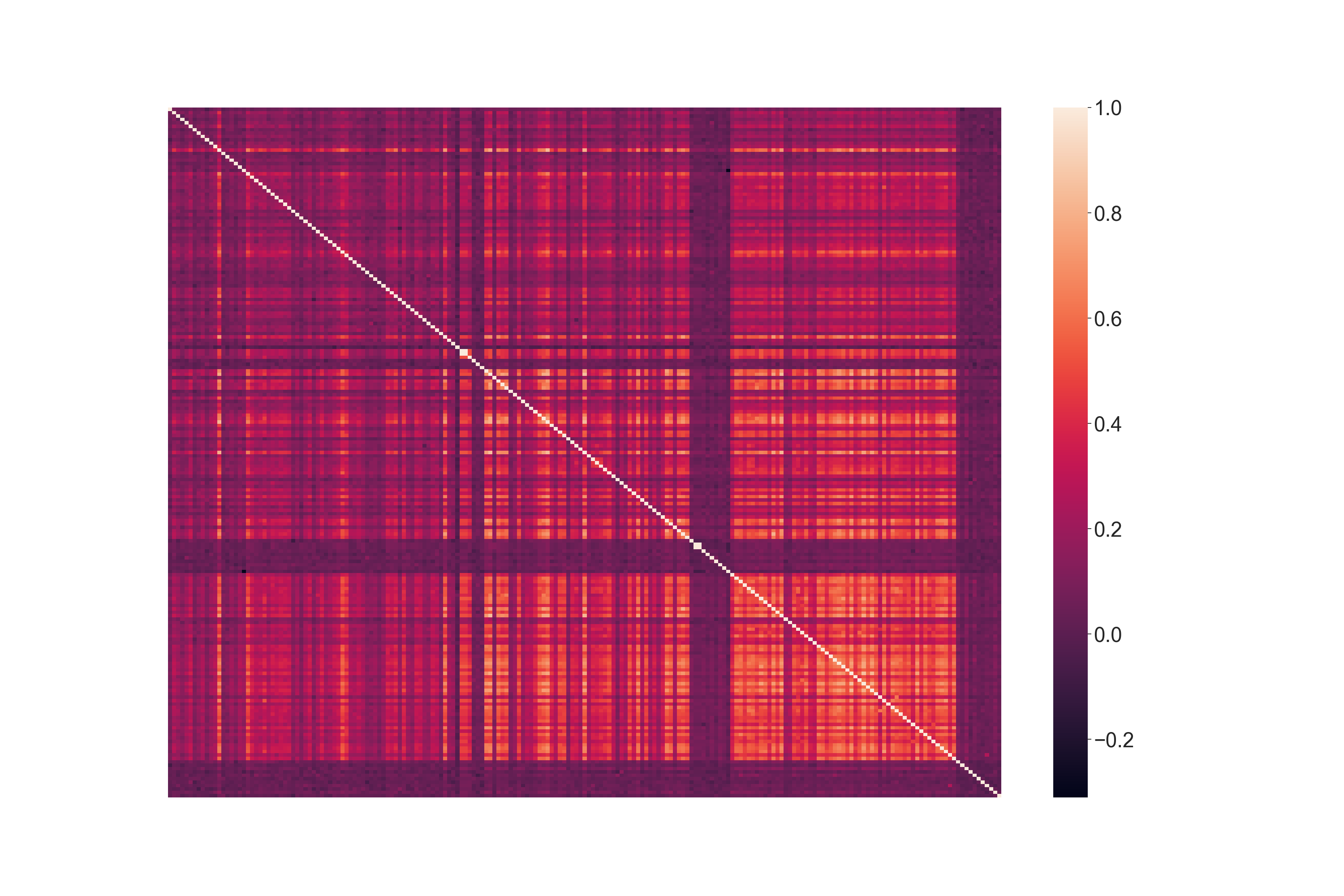}}
    \subfloat[Stock correlation]{\label{fig:cs_aff}\includegraphics[width=0.49\linewidth,trim={9cm 5cm 10cm 5cm},clip]{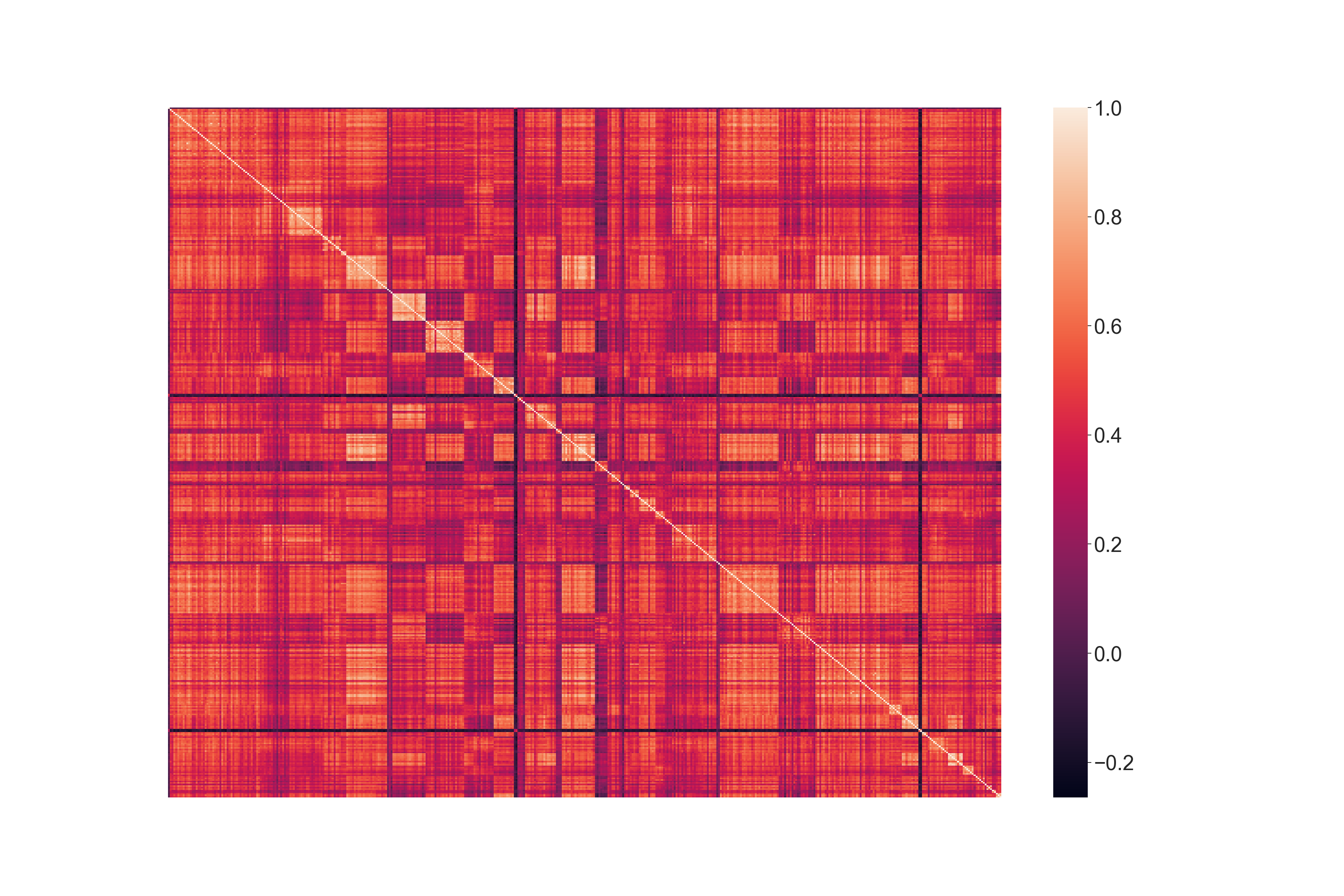}}\\
    %\par\medskip
    \subfloat[Cryptocurrencies mutual information]{\label{fig:mc_aff}\includegraphics[width=0.49\linewidth,trim={9cm 5cm 10cm 5cm},clip]{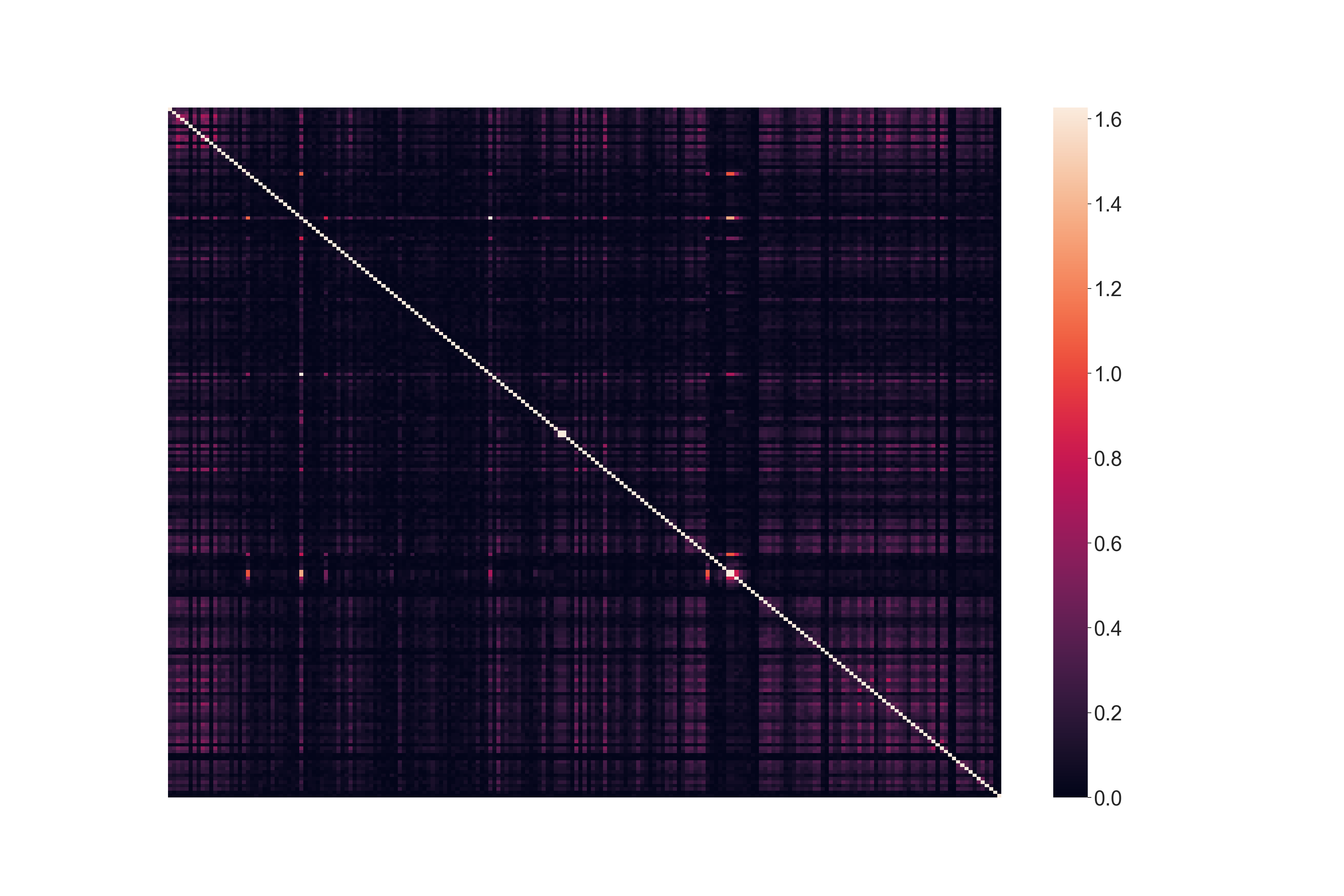}}
    \subfloat[Stock mutual information]{\label{fig:ms_aff}\includegraphics[width=0.49\linewidth,trim={9cm 5cm 10cm 5cm},clip]{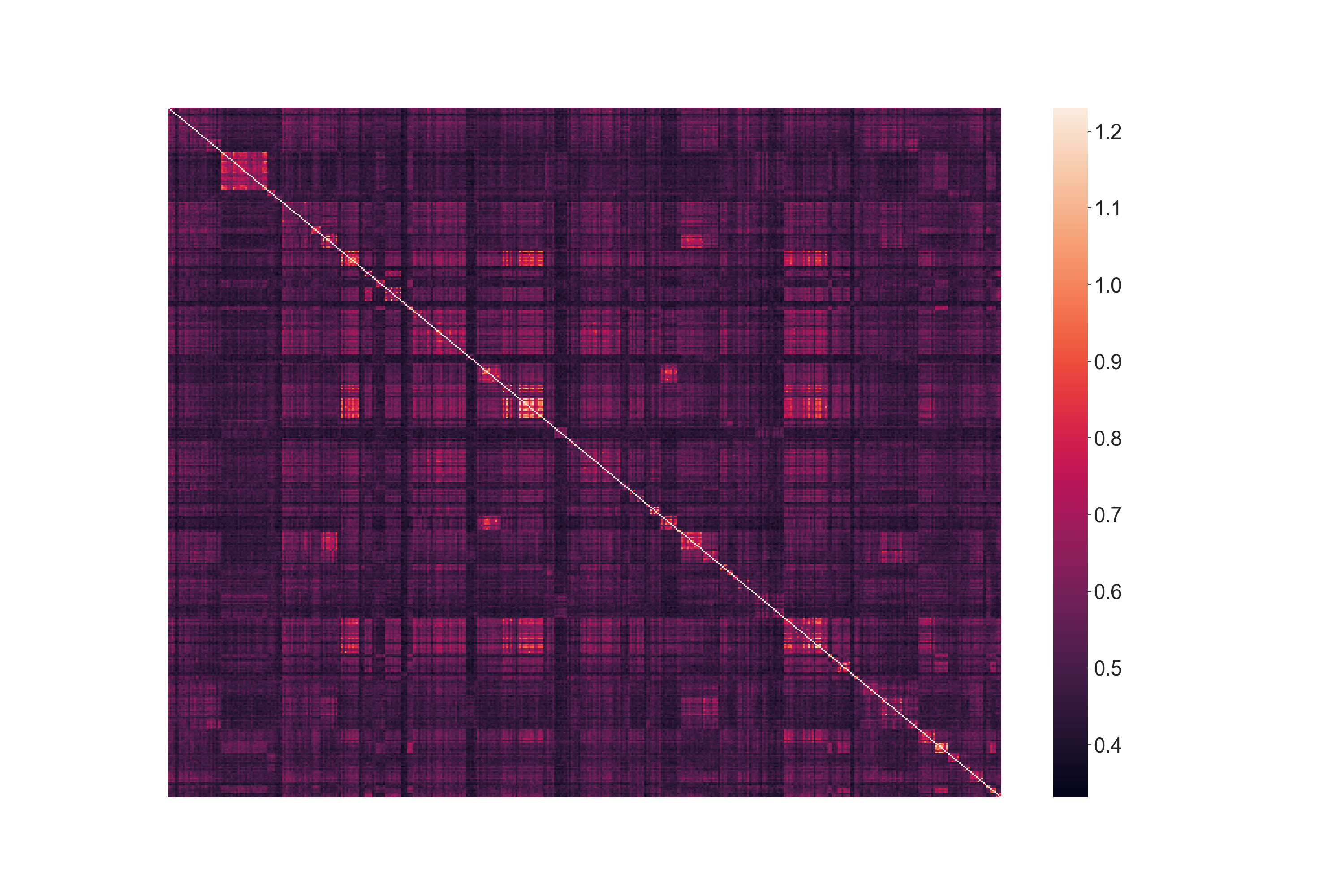}}
    \caption{Matrix visualizations of cryptocurrencies and stock markets derived from historical prices from January 2019 to September 2022, clustered by Affinity propagation. The matrices are derived from the Pearson correlation coefficient and mutual information calculated from the normalized log-returns of asset prices.}
    \label{fig:affinity_communities}
\end{figure}

In this section, we present the network communities identified using Louvain and Affinity propagation. 
The results with Louvain reveal the intricate community structures obtained using the MST market representation, while Affinity propagation provides an alternative viewpoint, illuminating additional patterns and dependencies in the data. The visualizations offer a fundamental insight into the community structure and relationships across the entire period of observation. However, the analysis of the clusters and communities is not our main goal in this study, as we are more focused on their utilization for portfolio construction.

The Louvain method was applied to the minimal spanning tree obtained from the distance matrices, as the direct application to the original correlation and mutual information matrices was not successful. Figure \ref{fig:louvain_communities} presents the communities detected using Louvain on both stocks and cryptocurrency data from the entire period. We observe that the communities identified using the mutual information relationship seem more logical than those with the Pearson correlation coefficient. In the network of stocks, the tree is deeper, and the communities contain more similar companies, while among the cryptocurrencies, the community hubs, such as \ac{BTC}, \ac{LTC}, and \ac{ETH}, seem more natural.

The affinity propagation algorithm performed well with negative weights, so the correlation matrix and the mutual information matrix were used in their original forms without any transformation. Figure \ref{fig:affinity_communities} presents the obtained communities in the stocks and cryptocurrency data from the entire period. Similarly, as in Figure \ref{fig:distribution_stock_and_crypto}, the values for cryptocurrencies are distributed closer to zero than for stocks.

%\afterpage{\clearpage}
\subsection{Stock portfolios}

In this section, we elaborate on the results for stock portfolios, starting with the whole observation period and then further investigating two critical periods, the beginning of the COVID-19 pandemic and the Russian invasion of Ukraine.

\subsubsection{Entire period}
\begin{figure}[hbt!]
    \centering
    \subfloat[All examined portfolios]{\label{fig:stock_all_obtained_portfolios}\includegraphics[width=0.49\linewidth]{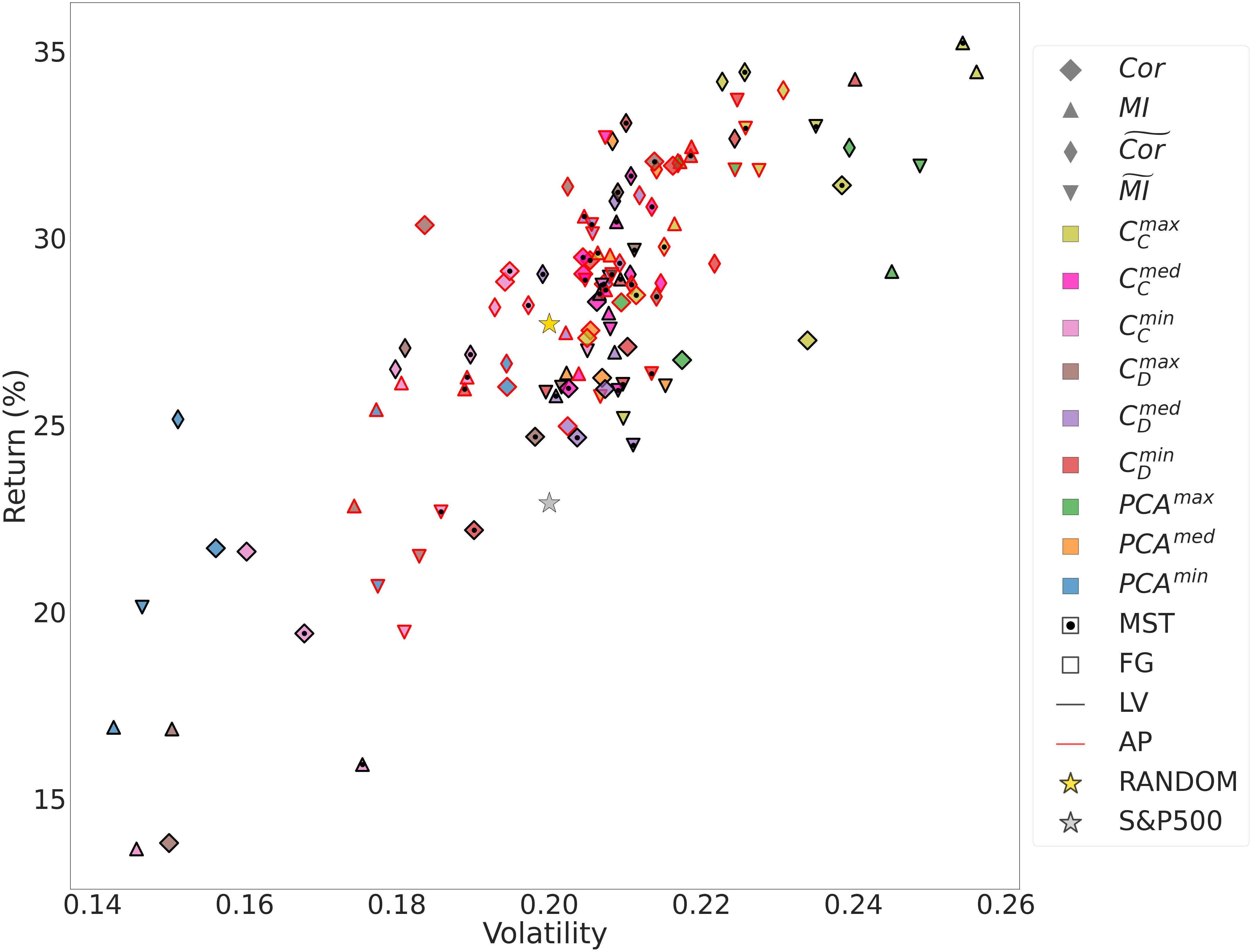}}
    \subfloat[Top 10 portfolios by return]{\label{fig:stock_best_by_return_stock_full_period}\includegraphics[width=0.49\linewidth]{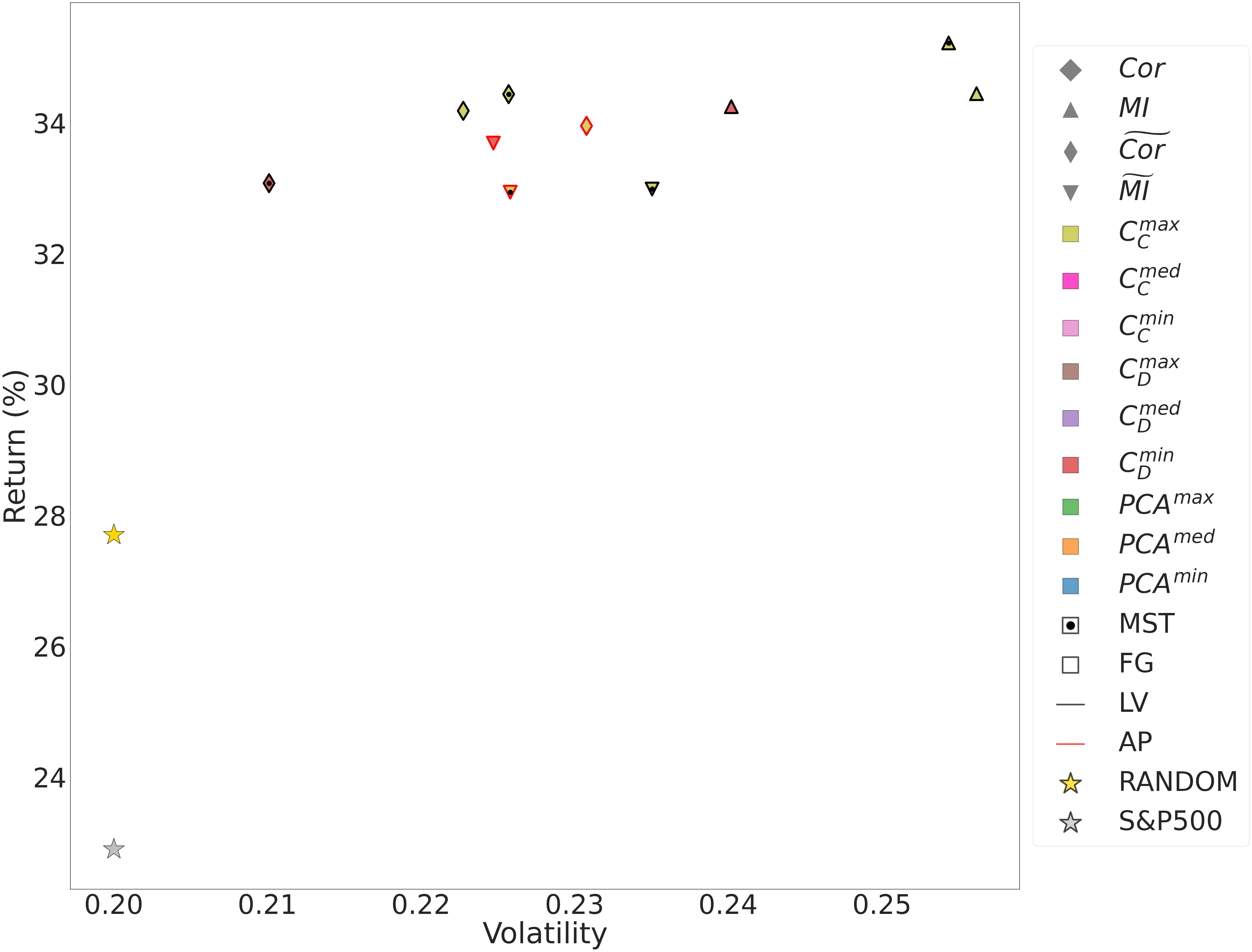}}\\
    \subfloat[Top 10 portfolios by volatility]{\label{fig:stock_best_by_volatility_stock_full_period}\includegraphics[width=0.49\linewidth]{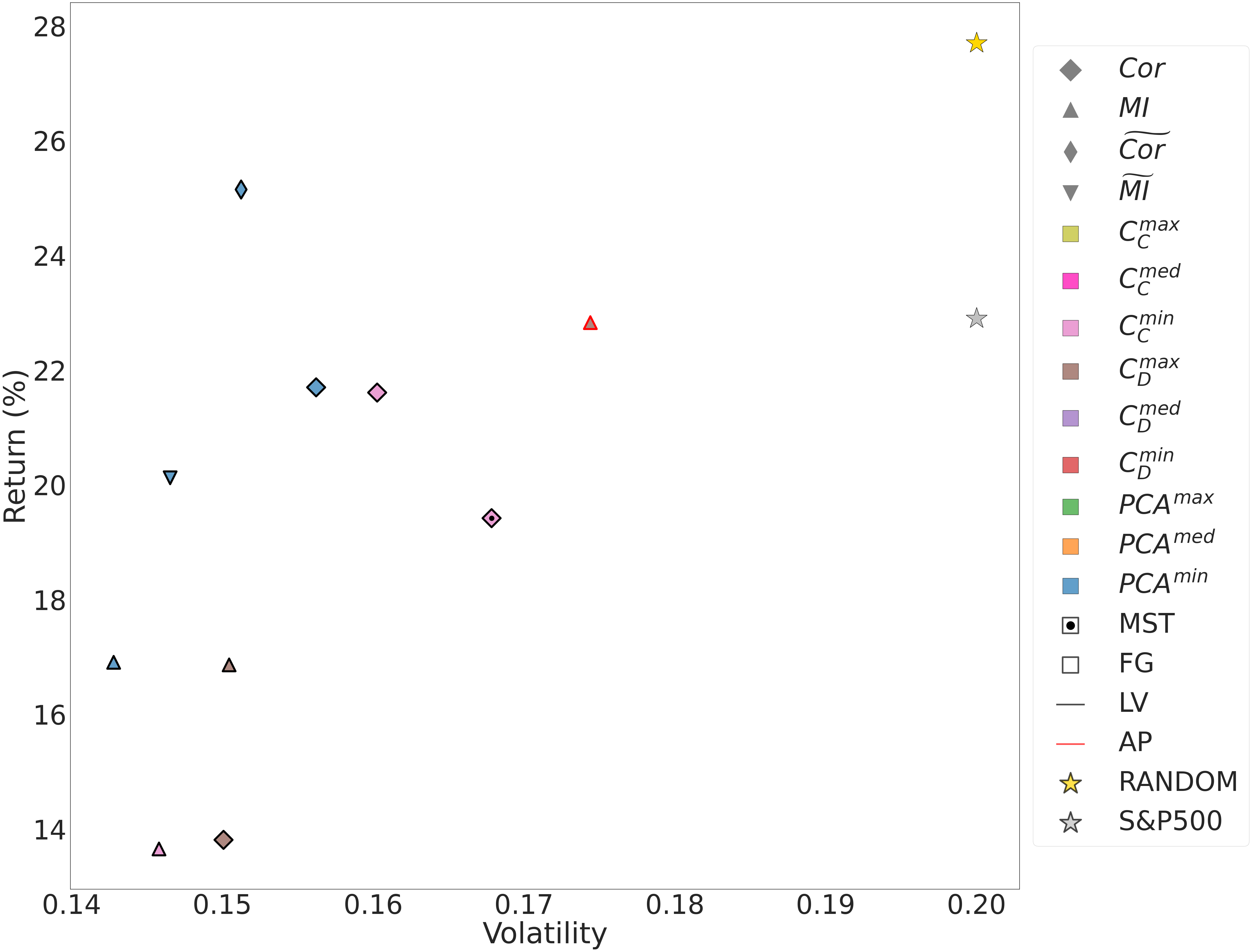}}
     \subfloat[Top 10 portfolios by return/volatility]{\label{fig:stock_best_by_return_volatility_ratio_stock_full_period}\includegraphics[width=0.49\linewidth]{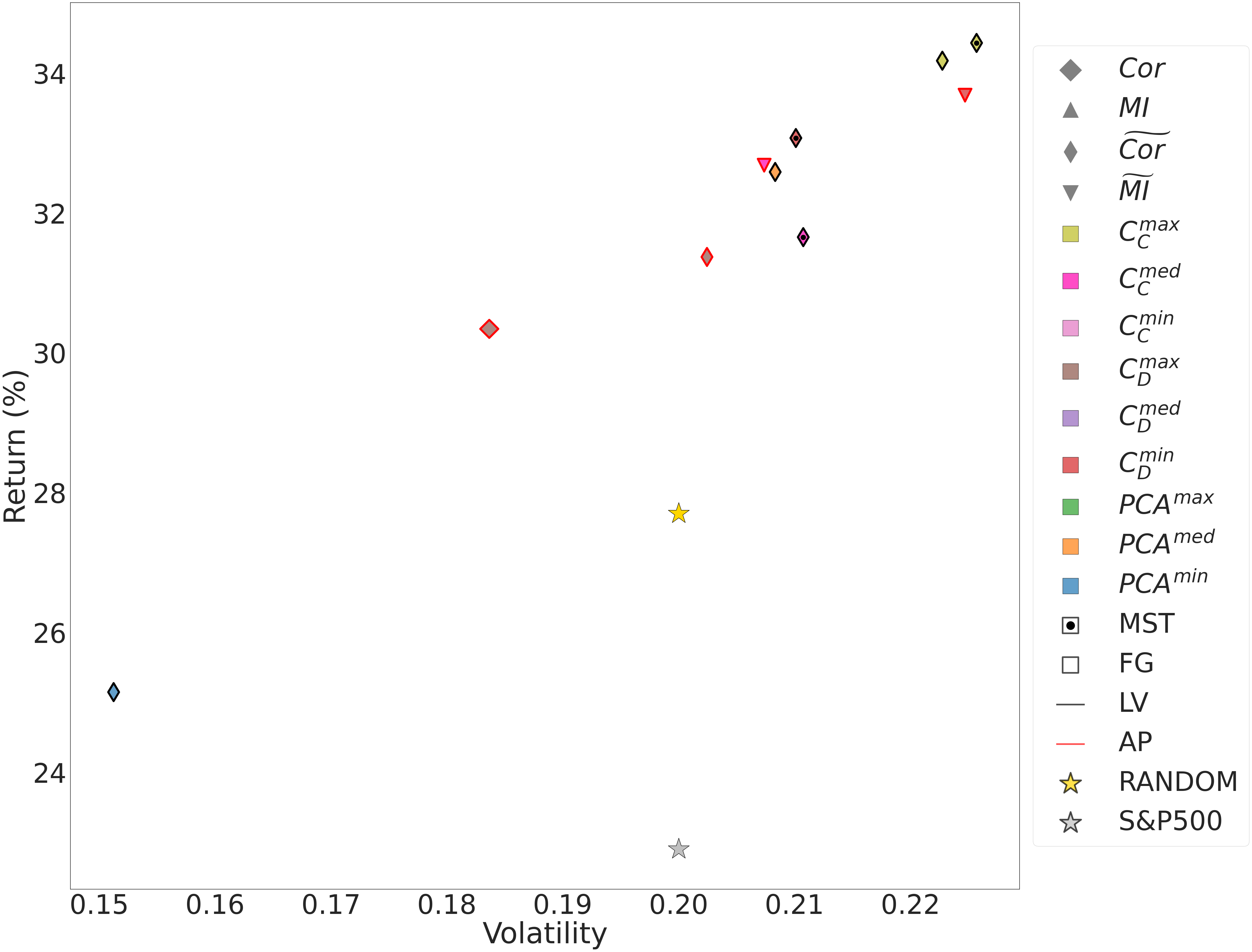}}
    \caption{Return vs. volatility of the examined investment portfolios composed of stocks during the entire observed period, from January 2019 to August 2022. The symbols represent: Cor – Correlation, MI – Mutual information, ($\sim$) – co-occurrence, $C_C$ – Closeness centrality, $C_D$ – Degree centrality, PCA – Principal component analysis, FG – Full graph, MST – Minimum spanning tree, LV – Louvain, and AP – Affinity propagation. A detailed description of all symbols is provided in Figure~\ref{fig:port_meth}.}
    \label{fig:stock_portfolios_full_period}
\end{figure}

First, in Figure \ref{fig:stock_portfolios_full_period}, we illustrate the returns of the generated portfolios as a function of their volatility for the entire period, which lays bare the intricate relationship between risk and return. The return/volatility ratio can be seen as a Sharpe ratio \cite{sharpe1994sharpe} with a zero risk-free rate, and it is of high importance in the assessment of portfolio strategies. 
A critical evaluation involves contrasting our portfolios with baselines, like \ac{RANDOM} and the \acs{SP500} index. Delving deeper, we scrutinized the top 10 portfolios, categorizing them based on three key metrics: return, volatility, and return/volatility. This analysis serves to highlight portfolios that not only offer high returns but also manage risk effectively. Portfolios excelling in these metrics potentially represent the most balanced and prudent investment choices within the observed period.

The top 10 portfolios in terms of return are shown in Figure \ref{fig:stock_best_by_return_stock_full_period}. We find that 7 out of 10 high-return portfolios are derived using the co-occurrence matrices, and they usually achieve lower volatility than the other. More importantly, our findings reveal that 7 out of the top 10 portfolios are a product of Louvain. This highlights its effectiveness in clustering financial assets in a manner that maximizes returns, suggesting its utility in identifying lucrative investment opportunities. The results for all possible portfolio construction strategies are given in Table~\ref{tab:stock_three_return_sme}. We can observe some patterns across different metrics used in portfolio construction. A notable trend is seen with the PCA metric, where portfolios comprising assets with minimal PCA values have lower returns compared to those with medial or maximal PCA values. 
%This suggests that while minimal PCA values might indicate lower risk, they are also associated with reduced returns. 
Similarly, portfolios built using the highest values of degree centrality (FG) usually show less favorable returns than those constructed around their minimal and medial values. We should note that in a distance network, a high degree centrality means that the assets are less related (more distant) from the other assets.
On the other hand, portfolios composed of assets with higher closeness centrality (MST) tend to have higher returns. With closeness centrality (FG) and degree centrality (MST), we have not observed clear trends. 

Figure \ref{fig:stock_best_by_volatility_stock_full_period} presents the 10 portfolios with the lowest volatility, where we have an equal number of portfolios coming from the mutual information matrix and the correlation matrix. Mirroring the trend observed in the high-return portfolios, 9 out of 10 low-volatility portfolios were derived using Louvain. This consistency underscores the algorithm's versatility in portfolio construction, irrespective of the primary objective, be it maximizing returns or minimizing volatility. Further analysis reveals a noteworthy correlation between the centrality of financial assets in the network and their volatility. Portfolios built with less central nodes positioned towards the network's periphery exhibit lower volatility. When we refer to peripheral or less central nodes in general, we mean nodes with lower PCA values, lower closeness centrality, and higher degree centrality on the full graph. For degree centrality calculated on MST, the notion of centrality is a little ambiguous because it will be higher for nodes having a larger distance from their neighbors, but the hubs have more links that make their degree larger. The inverse relationship between an asset's centrality and volatility is a crucial insight observed in several previous studies \cite{pozzi2013spread,onnela2002dynamic}. It suggests that peripheral financial assets, often overlooked in favor of more central ones, can offer a haven of stability in turbulent markets. Table  \ref{tab:stock_three_volatility_sme} corroborates this pattern, showcasing a range of portfolios along with their respective volatilities. These portfolios exemplify how lesser-known, peripheral financial assets can be instrumental in constructing lower-risk investment strategies. These findings not only validate our methodologies but also pave the way for a more nuanced understanding of risk management in portfolio construction, emphasizing the potential of peripheral assets in achieving stability. 

% Tables \ref{tab:stock_full_period_sharp_ratio} for investment portfolios composed of 25 stocks, Table \ref{tab:sharp_ratio_stock_15} for investment portfolios composed of 15 stocks, and Table \ref{tab:stock_sharp_ratio_35} for investment portfolios composed of 35 stocks show the results of the entire period analysis.

Figure \ref{fig:stock_best_by_return_volatility_ratio_stock_full_period} delivers a comprehensive view of portfolios distinguished by the best return/volatility ratio. These portfolios represent a balanced investment approach, blending desirable returns with manageable volatility. The analysis shows a trend that portfolios excelling in this ratio predominantly consist of financial assets located either in the middle or the periphery of the constructed network. This finding reinforces the previously noted direct relationship between the centrality of a financial asset and its volatility, suggesting that assets with moderate to low centrality can contribute significantly to a balanced portfolio. Furthermore, it is intriguing that 7 out of 10 of these well-balanced portfolios are constructed using the correlation co-occurrence matrix, which highlights the effectiveness of co-occurrence matrices in portfolio construction as they provide additional insights. 

To test the robustness of our investment portfolio generation process, we assessed both bigger and smaller investment portfolios, respectively including 35 and 15 stocks. The results for the investment portfolios of 35 stocks are consistent and follow the same trend as was found in the 25 stocks investment portfolios. These trends are especially clear in terms of investment portfolio volatility. Thus, in Table \ref{tab:converted_stock_volatility_data_35_stocks}, which shows the volatility of the 35 stocks investment portfolios, we can observe that in addition to following the pattern that portfolios that are more centrally located in the community are characterized by lower volatility, the deviations from the general trend are even smaller. The more consistent outcomes may be attributed to the larger number of financial assets, which leads to better smoothening. Table \ref{tab:converted_stock_volatility_data_15_stocks} displays the volatility of investment portfolios comprising 15 stocks, which follow a similar pattern as the larger portfolios but with slightly more pronounced deviations from the general trend, as can be expected. Similarly, the portfolio returns shown in Table \ref{tab:converted_stock_return_data_35_stocks} and Table \ref{tab:converted_stock_return_data_15_stocks} follow a similar trend as before.

Finally, in Table \ref{tab:stock_full_period_sharp_ratio}, Table \ref{tab:stock_sharp_ratio_35}, and Table \ref{tab:sharp_ratio_stock_15}, we show the Sharpe ratios. The interpretation of the Sharpe ratios is more challenging as it represents the interplay between the return and volatility which results in more variable values. Therefore, we have also included the column and row averages. One can argue that the Louvain method with the correlation co-occurrence matrix shows the best performance across almost all selection metrics.

\subsubsection{Beginning of the COVID-19 pandemic}  
\begin{figure}[hbt!]
    \centering
    \subfloat[All examined portfolios]{\label{fig:stock_all_obtained_portfolios_covid}\includegraphics[width=0.49\linewidth]{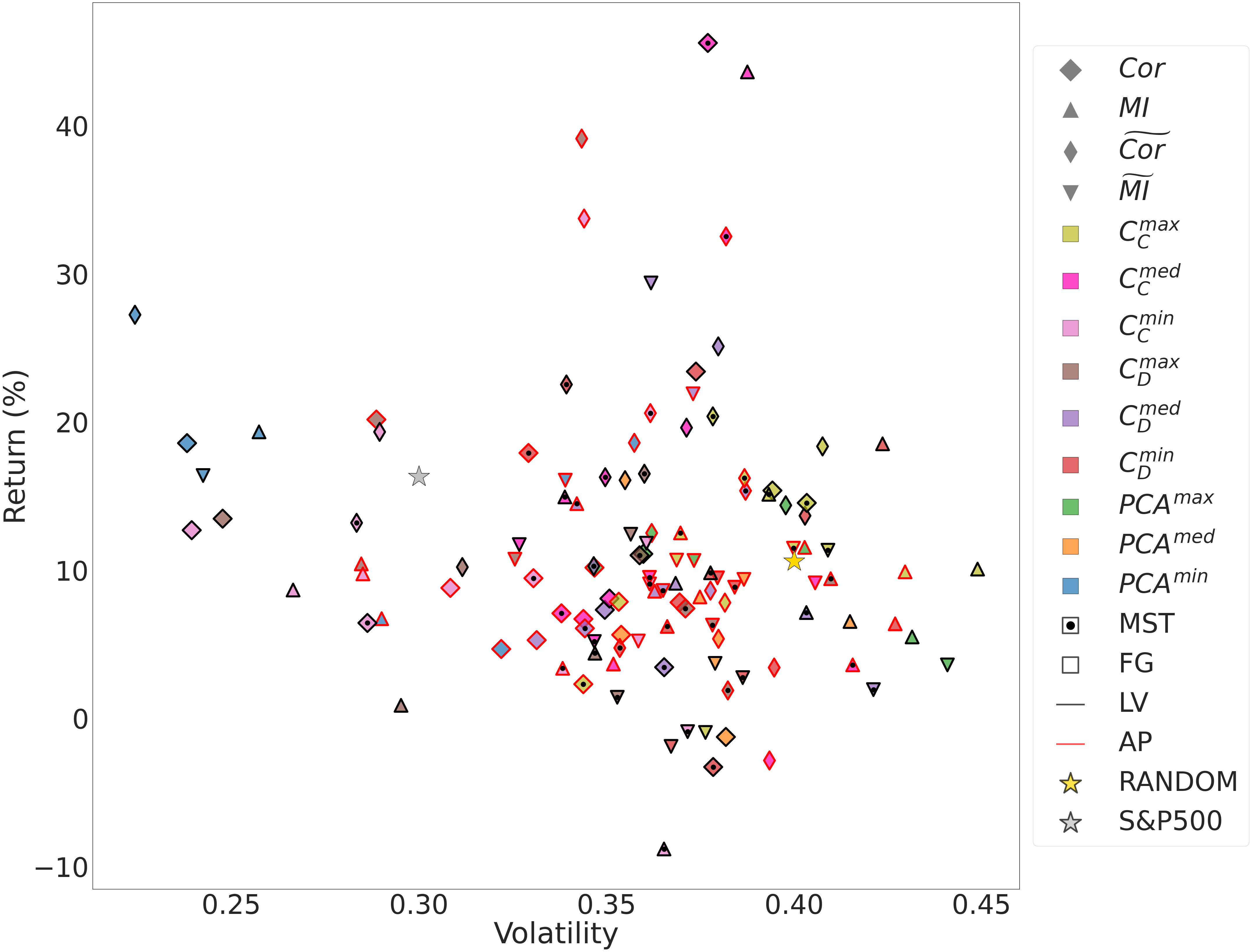}}
    \subfloat[Top 10 portfolios by return]{\label{fig:stock_best_by_return_stock_covid_period}\includegraphics[width=0.49\linewidth]{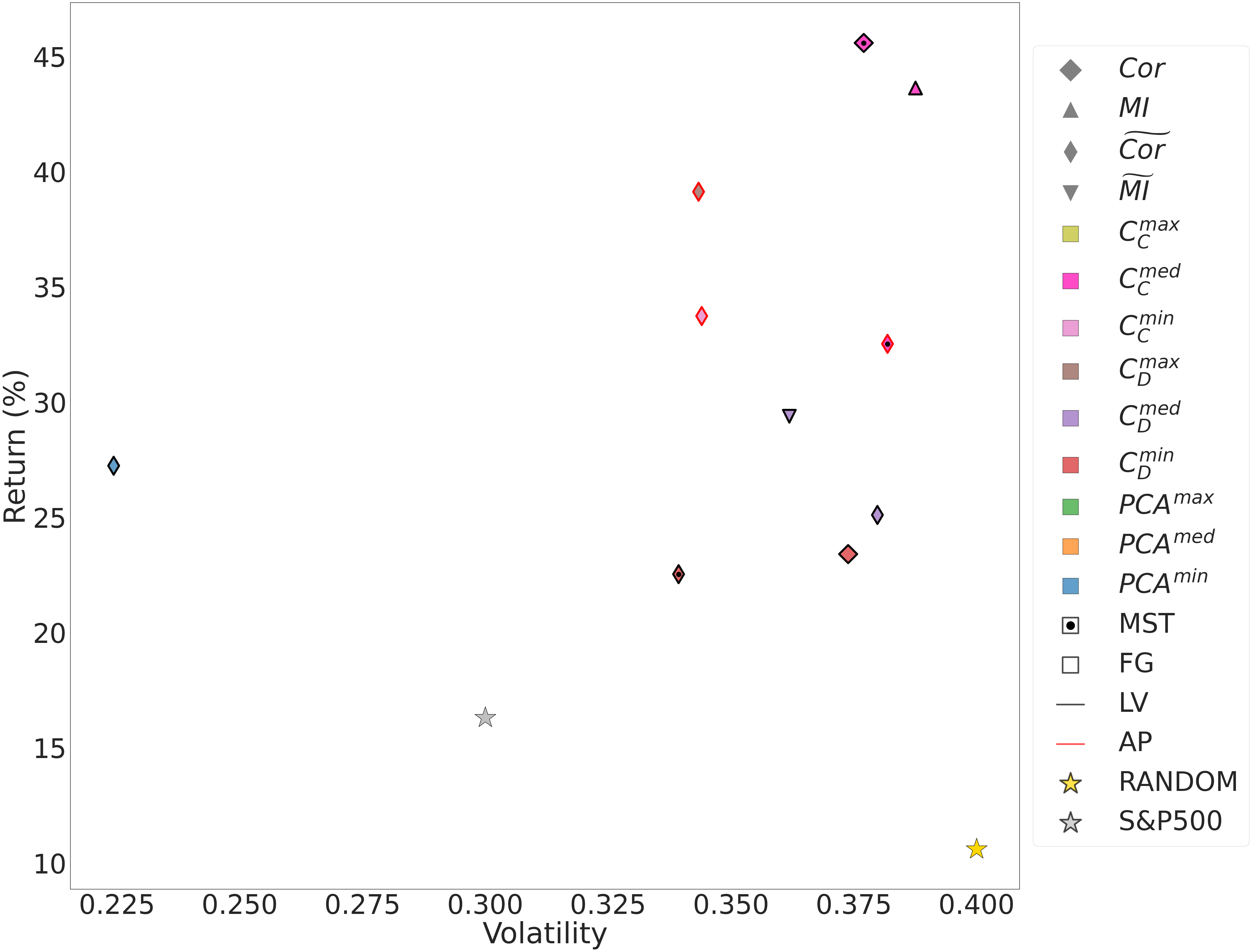}}\\
    \subfloat[Top 10 portfolios by volatility]{\label{fig:stock_best_by_volatility_stock_covid_period}\includegraphics[width=0.49\linewidth]{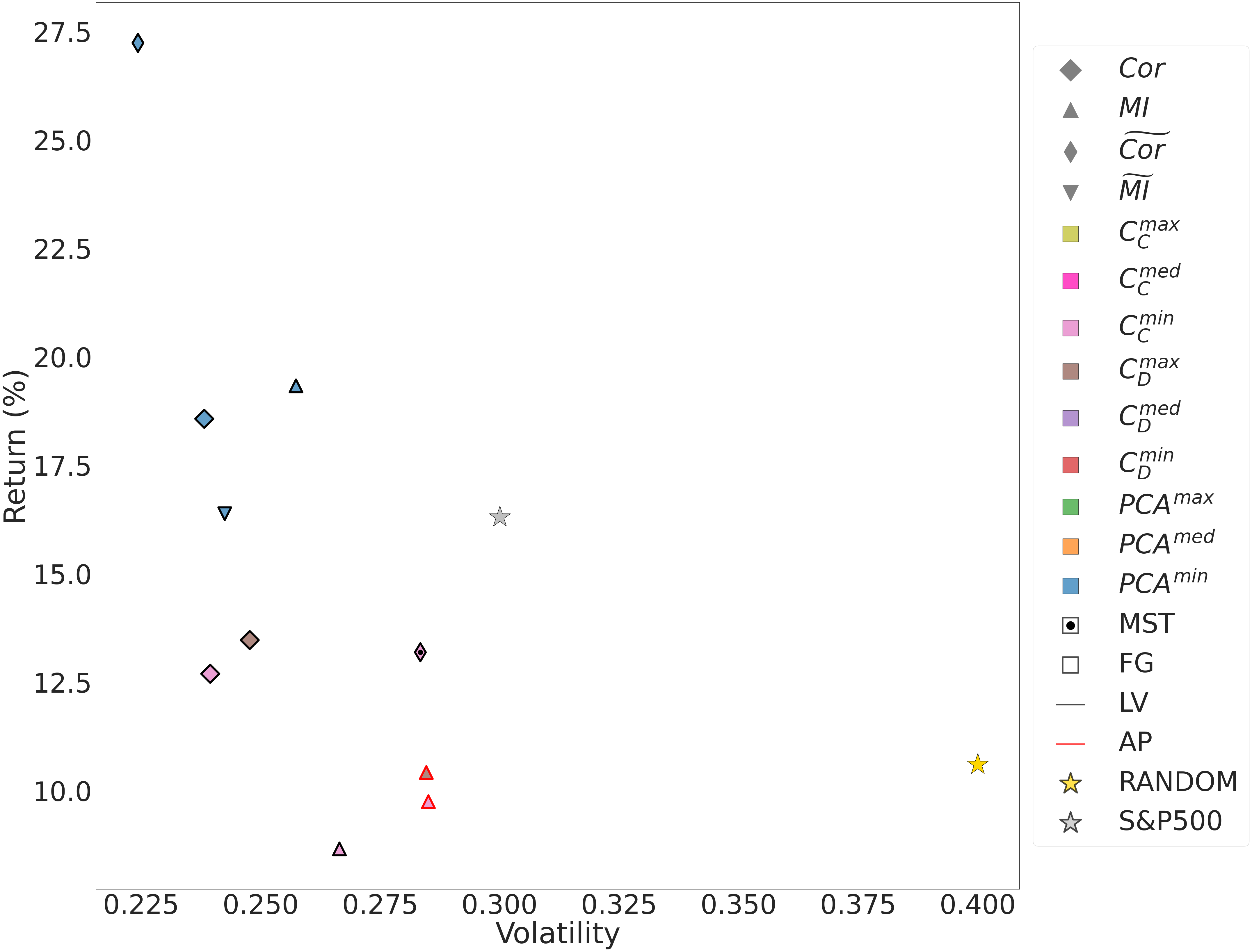}}
     \subfloat[Top 10 portfolios by return/volatility]{\label{fig:stock_best_by_return_volatility_ratio_stock_covid_period}\includegraphics[width=0.49\linewidth]{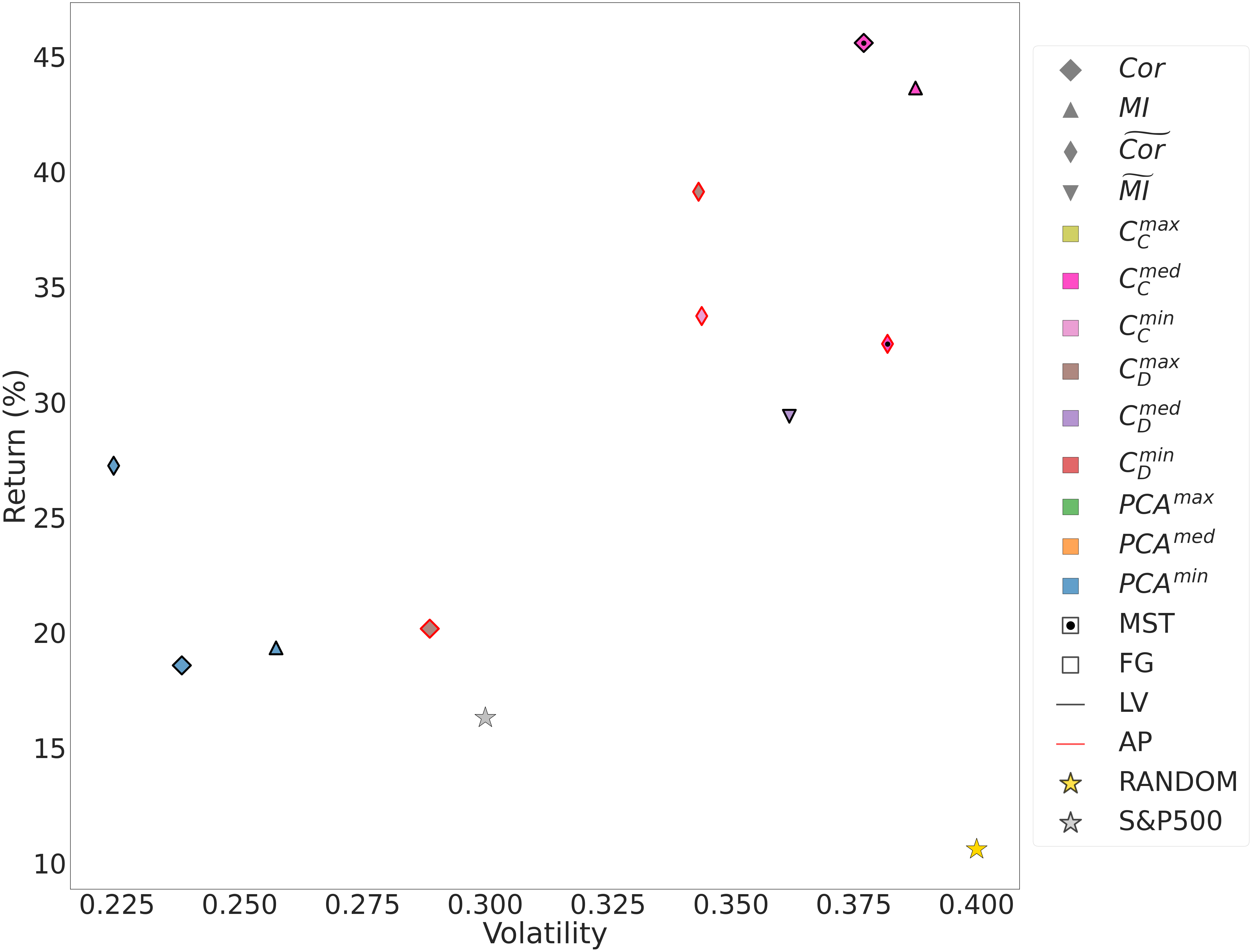}}
    \caption{Return vs. volatility of the examined stocks portfolios during the beginning of the COVID-19 pandemic, from January 2020 to December 2020. The symbols represent: Cor – Correlation, MI – Mutual information, ($\sim$) – co-occurrence, $C_C$ – Closeness centrality, $C_D$ – Degree centrality, PCA – Principal component analysis, FG – Full graph, MST – Minimum spanning tree, LV – Louvain, and AP – Affinity propagation. A detailed description of all symbols is provided in Figure~\ref{fig:port_meth}.}
    \label{fig:stock_portfolios_covid_period}
\end{figure}

For examining the effect of the beginning of the COVID-19 pandemic, we consider how portfolios trained on data from January 2019 until December 2019, perform during the period from January 2020 until December 2020 (see Figure \ref{fig:stock_portfolios_covid_period}).

According to the results from all considered portfolios, shown in Figure \ref{fig:stock_all_obtained_portfolios_covid}, we can notice that most of them are characterized by a relatively low or negative return, as well as high volatility. We can also note that a part of the portfolios that are built using the methods examined in this study still produce remarkable results compared to the baseline portfolios. The more detailed results given in Table \ref{tab:stock_covid_return}, Table \ref{tab:stock_covid_volatility}, and Table \ref{tab:stock_covid_sharp_ratio} show the return, volatility, and Sharpe ratio of all portfolios, which to some extent follow similar patterns as for the entire period. The majority of the portfolios are characterized by higher volatility than before, and more peripheral portfolios tend to provide lower volatility. On the other hand, more central portfolios generally achieve higher returns, except for portfolio selection based on PCA, where more peripheral portfolios typically yield both higher returns as well as lower volatility. It can be noticed that the Sharpe ratio for minimal PCA has the largest values on average, although it is the largest for Louvain with medial closeness (MST) and correlation.

To further examine the strategies for building portfolios in a crisis period such as the COVID-19 pandemic, we examine the top portfolios based on return, volatility and return/volatility. Figure \ref{fig:stock_best_by_return_stock_covid_period} shows the 10 best portfolios according to return, where it can be observed that all these portfolios achieve a higher return compared to the base portfolios and have lower volatility than the RANDOM portfolio. It is noticeable that the portfolio that achieves the highest return is built using Louvain and correlation with medial closeness (MST). Also, 6 out of 10 portfolios are built using the correlation co-occurrence matrix. Figure \ref{fig:stock_best_by_volatility_stock_covid_period} provides an overview of the portfolios characterized by the lowest volatility, which have lower volatility than the baseline portfolios and varying levels of return. One of the most interesting portfolios is the portfolio, which has the lowest volatility but achieves high return, which was created using the correlation co-occurrence matrix with Louvain and minimal PCA. Again, Louvain stands out with the largest participation, and all portfolios comprise more peripheral assets.

Figure \ref{fig:stock_best_by_return_volatility_ratio_stock_covid_period} presents the portfolios in the group characterized by the best return/volatility ratio. We can see that these portfolios have a better return than the baselines but not necessarily lower volatility. The portfolios that offer lower volatility also have lower returns, but still higher than the baseline portfolios. Again, the best portfolios in terms of return/volatility ratio are located in the middle or the periphery of the network, and the portfolio derived using the correlation co-occurrence matrix with Louvain and minimal PCA stands out. Figure \ref{fig:stock_best_by_return_volatility_ratio_stock_covid_time} shows the return over time of the top 5 portfolios by return/volatility ratio, where it can be seen how the baseline portfolios have a lower return during the entire period of observation.

\subsubsection{Russian invasion of Ukraine}
\begin{figure}[hbt!]
    \centering
    \subfloat[All examined portfolios]{\label{fig:stock_all_obtained_portfolios_war}\includegraphics[width=0.49\linewidth]{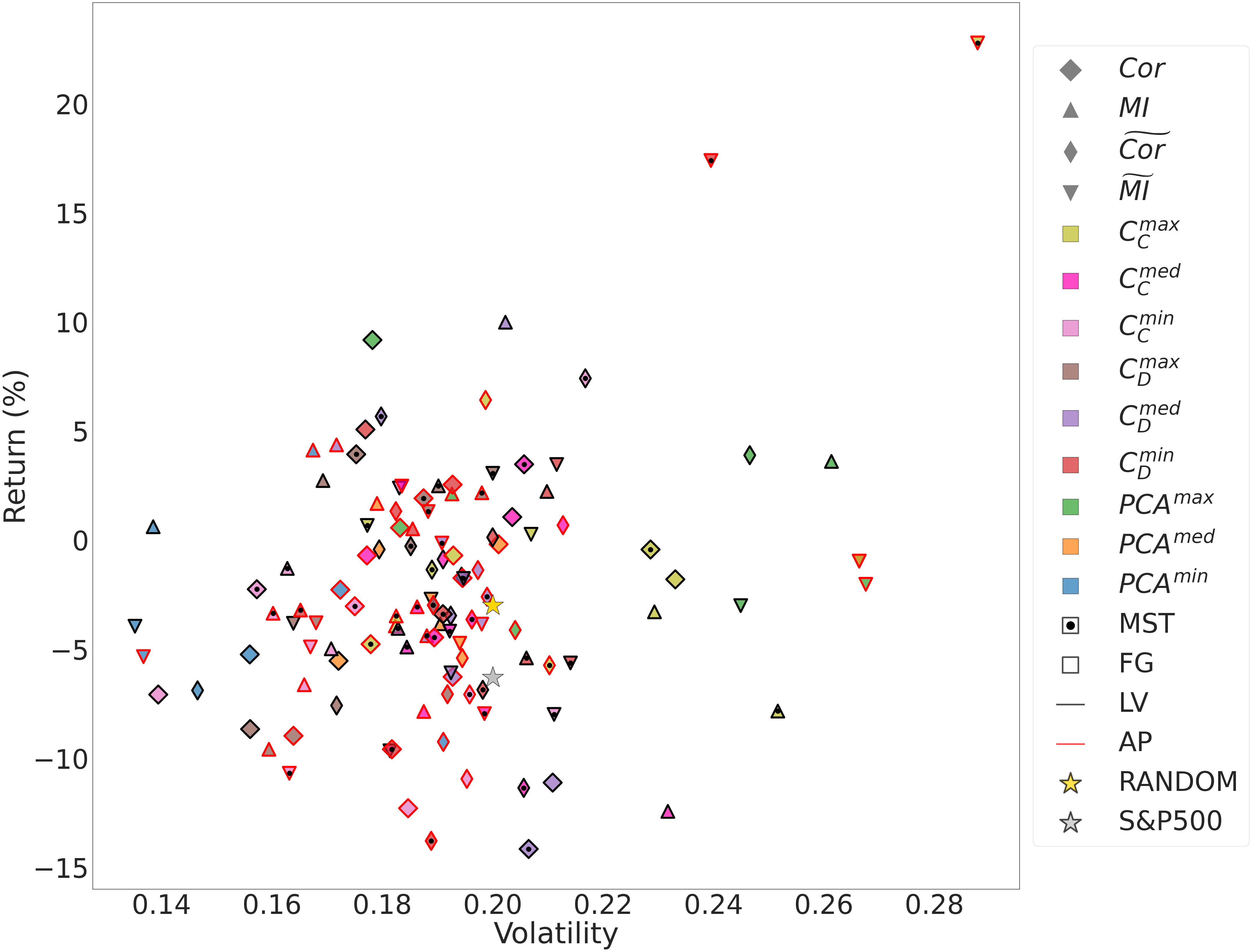}}
    \subfloat[Top 10 portfolios by return]{\label{fig:stock_best_by_return_stock_war_period}\includegraphics[width=0.49\linewidth]{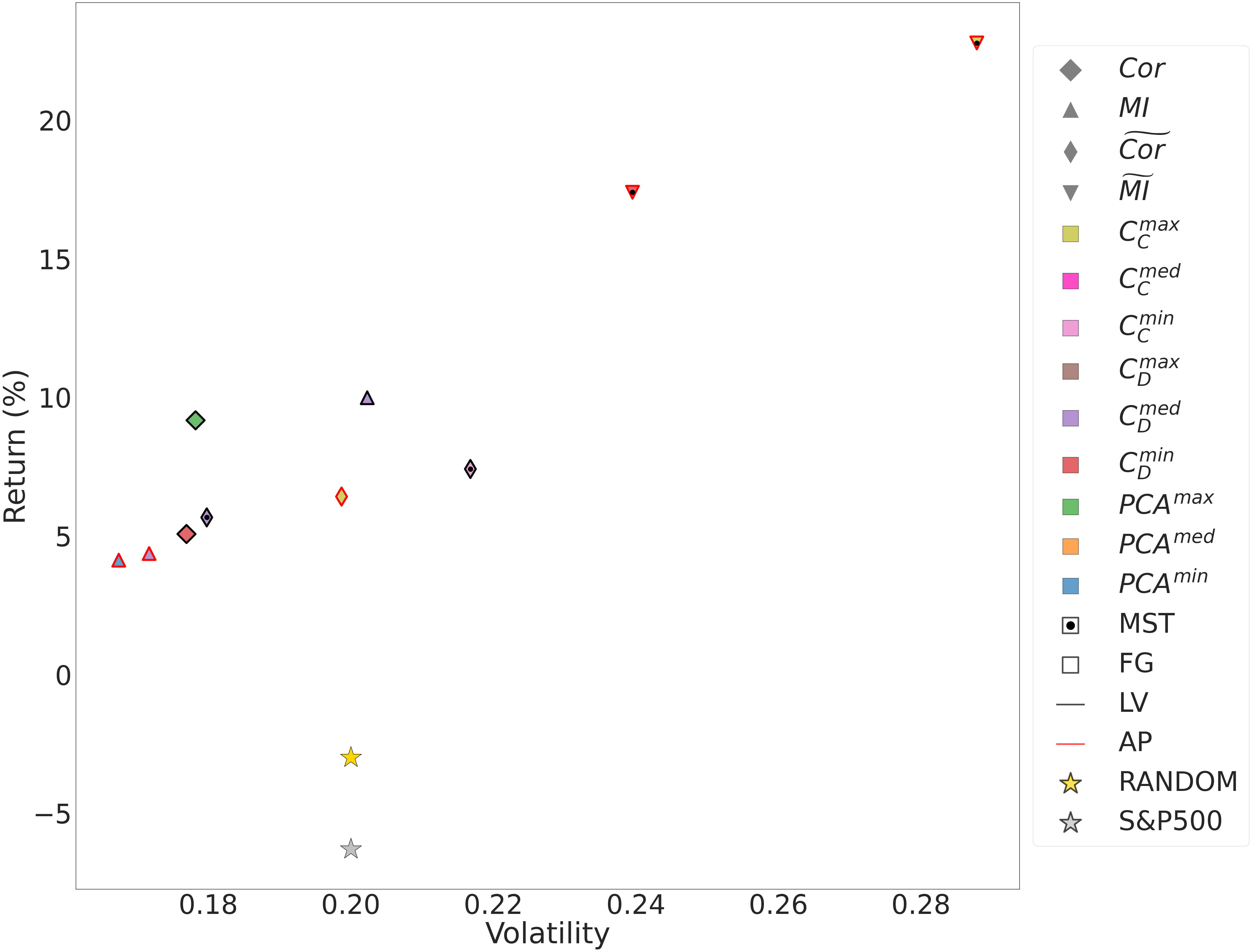}}\\
    \subfloat[Top 10 portfolios by volatility]{\label{fig:stock_best_by_volatility_stock_war_period}\includegraphics[width=0.49\linewidth]{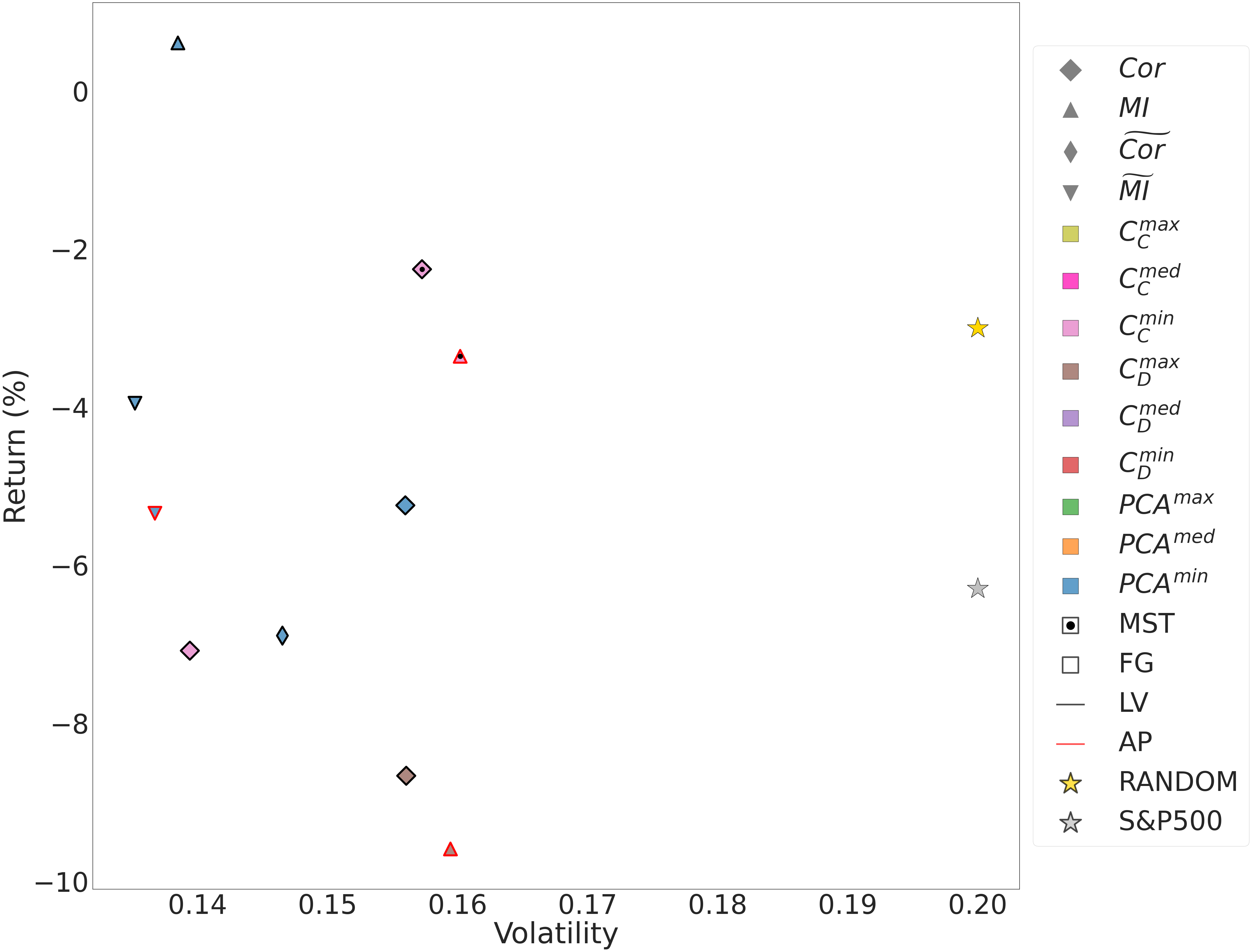}}
     \subfloat[Top 10 portfolios by return/volatility]{\label{fig:stock_best_by_return_volatility_ratio_stock_war_period}\includegraphics[width=0.49\linewidth]{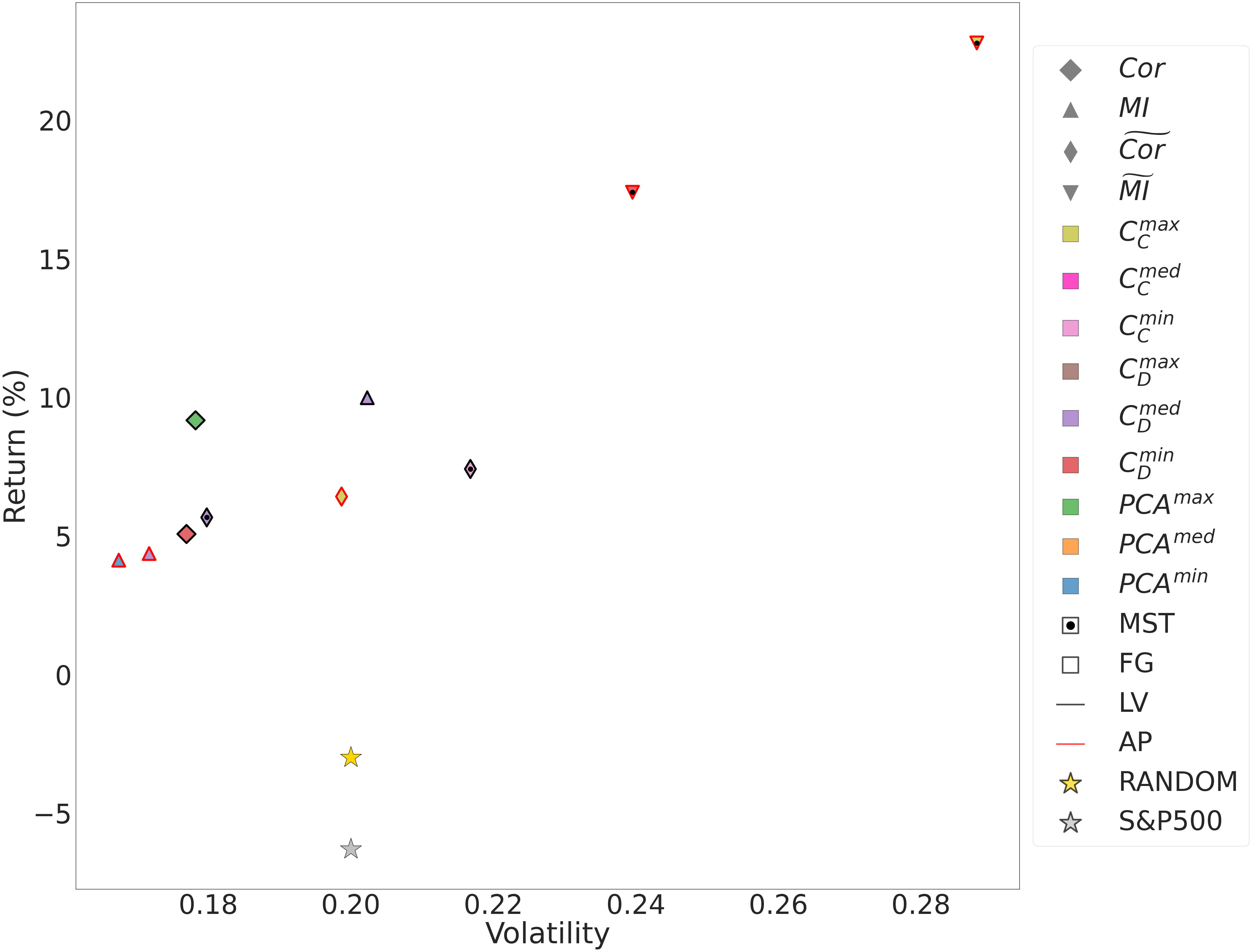}}
    \caption{Return vs. volatility of the examined stocks portfolios during the beginning of the Russian invasion of Ukraine, from August 2021 to July 2022. The symbols represent: Cor – Correlation, MI – Mutual information, ($\sim$) – co-occurrence, $C_C$ – Closeness centrality, $C_D$ – Degree centrality, PCA – Principal component analysis, FG – Full graph, MST – Minimum spanning tree, LV – Louvain, and AP – Affinity propagation. A detailed description of all symbols is provided in Figure~\ref{fig:port_meth}.}
    \label{fig:stock_portfolios_war_period}
\end{figure}

Another crisis period that was examined, is the start of the Russian invasion of Ukraine, from August 2021 to July 2022. Within this period, there is an overlap with the COVID-19 pandemic, so there are simultaneously two crises that wreaked havoc all over the world. Figure \ref{fig:stock_portfolios_war_period} shows a comprehensive overview of the constructed portfolios and their characteristics during this period.

\begin{figure}[hbt!]
    \centering
     \subfloat[Beginning of the COVID-19 pandemic]{\label{fig:stock_best_by_return_volatility_ratio_stock_covid_time}\includegraphics[width=0.49\linewidth]{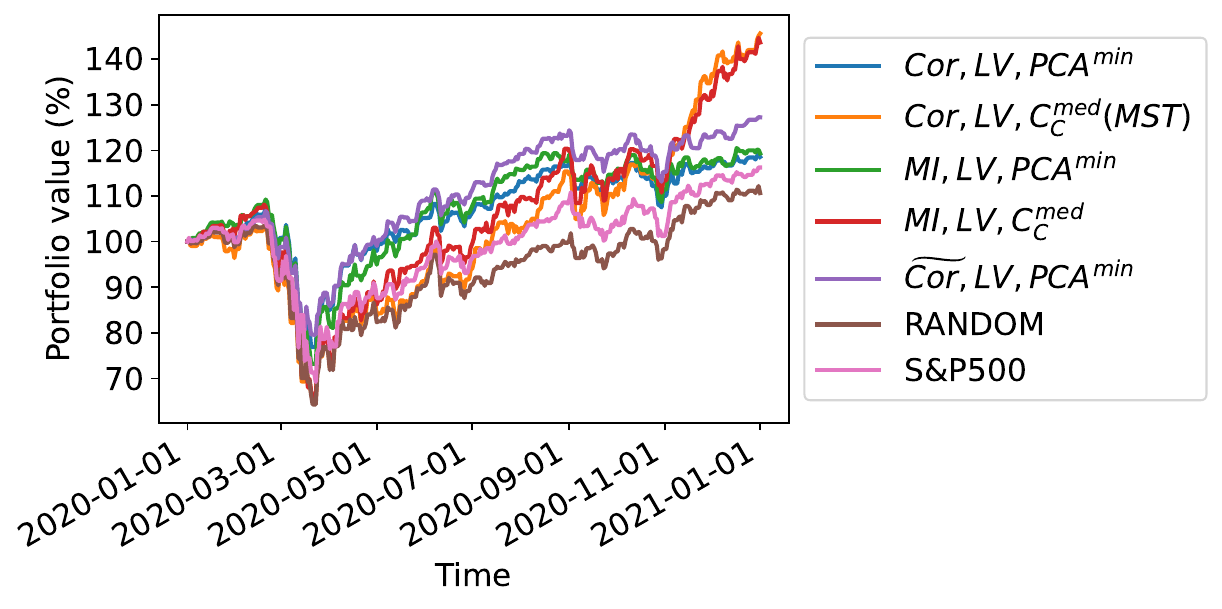}}
     \subfloat[Beginning of the Russian invasion of Ukraine]{\label{fig:stock_best_by_return_volatility_ratio_stock_war_time}\includegraphics[width=0.49\linewidth]{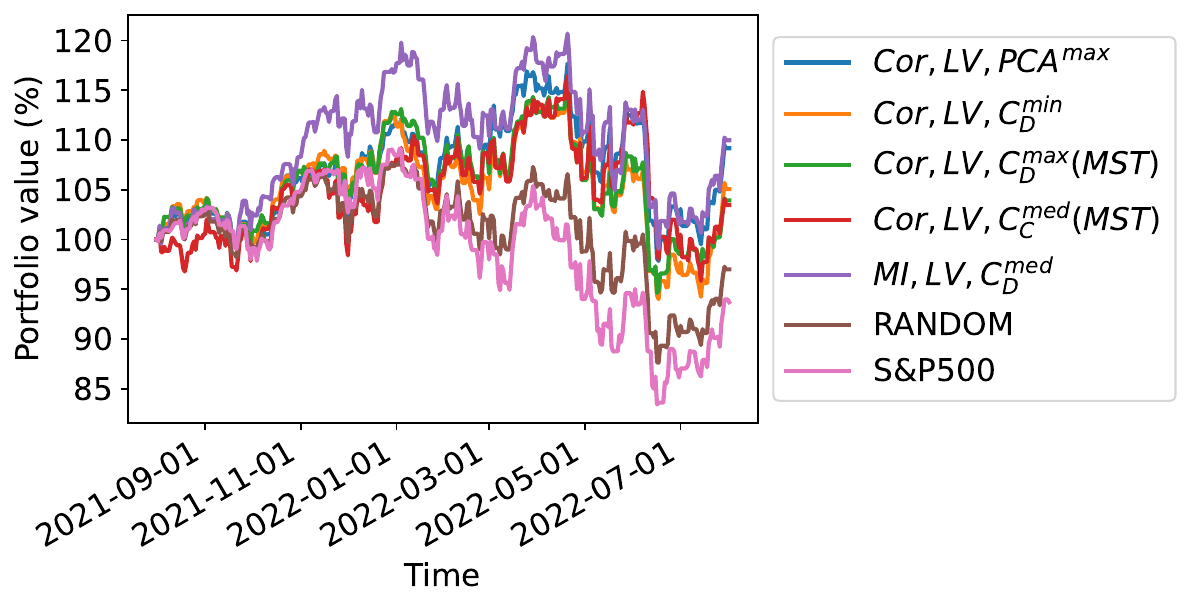}}
    \caption{Value in time of Top 5 stock portfolios by return/volatility. The symbols represent: Cor – Correlation, MI – Mutual information, ($\sim$) – co-occurrence, $C_C$ – Closeness centrality, $C_D$ – Degree centrality, PCA – Principal component analysis, FG – Full graph, MST – Minimum spanning tree, LV – Louvain, and AP – Affinity propagation. A detailed description of all symbols is provided in Figure~\ref{fig:port_meth}.}
    \label{fig:stock_best_by_return_volatility_ratio_stock_time}
\end{figure}

First, Figure \ref{fig:stock_all_obtained_portfolios_war} shows the returns vs. volatility of all considered portfolios. Since this is a period where two crises overlap, the majority of portfolios are characterized by low or negative returns. Nevertheless, a small part of the portfolios have a large positive return while also having a high volatility. We can also notice that some of the portfolios are characterized by a higher return compared to the baseline portfolios. As a result of the overall chaos resulting from the COVID-19 crisis and the Russian invasion of Ukraine, we can see in Table \ref{tab:stock_war_return} that the pattern that was present in the previous examinations has been largely disrupted in some of the measures. We can notice that the portfolios that are more centrally located within the network are not characterized by a high return in all cases. 
A pattern disruption is also noticeable in the volatilities shown in Table \ref{tab:stock_war_volatility}, although still, the portfolios consisting of less central assets have lower volatility in most cases. The Sharpe ratio given in Table \ref{tab:stock_war_sharp_ratio}, indicates that more central portfolios perform better in general.

Figure \ref{fig:stock_best_by_return_stock_war_period} presents the top 10 portfolios that are characterized by the highest returns. It is noticeable that 6 out of 10 portfolios are characterized by higher returns and lower volatility than the baseline portfolios. In this case, AP contributes to forming 5 out of 10 portfolios, including the two with the largest return. Figure \ref{fig:stock_best_by_volatility_stock_war_period} shows the portfolios characterized with the lowest volatility. It is interesting to note that almost all portfolios characterized by low volatility are characterized by a very small or negative return, and all of them have lower volatility compared to the baseline portfolios. As the most stable portfolio during such a crisis period, the portfolio that was built using the mutual information matrix in combination with the Louvain algorithm and the minimum value of the principal component analysis metric stands out as the only one yielding a positive return. Still, the portfolios consisting of less central assets, with lower closeness centrality, larger degree centrality, and minimal PCA, have the lowest volatility. At the very end, the portfolios with the best return/volatility ratio presented in Figure \ref{fig:stock_best_by_return_volatility_ratio_stock_war_period} completely overlap with those with the best return. In addition, Figure \ref{fig:stock_best_by_return_volatility_ratio_stock_war_time} shows the return over time of the top 5 portfolios by return/volatility ratio. Again, despite the crisis period in which the analyses were performed, the selected portfolios are characterized by a better return during the entire period compared to the baseline portfolios.

%\afterpage{\clearpage}

\subsection{Cryptocurrency portfolios}

\subsubsection{Entire period}

\begin{figure}[hbt!]
    \centering
    \subfloat[All examined portfolios]{\label{fig:crypto_all_obtained_portfolios}\includegraphics[width=0.49\linewidth]{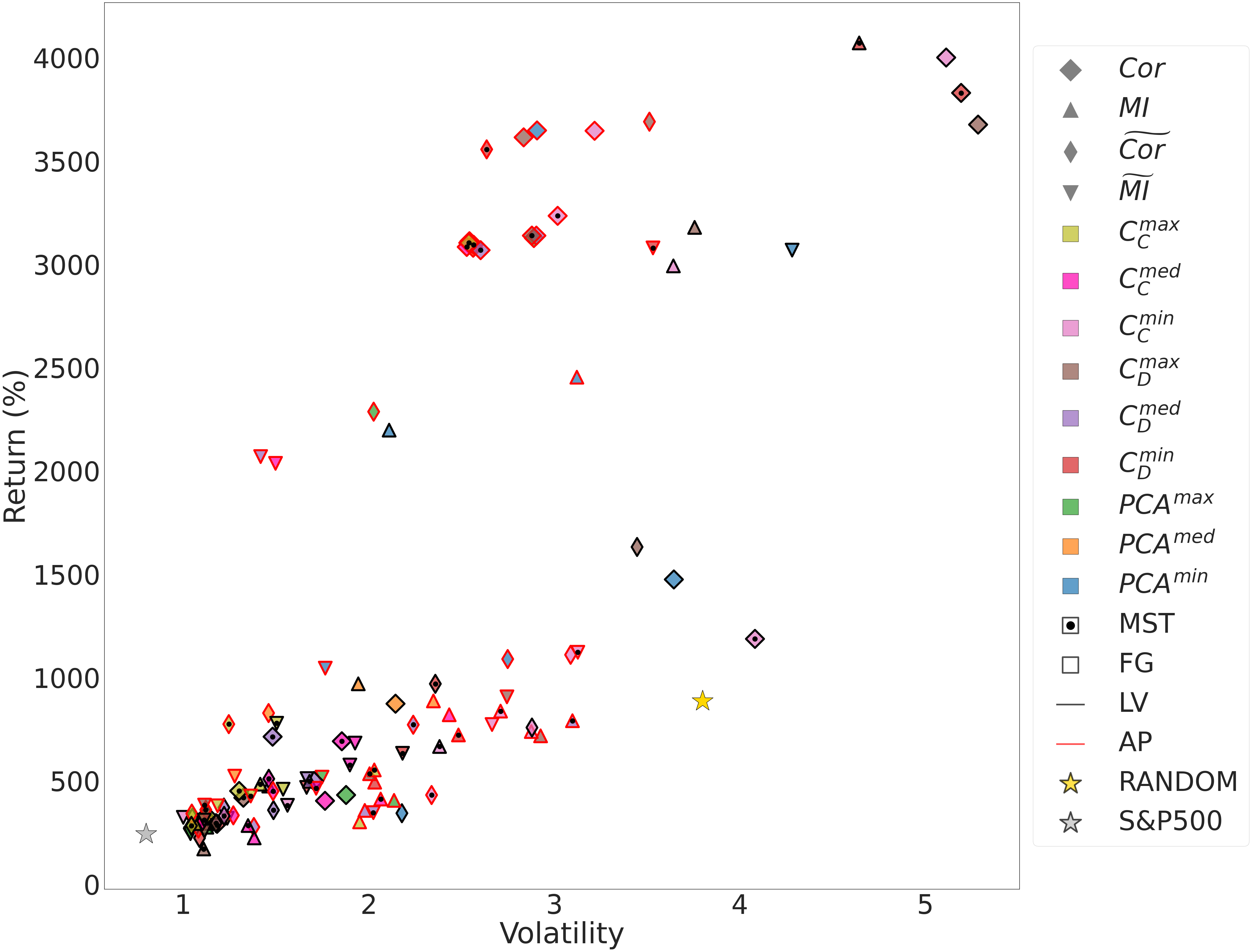}}
    \subfloat[Top 10 portfolios by return]{\label{fig:crypto_best_by_return_stock_full_period}\includegraphics[width=0.49\linewidth]{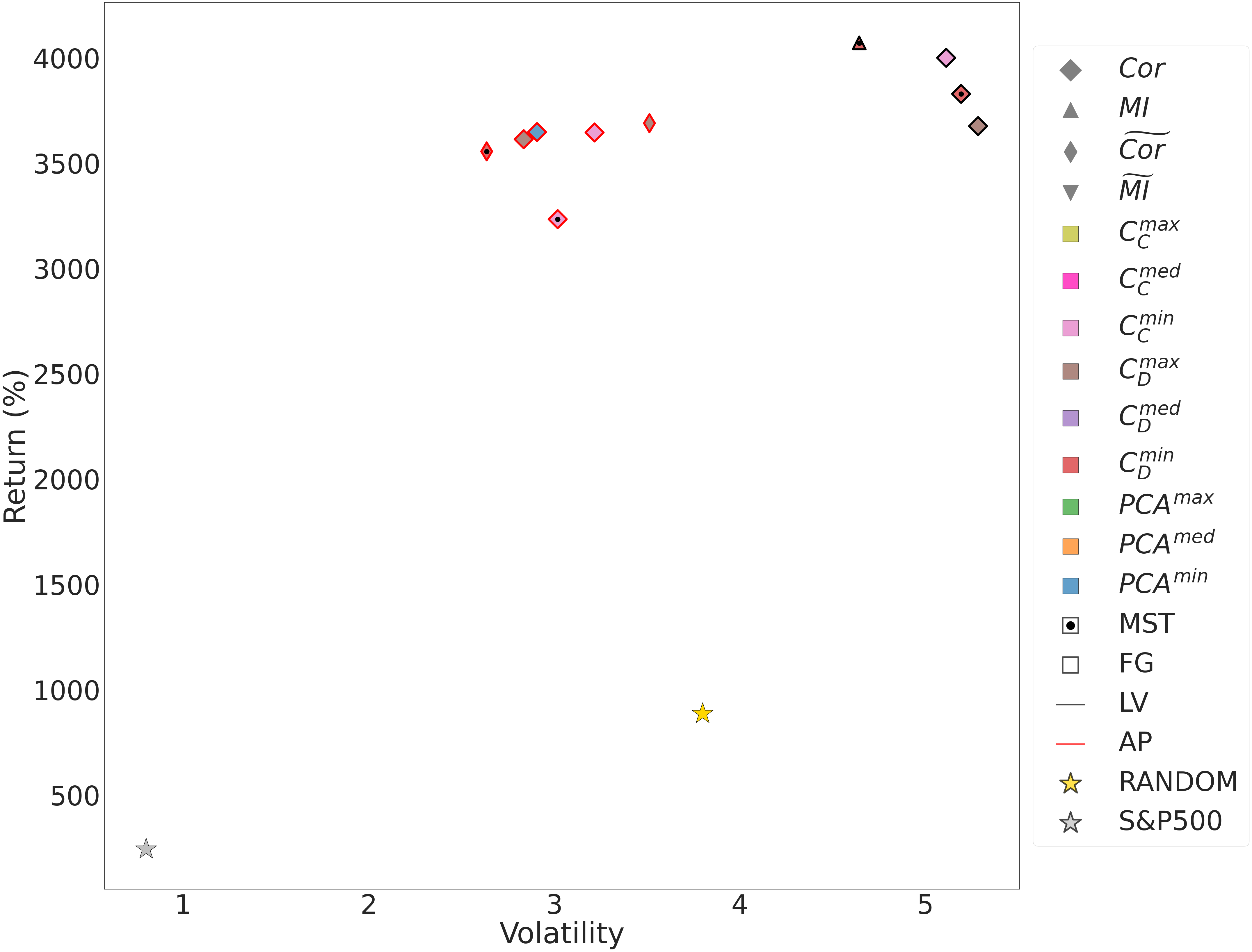}}\\
    \subfloat[Top 10 portfolios by volatility]{\label{fig:crypto_best_by_volatility_stock_full_period}\includegraphics[width=0.49\linewidth]{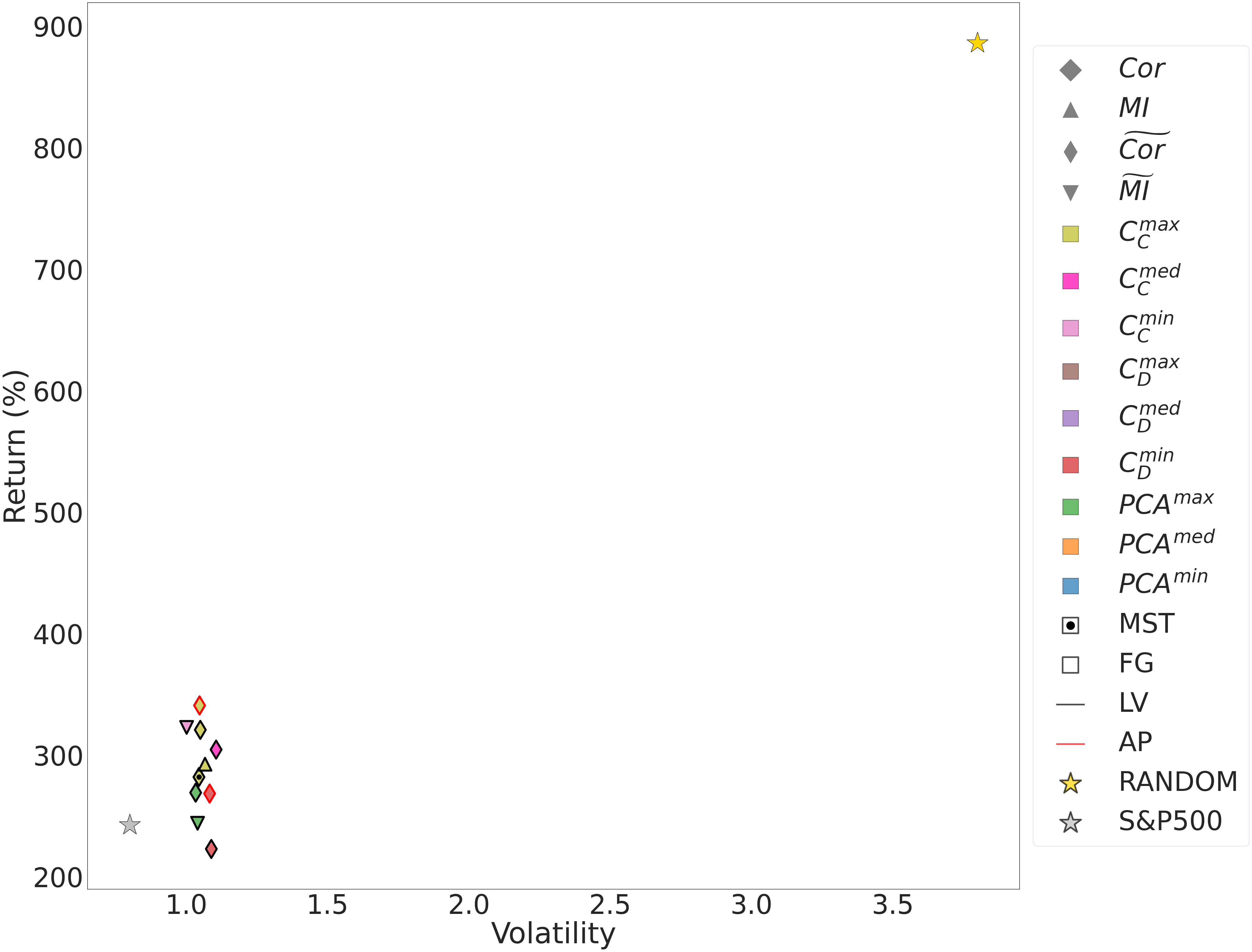}}
    \subfloat[Top 10 portfolios by return/volatility]{\label{fig:crypto_best_by_return_volatility_ratio_stock_full_period}\includegraphics[width=0.49\linewidth]{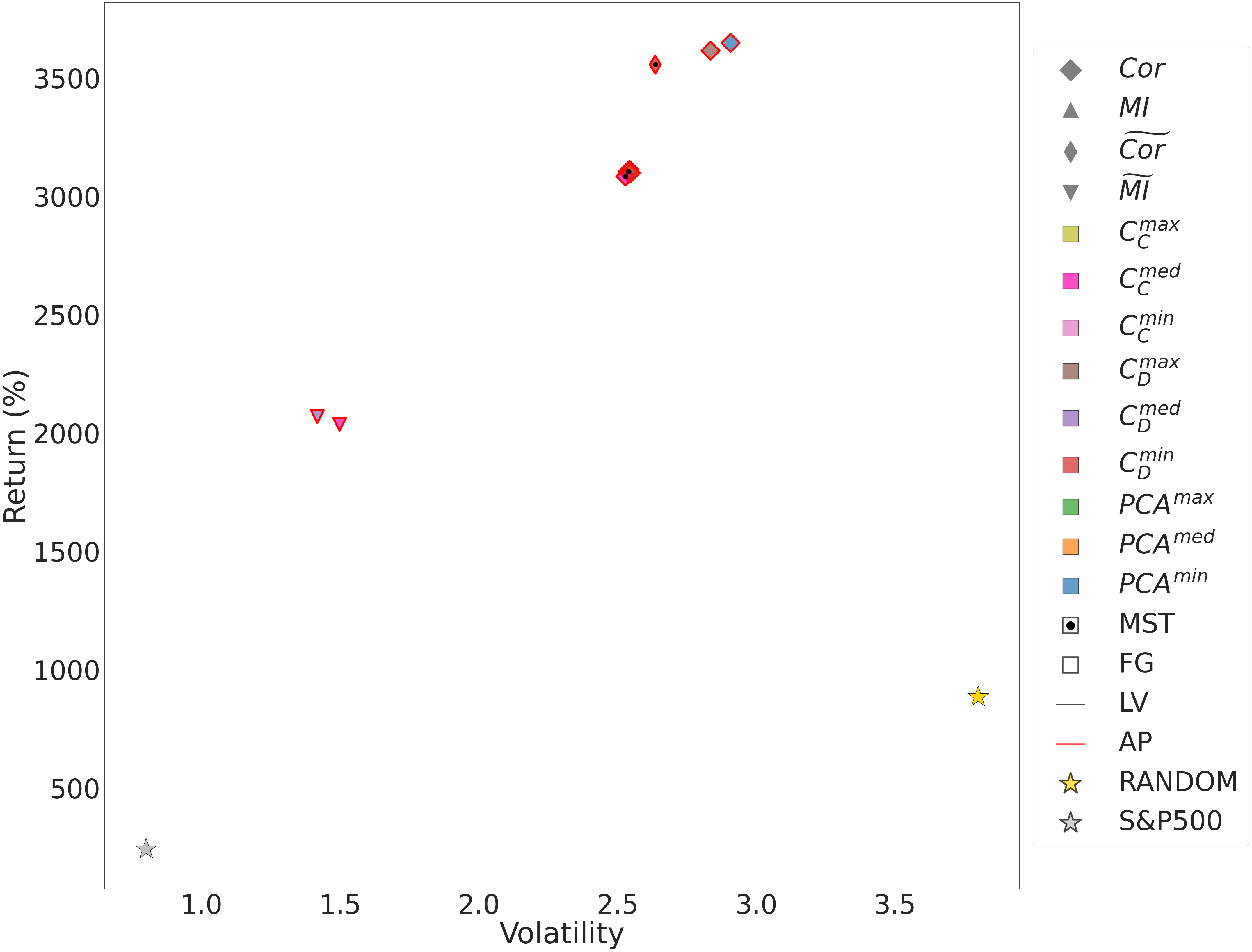}}\\
    \caption{Return vs. volatility of the examined investment portfolios composed of cryptocurrencies during the period from January 2019 to September 2022. The symbols represent: Cor – Correlation, MI – Mutual information, ($\sim$) – co-occurrence, $C_C$ – Closeness centrality, $C_D$ – Degree centrality, PCA – Principal component analysis, FG – Full graph, MST – Minimum spanning tree, LV – Louvain, and AP – Affinity propagation. A detailed description of all symbols is provided in Figure~\ref{fig:port_meth}.}
    \label{fig:crypto_portfolios_full_period}
\end{figure}

Figure \ref{fig:crypto_all_obtained_portfolios} shows the portfolios made up of cryptocurrencies across the full analysis period, from January 2019 to September 2022, whereas Table~\ref{tab:crypto_three_return_sme} provides more details about the portfolio returns and their standard errors. As baseline portfolios, we use a \acf{RANDOM} and a portfolio composed of all \acs{TOP 203} assets weighted by market cap. In comparison to the stock portfolios, we find that the distribution of cryptocurrency portfolios is not as smooth -- there is a larger variance in the level of return among the portfolios for a fixed level of volatility. We also observe that portfolios consisting of cryptocurrencies that are positioned more centrally within the network are usually associated with a lower return, whereas portfolios composed of assets located on the periphery have higher returns. Again, we should keep in mind that, generally speaking, more central assets are those with larger closeness centrality, larger PCA, or lower degree centrality on the full graph. We note that the crypto market is unregulated, and therefore, we would expect portfolio patterns that are less pronounced than those observed in portfolios comprising stocks. A similar situation appears in the volatility of portfolios (see Table \ref{tab:crypto_three_volatility_sme}). The portfolios composed of currencies that are positioned more centrally within the network show lower levels of volatility, while those that lie on the periphery of the network demonstrate a higher return. The Sharpe ratio for the entire period is given in Table \ref{tab:crypto_sharp_ratio_full_period}.

Figure \ref{fig:crypto_best_by_return_stock_full_period} shows the top 10 portfolios with the highest returns. We notice that all these portfolios have higher returns than the baseline portfolios but are also characterized by higher volatility. It can be observed that 7 out of the top 10 portfolios, according to returns, are obtained using the correlation matrix. We can also see that all portfolios are composed of more peripheral assets with minimal closeness centrality, minimal PCA, minimal degree centrality on MST or maximal degree centrality on FG. Interestingly, 6 out of 10 portfolios are calculated using AP, and they all have lower volatility compared to the other 4 obtained by Louvain, but also have a lower return on average.

Figure \ref{fig:crypto_best_by_volatility_stock_full_period} plots the top 10 portfolios characterized with the lowest volatility. In this case, we observe that the portfolios are characterized by similar volatility but offer different returns at the same time. Also, the returns of these portfolios are relatively low. These returns are higher than those of the TOP 203 portfolio but lower than those of the RANDOM portfolio. Similarly as before, the top 10 portfolios with the lowest volatility are most often built with the correlation matrix. In contrast, now most of the portfolios are built using Louvain and more central assets. 

Figure \ref{fig:crypto_best_by_return_volatility_ratio_stock_full_period} shows the top 10 portfolios with the best return/volatility ratio. All these portfolios have higher returns compared to the baselines, and they are all made using affinity propagation, most often in combination with the correlation matrix. %These portfolios are constituted of currencies are located more central in the cryptocurrency network.

To assess the resilience of our cryptocurrency portfolio creation methodology, we examined portfolios including 30 and 10 cryptocurrencies. The outcomes for 30 cryptocurrency portfolios correspond with those of 20 cryptocurrency portfolios, especially regarding volatility. Table \ref{tab:crypto_volatility_data_30_crypto} illustrates that portfolios including a greater proportion of central assets have reduced volatility, accompanied by diminished departures from the trend attributable to the increased asset count. Likewise, Table \ref{tab:crypto_volatility_data_10_crypto} displays findings for 10 cryptocurrency portfolios, adhering to the same trend but exhibiting somewhat bigger variances, as anticipated. The portfolio returns shown in Table \ref{tab:crypto_return_data_30_crypto} and Table \ref{tab:crypto_return_data_10_crypto} follow a similar trend as before.

Finally, in Table \ref{tab:crypto_sharp_ratio_full_period}, Table \ref{tab:crypto_sharp_ratio_30_coins}, and Table \ref{tab:crypto_10_sharp_ratio}, we have shown the Sharpe ratios calculated using the USA treasury bill rate as a risk-free rate. The interpretation of the Sharpe ratios is even more challenging than for stock portfolios. One can say that, in general, Affinity propagation yields a better Sharpe ratio than Louvain. Interestingly, degree centrality (FG) and closeness centrality (MST) with maximum values frequently perform better than the others, although they have opposite centrality natures. 

%Based on the results from the cryptocurrency portfolio sharp ratio, it is clear that Louvain consistently delivers more stable Sharpe ratios, which is similar to its performance in stock portfolios. On the other hand, Affinity Propagation (AP) creates higher-return portfolios, making it more appropriate for high-risk, high-reward strategies.}

\subsubsection{Beginning of the COVID-19 pandemic}

\begin{figure}[hbt!]
    \centering
    \subfloat[All examined portfolios]{\label{fig:crypto_all_obtained_portfolios_covid}\includegraphics[width=0.49\linewidth]{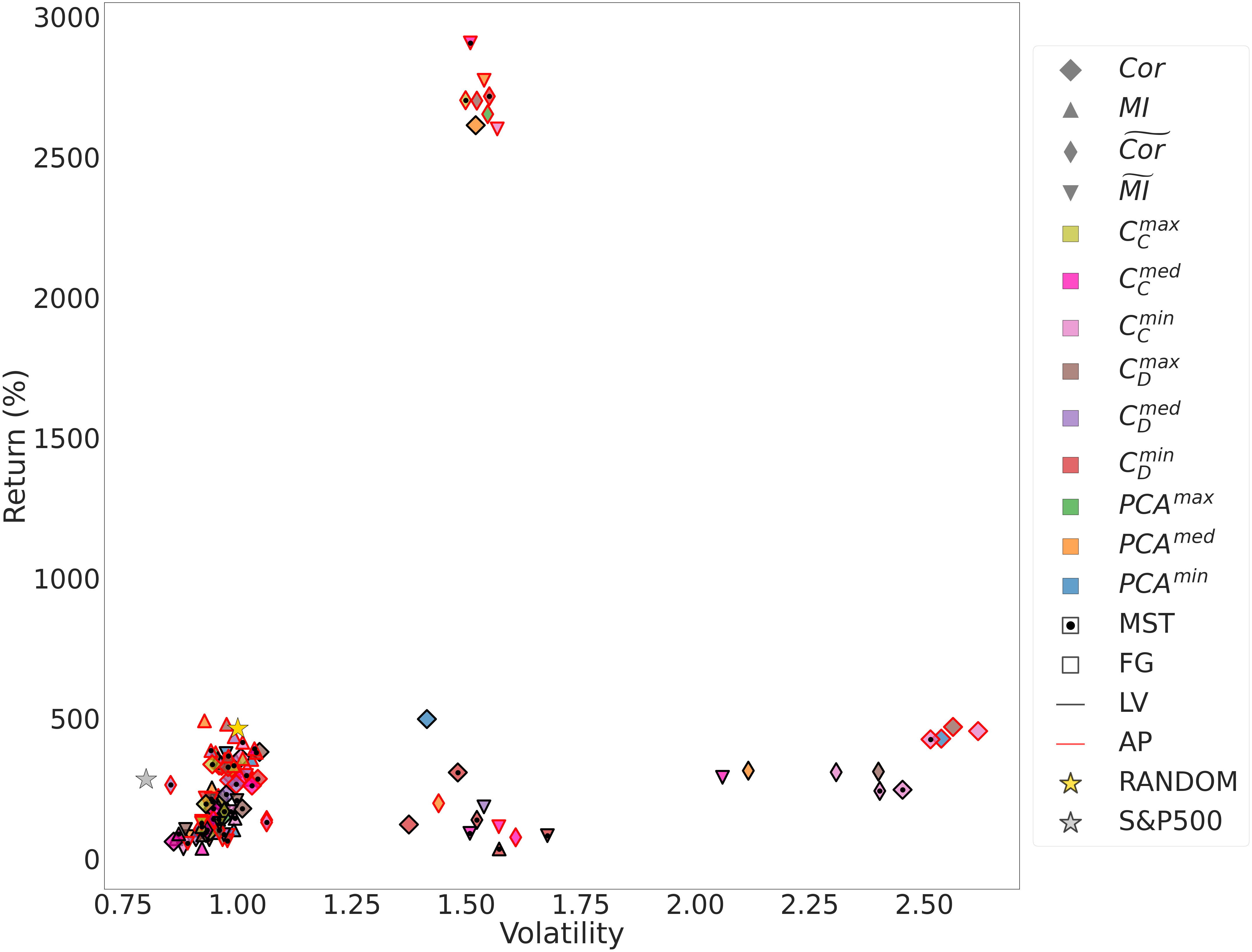}}
    \subfloat[Top 10 portfolios by return]{\label{fig:crypto_best_by_return_stock_covid_period}\includegraphics[width=0.49\linewidth]{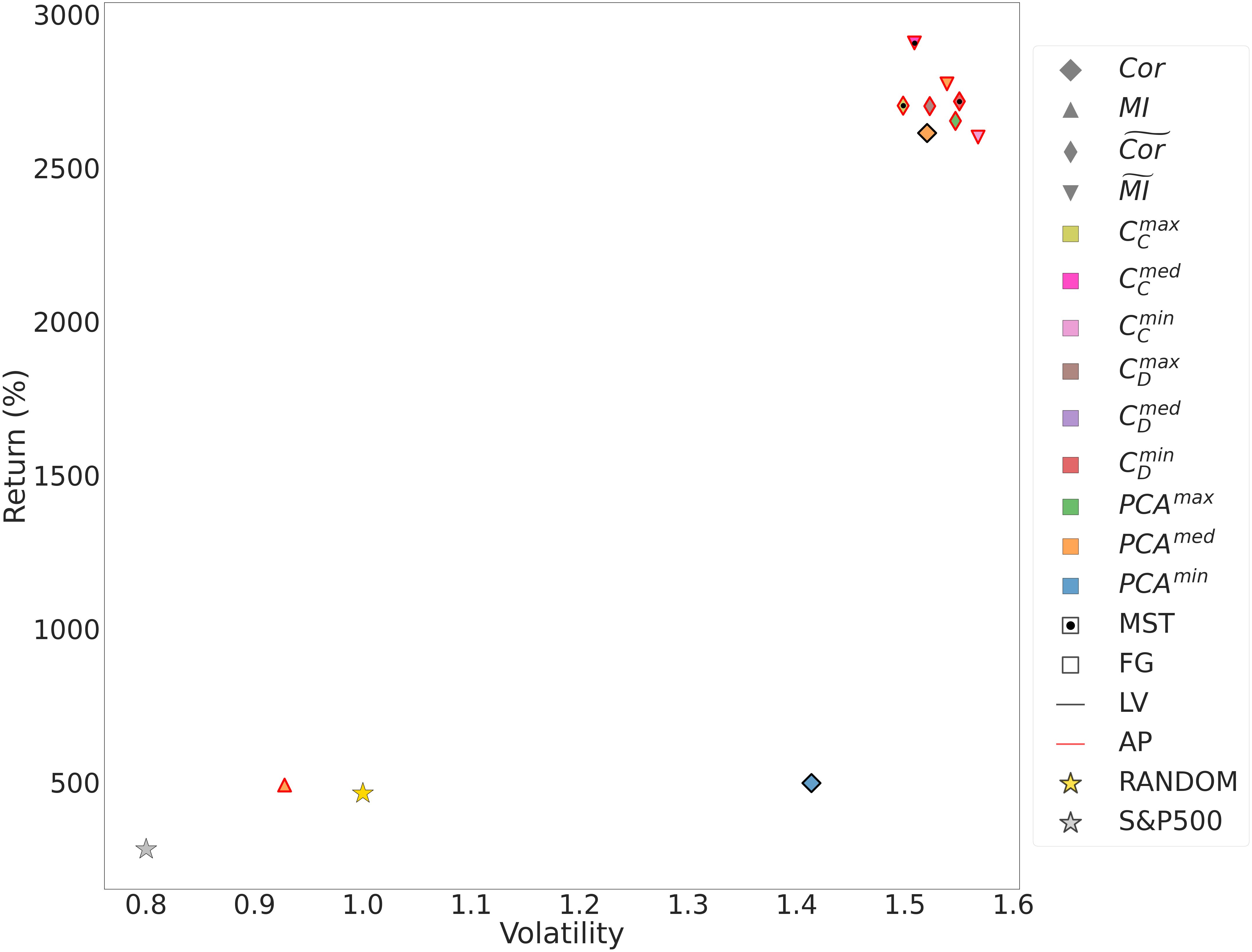}}\\
    \subfloat[Top 10 portfolios by volatility]{\label{fig:crypto_best_by_volatility_stock_covid_period}\includegraphics[width=0.49\linewidth]{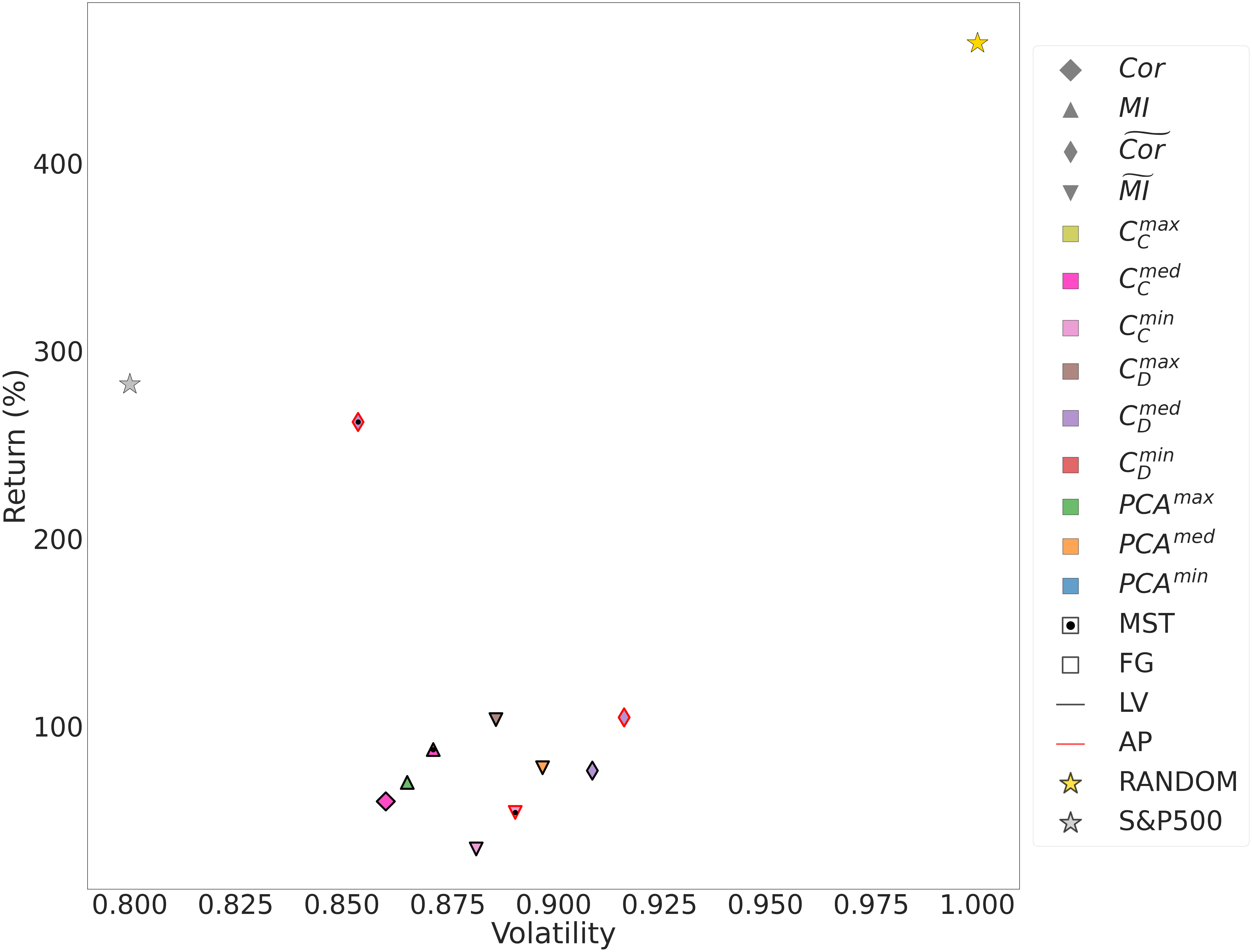}}
    \subfloat[Top 10 portfolios by return/volatility]{\label{fig:crypto_best_by_return_volatility_ratio_stock_covid_period}\includegraphics[width=0.49\linewidth]{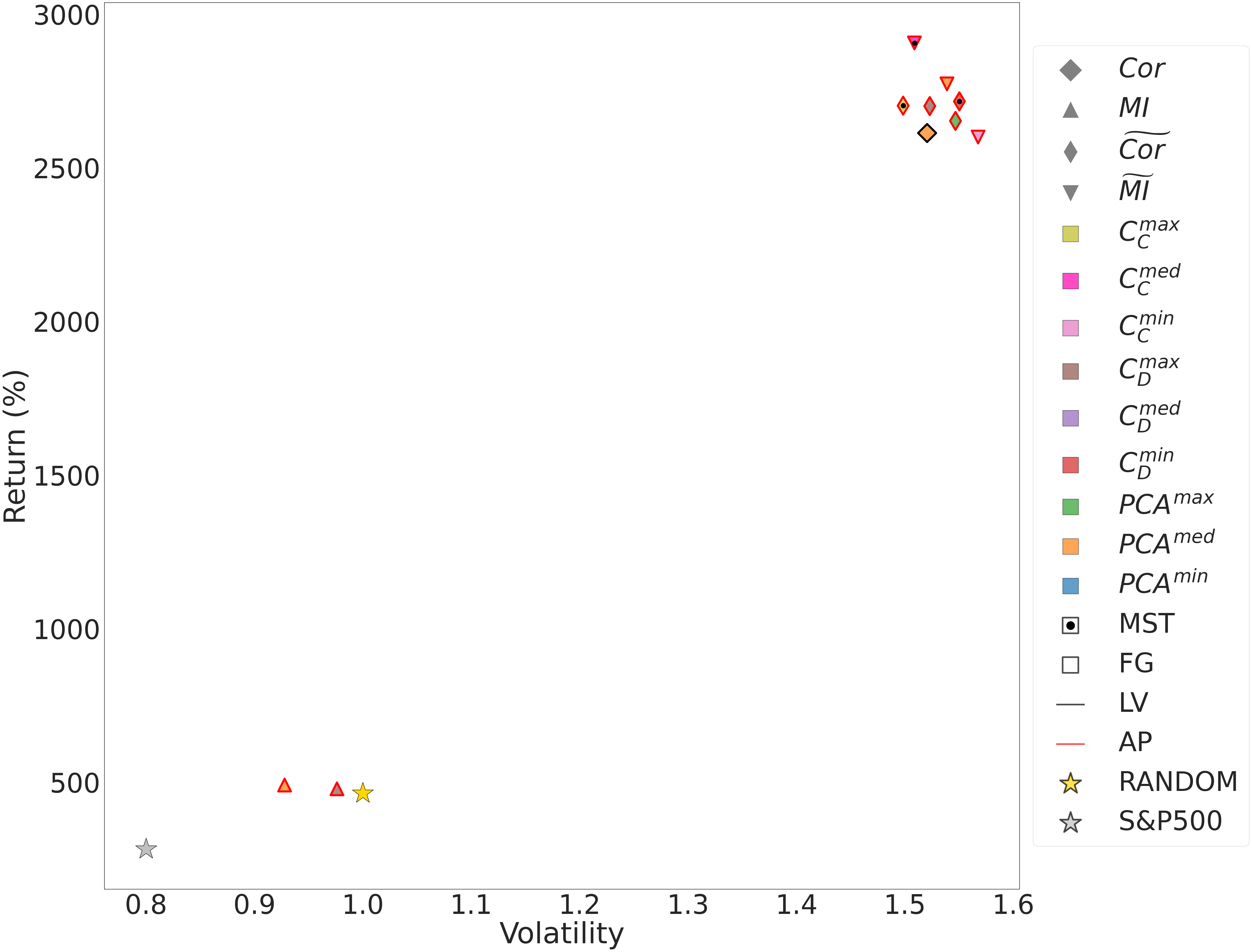}}
    \caption{Return vs. volatility of the examined cryptocurrency portfolios during the COVID-19 pandemic, from January 2020 to December 2020. The symbols represent: Cor – Correlation, MI – Mutual information, ($\sim$) – co-occurrence, $C_C$ – Closeness centrality, $C_D$ – Degree centrality, PCA – Principal component analysis, FG – Full graph, MST – Minimum spanning tree, LV – Louvain, and AP – Affinity propagation. A detailed description of all symbols is provided in Figure~\ref{fig:port_meth}.}
    \label{fig:crypto_portfolios_covid_period}
\end{figure}

Figure \ref{fig:crypto_portfolios_covid_period} presents the results for cryptocurrency portfolios built using data from the COVID-19 period. Specifically, Figure \ref{fig:crypto_all_obtained_portfolios_covid} plots all generated portfolios, highlighting the diversity among them. These findings are detailed in Table \ref{tab:crypto_covid_return}, which shows the returns for each portfolio. Here, we do not observe a specific trend in the relationship between centrality and returns, as there are situations with the highest returns for all of the maximal, medial, or minimal ranges. Table \ref{tab:crypto_covid_volatility} provides an overview of each portfolio's volatility during the COVID-19 period. The volatility patterns mirror the unconventional patterns seen in the returns. Finally, the Sharpe ratio is given in \ref{tab:crypto_sharp_ratio_covid_period}.

Figure \ref{fig:crypto_best_by_return_stock_covid_period} shows the top ten portfolios with the highest return during the COVID-19 period. These portfolios have higher returns and volatility than the baseline portfolios. Differently, from the observation for the full period, now the top 10 portfolios are similarly distributed among the correlation and mutual information approaches. Notably, affinity propagation is used in 8 out of 10 high-return portfolios, whereas we do not find any impact of centrality or PCA on the portfolios with the highest return, as all options appear in the top 10 portfolios. 

Figure \ref{fig:crypto_best_by_volatility_stock_covid_period} focuses on portfolios with the lowest volatility. These are positioned between the volatilities of the two baseline portfolios. The TOP 203 portfolio shows lower volatility than the discovered portfolios, whereas the RANDOM portfolio is more volatile. 6 out of 10 portfolios are constructed using mutual information, and 9 out of 10 portfolios are constructed with central or medial assets.

Figure \ref{fig:crypto_best_by_return_volatility_ratio_stock_covid_period} presents the top ten portfolios with the optimal return/volatility ratio during the COVID-19 crisis, where most are the same as the top 10 portfolios by return. All portfolios have a higher return-to-volatility ratio than the baseline portfolios. Also, 6 out of the 10 portfolios are identified using affinity propagation, while most portfolios consist of central or medial financial assets.

\subsubsection{Russian invasion of Ukraine}

\begin{figure}[hbt!]
    \centering
    \subfloat[All examined portfolios]{\label{fig:crypto_all_obtained_portfolios_war}\includegraphics[width=0.49\linewidth]{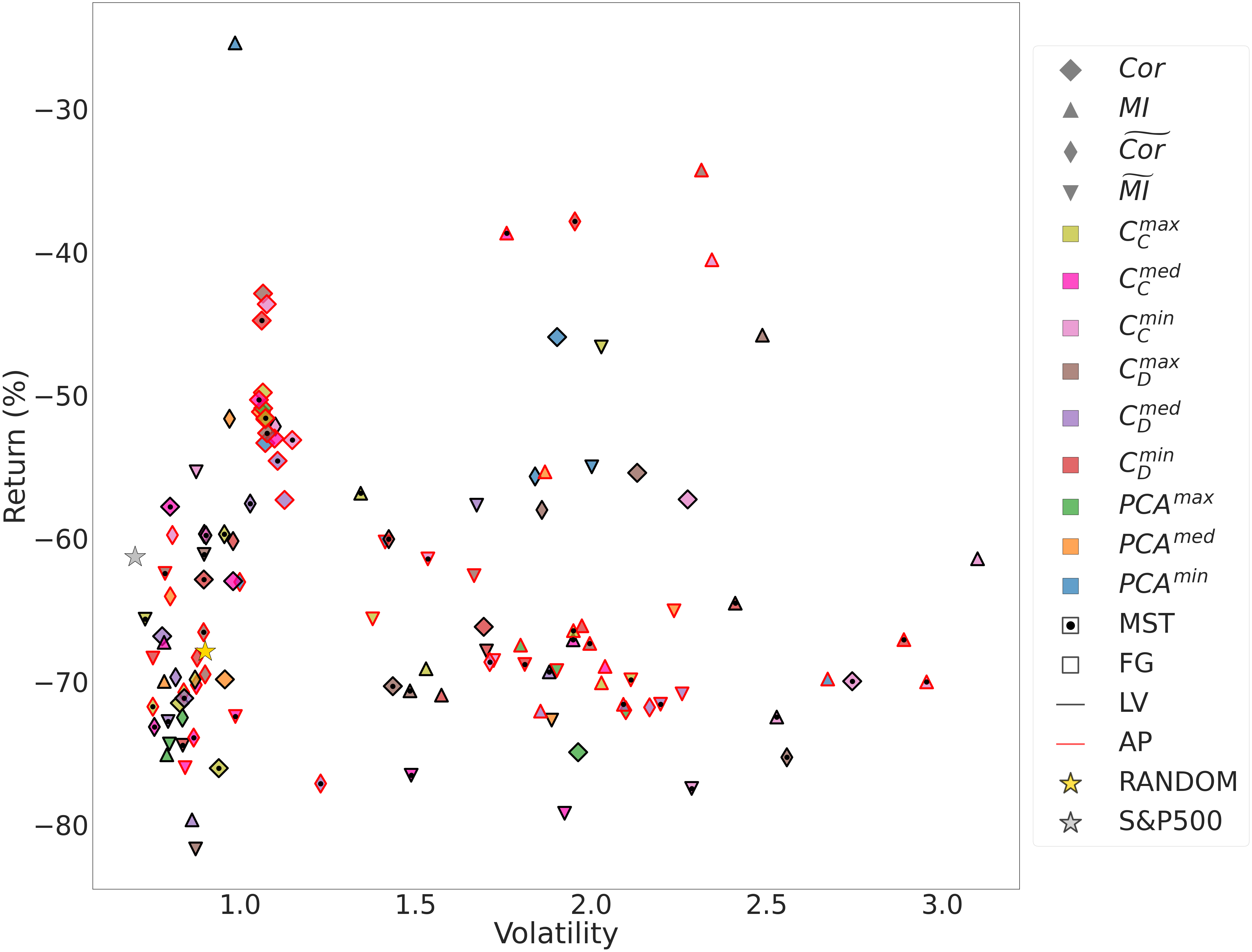}}
    \subfloat[Top 10 portfolios by return]{\label{fig:crypto_best_by_return_stock_war_period}\includegraphics[width=0.49\linewidth]{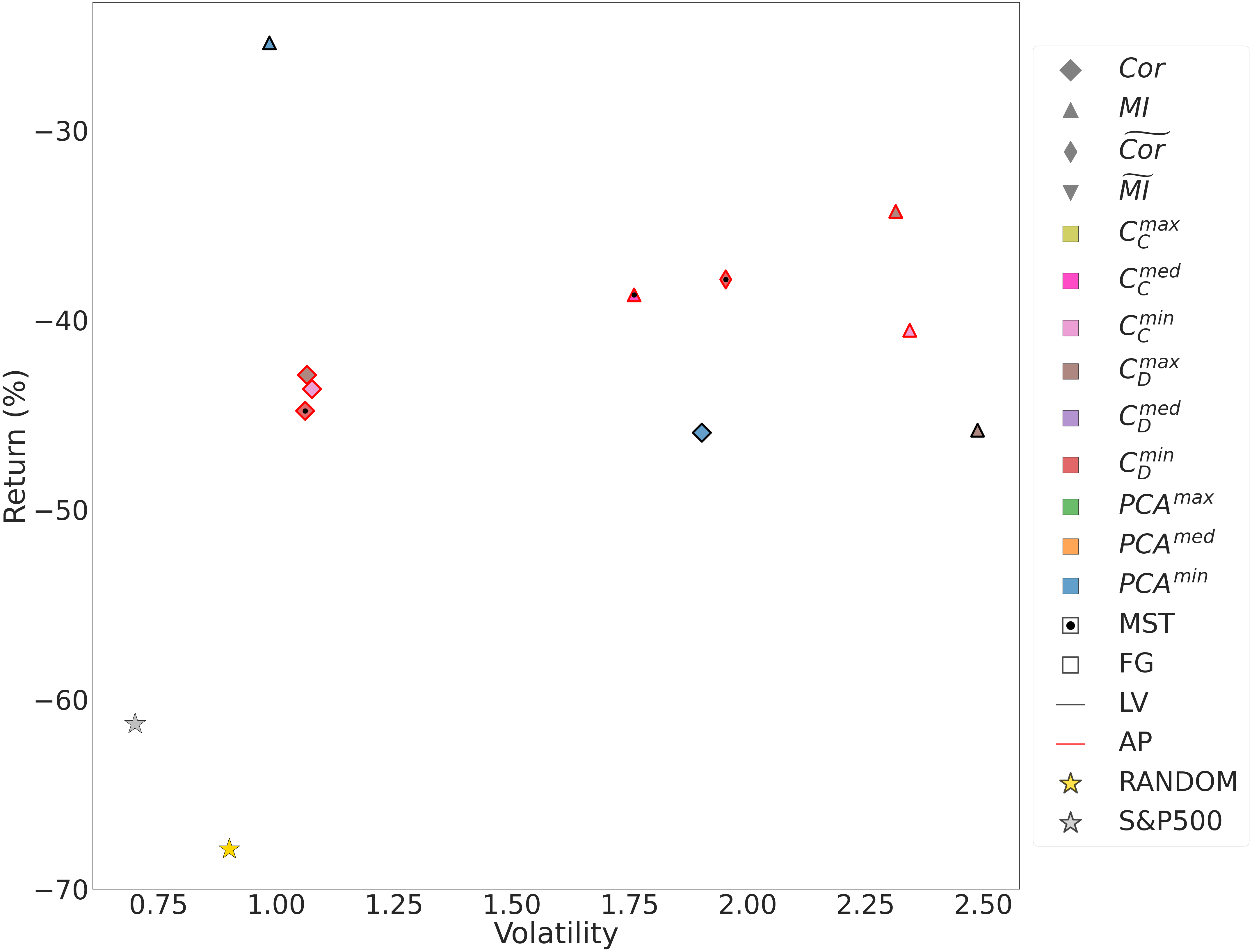}}\\
    \subfloat[Top 10 portfolios by volatility]{\label{fig:crypto_best_by_volatility_stock_war_period}\includegraphics[width=0.49\linewidth]{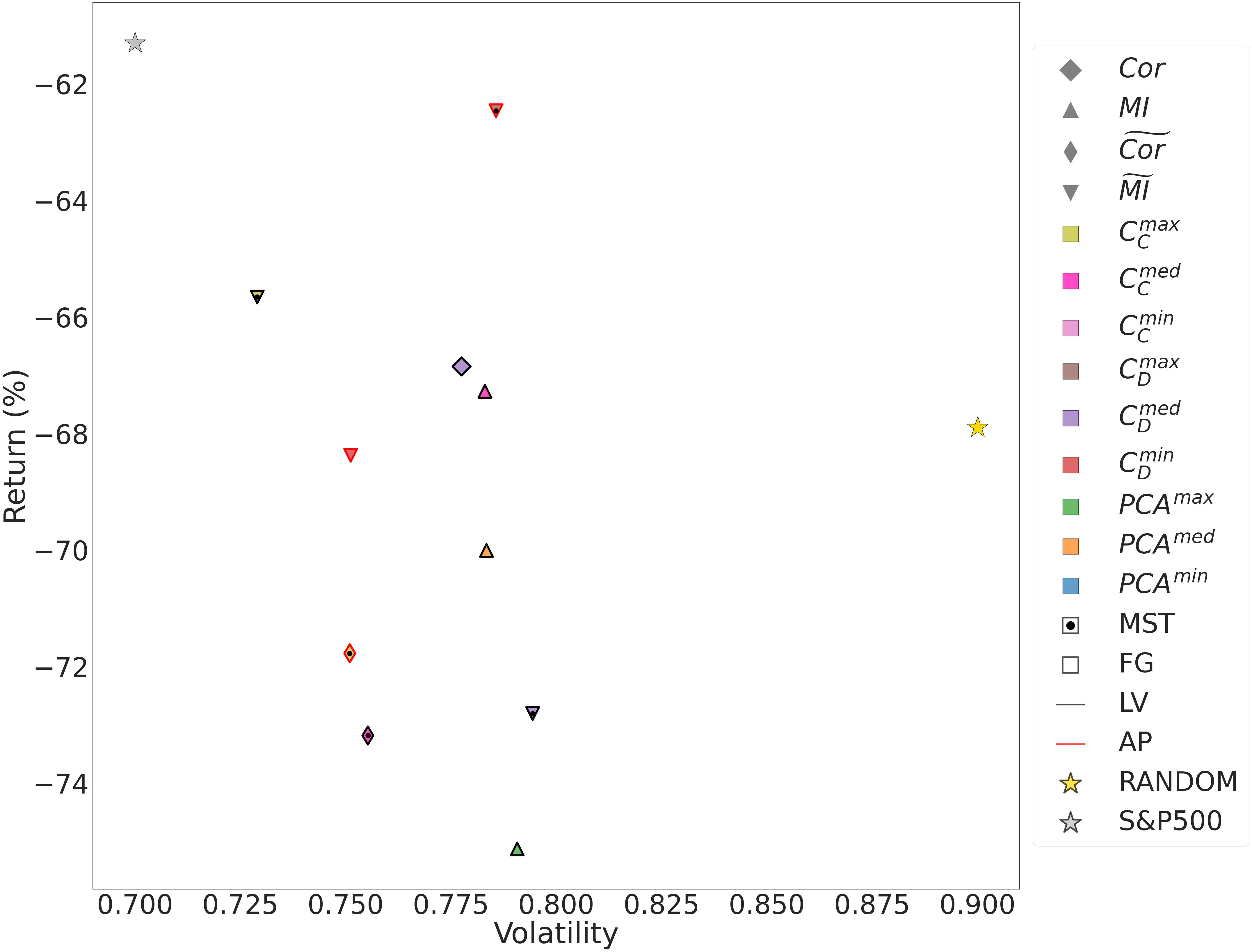}}
    \subfloat[Top 10 portfolios by return/volatility]{\label{fig:crypto_best_by_return_volatility_ratio_stock_war_period}\includegraphics[width=0.49\linewidth]{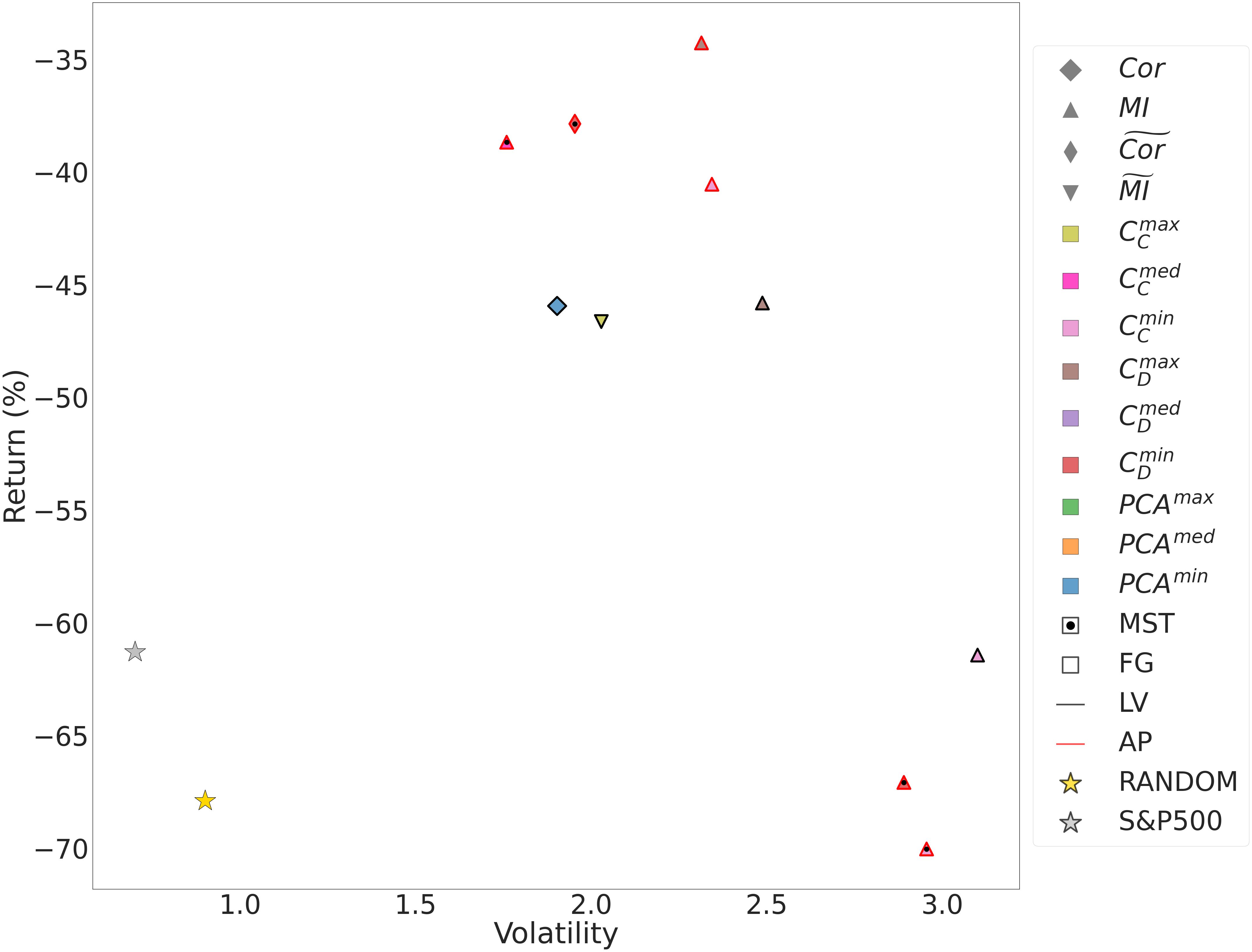}}
    \caption{Return vs. volatility of cryptocurrency portfolios during the beginning of the Russian invasion of Ukraine, from September 2021 to August 2022. The symbols represent: Cor – Correlation, MI – Mutual information, ($\sim$) – co-occurrence, $C_C$ – Closeness centrality, $C_D$ – Degree centrality, PCA – Principal component analysis, FG – Full graph, MST – Minimum spanning tree, LV – Louvain, and AP – Affinity propagation. A detailed description of all symbols is provided in Figure~\ref{fig:port_meth}.}
    \label{fig:crypto_portfolios_war_period}
\end{figure}

Figure \ref{fig:crypto_portfolios_war_period} shows the analysis of cryptocurrency portfolios during the Russian invasion of Ukraine. A complete look at the generated portfolios for this period is shown in Figure \ref{fig:crypto_all_obtained_portfolios_war}, where it is noticeable that all portfolios have negative returns. 
%The majority of the portofolios are located on the left side of Figure \ref{fig:crypto_all_obtained_portfolios_war}, indicating low volatility. This trend suggests that the invasion probably disrupted cryptocurrency markets, affecting portfolios that might otherwise have performed well. 
The returns, volatility, and Sharpe ratio of these portfolios are given in Table \ref{tab:crypto_war_return}, Table \ref{tab:crypto_war_volatility} and \ref{tab:crypto_sharp_ratio_war_period}, respectively.

\begin{figure}[hbt!]
   \centering
    \subfloat[Beginning of the COVID-19 pandemic]{\label{fig:crypto_best_by_return_volatility_ratio_stock_covid_time}\includegraphics[width=0.49\linewidth]{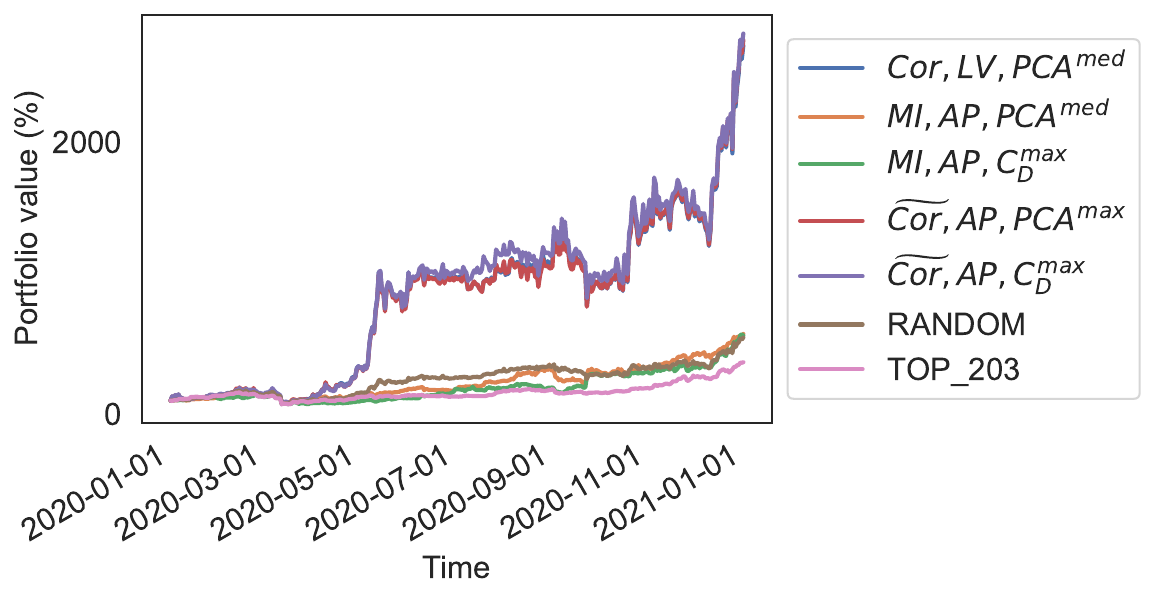}}
    \subfloat[Beginning of the Russian invasion of Ukraine]{\label{fig:crypto_best_by_return_volatility_ratio_stock_war_time}\includegraphics[width=0.49\linewidth]{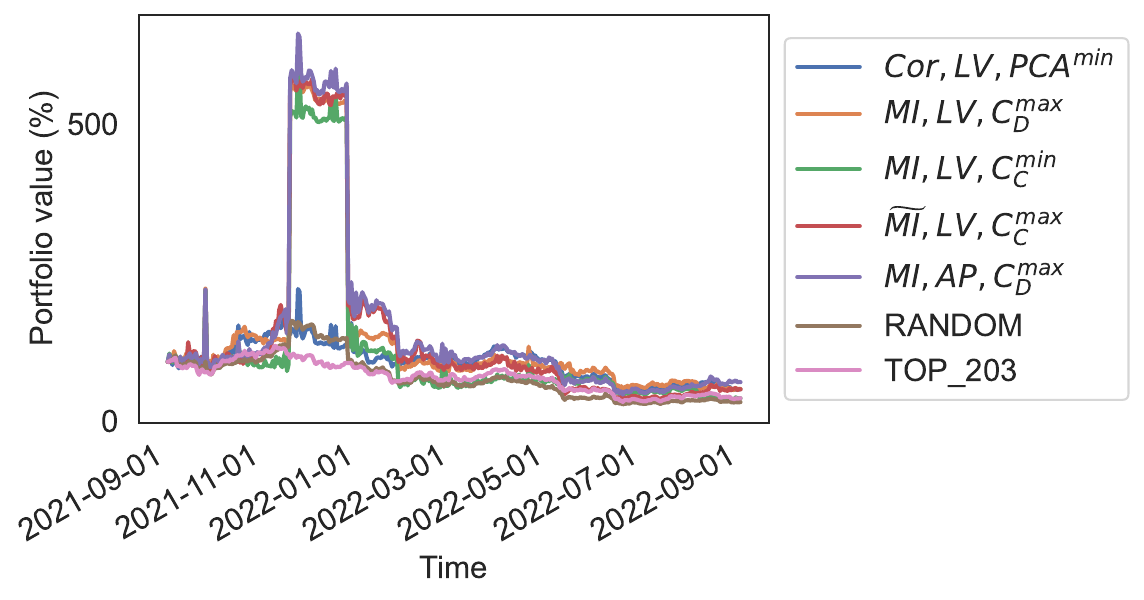}}
    \caption{Value in time of Top 5 cryptocurrency portfolios by return/volatility. The symbols represent: Cor – Correlation, MI – Mutual information, ($\sim$) – co-occurrence, $C_C$ – Closeness centrality, $C_D$ – Degree centrality, PCA – Principal component analysis, FG – Full graph, MST – Minimum spanning tree, LV – Louvain, and AP – Affinity propagation. A detailed description of all symbols is provided in Figure~\ref{fig:port_meth}.}
\end{figure}

The top 10 cryptocurrency portfolios with the highest return during this period are presented in Figure \ref{fig:crypto_best_by_return_stock_war_period}. All of these portfolios have higher returns and volatility than the baseline portfolios. Moreover, 7 out of 10 of the portfolios were obtained using affinity propagation. Figure \ref{fig:crypto_best_by_volatility_stock_war_period} shows the ten portfolios with the lowest volatility. Noticeably, 7 portfolios were generated using Louvain, and 7 were derived using network representations from the mutual information matrix. 

Finally, Figure \ref{fig:crypto_best_by_return_volatility_ratio_stock_war_period} examines the top 10 portfolios by return/volatility ratio. It can be seen that 8 out of the 10 portfolios were generated using the mutual information matrix (7 with the original and 1 with the co-occurrence). We should note that the top 10 portfolios are positioned in the upper right corner, unlike previously for portfolios with positive returns where they are positioned in the upper left corner. This atypical indication of the return/volatility ratio was also observed for the Sharpe ratio in \cite{mcleod2004interpreting} where the authors have given a probabilistic explanation that these portfolios can still be considered as best performing because they provide the best chance of achieving positive results, although sometimes they could even have both lower return and higher volatility than others. At the very end, Figure \ref{fig:crypto_best_by_return_volatility_ratio_stock_war_time} provides a detailed overview of the portfolio value of the top 5 portfolios by return/volatility. It can be noticed that four out of five portfolios experienced a significant increase in returns at the start of November 2021, followed by a significant decrease in returns at the start of January 2022. A closer inspection showed that these events were caused by variations in a cryptocurrency known as \emph{Asch}. This cryptocurrency's price increased by 20,000\% at the beginning of November 2021 but decreased significantly by the end of December 2021.

\section{Conclusion}\label{sec:con}

In this study, we explored network-based approaches to portfolio diversification that could enhance traditional investment strategies. Our methodology involved analyzing both linear and non-linear relationships between financial assets using Pearson correlation coefficient and mutual information, transforming these metrics into distance matrices, constructing network representations, and applying community detection algorithms to identify distinct market segments. From these segments, we selected assets based on a combination of statistical and network criteria, resulting in 120 different portfolio construction strategies.

Our analysis suggests that the behavior of network-based strategies might differ between asset classes. For instance, in cryptocurrency portfolios, assets with higher centrality tend to exhibit lower volatility—a trend that contrasts with the behavior observed in stock portfolios but aligns with previous literature~\cite{onnela2002dynamic,pozzi2013spread}. Additionally, while many network-based strategies appear to outperform baseline portfolios for stocks, their effectiveness for cryptocurrencies is less clear, reflecting a higher variability in performance. Notably, the behavior of these strategies also varied during periods of market stress, such as during the COVID-19 pandemic and the Russian invasion of Ukraine, which disrupted more stable patterns seen in calmer periods. These results corroborate recent findings showing the rising volatility during the pandemic~\cite{he2020impact,belhassine2021contagion,yousfi2024pandemic}.

Despite these results, our study has several important limitations. First, the complexity and unpredictability of the cryptocurrency market pose significant challenges, as trends in this market are less consistent compared to stocks. Second, the overlapping crises examined in our study presented unique challenges that may not be fully addressed by our methodology, warranting further investigation. Third, our reliance on equally weighted portfolio construction may not reflect optimal real-world investment strategies. Finally, we did not explore data preprocessing techniques—such as regularization to stabilize asset relationship estimates and random matrix filtering to reduce noise from finite sample effects—which might improve the robustness of the constructed correlation matrices and network representations. However, we believe that the minimum spanning tree representation and the way the clustering and centrality metrics are applied already filter out most of the spurious asset relationships. Future work should focus on developing more sophisticated methodologies that capture both linear and non-linear dependencies between financial instruments, refine asset weighting schemes, and incorporate effective data preprocessing strategies to better reflect market conditions and investor preferences. 

There are also several other potential directions for future work. First, one can consider the application of the Mahalanobis distance, which is a scale-invariant measure that can be particularly helpful in turmoil periods \cite{ahelegbey2022netvix}. However, in this study, we have focused on the application of correlation and mutual information, which we believe can be more insightful for portfolio diversification as it also captures nonlinear dependencies. Nevertheless, the application of the Mahalanobis distance is an interesting avenue for future research. Second, one can particularly focus on the robustness of the portfolio approaches to extreme events and attacks or even develop diversification approaches that are particularly designed for that. Lastly, having in mind the outcomes of our study, mixed portfolios can be constructed consisting of network-based approaches of diverse asset types, e.g., stocks and cryptocurrencies.

In summary, our network-based approach offers a framework for evaluating portfolio performance that highlights the complex dynamics underlying financial markets. While our findings contribute to the literature on portfolio diversification, they also emphasize the need for further refinement of these strategies to enhance their robustness and practical relevance in both conventional and digital asset markets.

\section*{List of abbreviations}
\begin{acronym}[TDMA] % The argument is the widest acronym in the list
    \acro{AP}{Affinity propagation}
    \acro{BTC}{Bitcoin}
    \acro{Cor}{Pearson correlation coefficient}
    \acro{cCor}[$\widetilde{\mathrm{Cor}}$]{Co-occurrence Pearson correlation coefficient}
    \acro{DBHT}{Directed bubble hierarchical tree}
    \acro{ETH}{Etherium}
    \acro{FG}{Full graph}
    \acro{LTC}{LiteCoin}
    \acro{LV}{Louvain}
    \acro{MI}{Mutual information}
    \acro{cMI}[$\widetilde{\mathrm{MI}}$]{Co-occurrence mutual information}
    \acro{MST}{Minimal spanning tree}
    \acro{PMFG}{Planar maximally filtered graph}
    \acro{PCA}{Principal component analysis}
    \acro{RANDOM}{Portfolio comprising randomly selected assets}
    \acro{SP500}[S\&P 500]{Standard and Poor's 500}
    \acro{DJIA}{Dow Jones Industrial Average}
    \acro{TOP 203}{Top 203 cryptocurrencies}
\end{acronym}
\backmatter

\section*{Declarations}

\bmhead*{Availability of data and materials}
%The data used in this study is publicly available and anyone can use it to reproduce the results. We collected the stocks data from \href{www.finnhub.io}{www.finnhub.io}, and the cryptocurrencies data from \href{www.coinmarketcap.com}{www.coinmarketcap.com}, but it is available through other sources as well.
The datasets supporting the conclusions of this article are available at FinnHub \href{https://www.finnhub.io}{https://www.finnhub.io} and CoinMarketCap \href{https://www.coinmarketcap.com}{https://www.coinmarketcap.com}, as well as through many other sources.

\bmhead*{Competing interests}
The authors declare that they have no competing interests.

\bmhead*{Funding}
This work was partially funded by the Faculty of Computer Science and Engineering, Ss. Cyril and Methodius University in Skopje, North Macedonia, under project PM-NSML, and the Austrian Research Promotion Agency (FFG) under project AMALFI - 898883.

\bmhead*{Author's contributions}
D.K. wrote the code and prepared most of the methods, data and results sections. I.M. co-designed the study, contributed with expertise, supervised the coding, and edited the manuscript. V.S. wrote the introduction and the conclusion, contributed with expertise, and edited the manuscript. M.M. designed the study, contributed with expertise, wrote the related work section and large parts of the Methods and Data, and Results sections, supervised the writing and coding, and edited the manuscript. All authors read and approved the final manuscript.

\bmhead*{Acknowledgements}
We would like to thank Sebastian Krawczyk for his useful comments.

\clearpage

\begin{appendices}

\section{Conventional stock portfolios}

% Before the first table
\renewcommand{\arraystretch}{0.7}

\begin{table}[!htbp]
\centering
\resizebox{\textwidth}{!}{

}
\caption{Mean return (\%) $\pm$ standard error of portfolios of 25 stocks (01.2019 -- 08.2022).}
\label{tab:stock_three_return_sme}

\end{table}

% ----------------------------

\begin{table}[!htbp]
\centering
\resizebox{\textwidth}{!}{
%
}
\caption{Mean annualized log return volatility $\pm$ standard error of portfolios of 25 stocks (1.2019 -- 8.2022).}
\label{tab:stock_three_volatility_sme}
\end{table}

\begin{table}[!htbp]
\centering
\resizebox{\textwidth}{!}{
%
}
\caption{Mean return (\%) $\pm$ standard error of portfolios of 35 stocks (1.2019 -- 8.2022).}
\label{tab:converted_stock_return_data_35_stocks}
\end{table}

% Volatility test

\begin{table}[!htbp]
\centering
\resizebox{\textwidth}{!}{
%
}
\caption{Mean annualized log return volatility $\pm$ standard error of portfolios of 35 stocks (1.2019 -- 8.2022).}
\label{tab:converted_stock_volatility_data_35_stocks}
\end{table}

% ------------------

% Return 15 test

\begin{table}[!htbp]
\centering
\resizebox{\textwidth}{!}{
%
}
\caption{Mean return (\%) $\pm$ standard error of portfolios of 15 stocks (1.2019 -- 8.2022).}
\label{tab:converted_stock_return_data_15_stocks}
\end{table}

% Volatility 15 test

\begin{table}[!htbp]
\centering
\resizebox{\textwidth}{!}{
%
}
\caption{Mean annualized log return volatility $\pm$ standard error of portfolios of 15 stocks (1.2019 -- 8.2022).}
\label{tab:converted_stock_volatility_data_15_stocks}
\end{table}

% ---------------------------

% SHARP RATIO
\begin{table}[!htbp]
\centering
\footnotesize
%\resizebox{\textwidth}{!}{
{
%
} \\
\cmidrule{3-10}
& & \textit{C} & \textit{MI} & $\widetilde{\mathit{Cor}}$ & $\widetilde{\mathit{MI}}$ & \textit{C} & \textit{MI} & $\widetilde{\mathit{Cor}}$ & $\widetilde{\mathit{MI}}$ & \\
\midrule
& max & 1.36 & 1.36 & 1.48 & 1.46 & 1.35 & 1.57 & 1.52 & 1.49 & 1.45 \\ 
& med & 1.34 & 1.45 & 1.60 & 1.26 & 1.44 & 1.51 & 1.59 & 1.34 & 1.44 \\ 
\multirow{-3}{*}{\textbf{PCA}} & min & 1.36 & 1.21 & 1.60 & 1.40 & 1.33 & 1.51 & 1.44 & 1.19 & 1.38 \\ 
\midrule
& max & 0.89 & 1.26 & 1.60 & 1.43 & 1.62 & 1.35 & 1.55 & 1.24 & 1.37 \\ 
& med & 1.37 & 1.40 & 1.53 & 1.46 & 1.29 & 1.47 & 1.60 & 1.54 & 1.46 \\ 
\multirow{-3}{*}{\textbf{Degree (FG)}} & min & 1.36 & 1.49 & 1.52 & 1.53 & 1.56 & 1.63 & 1.34 & 1.60 & 1.50 \\ 
\midrule
& max & 1.28 & 1.46 & 1.60 & 1.34 & 1.40 & 1.52 & 1.51 & 1.49 & 1.45 \\ 
& med & 1.47 & 1.44 & 1.43 & 1.40 & 1.47 & 1.40 & 1.45 & 1.73 & 1.47 \\  
\multirow{-3}{*}{\textbf{Closeness (FG)}} & min & 1.22 & 0.99 & 1.60 & 1.38 & 1.46 & 1.51 & 1.55 & 1.16 & 1.36 \\ 
\midrule
& max & 1.40 & 1.49 & 1.58 & 1.53 & 1.58 & 1.62 & 1.40 & 1.52 & 1.52 \\  
& med & 1.35 & 1.41 & 1.54 & 1.30 & 1.45 & 1.60 & 1.48 & 1.60 & 1.47 \\ 
\multirow{-3}{*}{\textbf{Degree (MST)}} & min & 1.25 & 1.46 & 1.64 & 1.33 & 1.47 & 1.44 & 1.44 & 1.33 & 1.42 \\ 
\midrule
& max & 1.35 & 1.39 & 1.52 & 1.44 & 1.35 & 1.48 & 1.48 & 1.57 & 1.45 \\ 
& med & 1.41 & 1.55 & 1.55 & 1.35 & 1.50 & 1.50 & 1.53 & 1.51 & 1.49 \\ 
\multirow{-3}{*}{\textbf{Closeness (MST)}} & min & 1.20 & 0.95 & 1.58 & 1.50 & 1.50 & 1.44 & 1.51 & 1.30 & 1.37 \\  
\midrule
\textbf{Column Average} & & 1.31 & 1.35 & 1.56 & 1.41 & 1.45 & 1.50 & 1.49 & 1.44 & \textbf{1.44} \\  
\bottomrule
\end{tabular}
}
\caption{Sharpe ratio of portfolios of 25 stocks (1.2019 -- 8.2022).}
\label{tab:stock_full_period_sharp_ratio}
\end{table}

\begin{table}[!htbp]
\centering
\footnotesize
%\resizebox{\textwidth}{!}{
{

\begin{tabular}{ccccccccccc}
\toprule
\multirow{2}{*}{} & \multirow{2}{*}{} & \multicolumn{4}{c}{\textbf{Louvain}} & \multicolumn{4}{c}{\textbf{Affinity propagation}} & \multirow{2}{*}{\begin{tabular}{@{}c@{}}\textbf{Row}\\\textbf{Average}\end{tabular}} \\
\cmidrule{3-10}
& & \textit{C} & \textit{MI} & $\widetilde{\mathit{Cor}}$ & $\widetilde{\mathit{MI}}$ & \textit{C} & \textit{MI} & $\widetilde{\mathit{Cor}}$ & $\widetilde{\mathit{MI}}$ & \\
\midrule
& max & 1.41 & 1.34 & 1.50 & 1.47 & 1.45 & 1.50 & 1.57 & 1.50 & 1.47 \\  
& med & 1.37 & 1.47 & 1.57 & 1.31 & 1.49 & 1.47 & 1.55 & 1.42 & 1.46 \\  
\multirow{-3}{*}{\textbf{PCA}} & min & 1.44 & 1.35 & 1.64 & 1.29 & 1.37 & 1.47 & 1.44 & 1.30 & 1.41 \\  
\midrule
& max & 1.14 & 1.36 & 1.62 & 1.44 & 1.68 & 1.42 & 1.60 & 1.25 & 1.44 \\  
& med & 1.41 & 1.40 & 1.57 & 1.42 & 1.34 & 1.40 & 1.53 & 1.60 & 1.46 \\  
\multirow{-3}{*}{\textbf{Degree (FG)}} & min & 1.42 & 1.49 & 1.53 & 1.51 & 1.55 & 1.53 & 1.33 & 1.58 & 1.49 \\  
\midrule
& max & 1.34 & 1.44 & 1.56 & 1.35 & 1.48 & 1.43 & 1.49 & 1.53 & 1.45 \\  
& med & 1.48 & 1.49 & 1.46 & 1.43 & 1.47 & 1.40 & 1.47 & 1.68 & 1.48 \\  
\multirow{-3}{*}{\textbf{Closeness (FG)}} & min & 1.34 & 1.10 & 1.55 & 1.43 & 1.56 & 1.49 & 1.54 & 1.19 & 1.40 \\  
\midrule
& max & 1.36 & 1.43 & 1.56 & 1.50 & 1.55 & 1.52 & 1.36 & 1.52 & 1.47 \\  
& med & 1.43 & 1.49 & 1.62 & 1.44 & 1.53 & 1.55 & 1.54 & 1.46 & 1.51 \\  
\multirow{-3}{*}{\textbf{Degree (MST)}} & min & 1.38 & 1.39 & 1.60 & 1.37 & 1.53 & 1.50 & 1.57 & 1.36 & 1.46 \\  
\midrule
& max & 1.41 & 1.41 & 1.54 & 1.45 & 1.40 & 1.43 & 1.47 & 1.59 & 1.46 \\  
& med & 1.47 & 1.58 & 1.53 & 1.43 & 1.54 & 1.46 & 1.51 & 1.54 & 1.51 \\  
\multirow{-3}{*}{\textbf{Closeness (MST)}} & min & 1.34 & 0.99 & 1.58 & 1.46 & 1.56 & 1.50 & 1.50 & 1.35 & 1.41 \\  
\midrule
\textbf{Column Average} & & 1.38 & 1.38 & 1.56 & 1.42 & 1.50 & 1.47 & 1.50 & 1.46 & \textbf{1.46} \\  
\bottomrule
\end{tabular}
}
\caption{Sharpe ratio of portfolios of 35 stocks (1.2019 -- 8.2022).}
\label{tab:stock_sharp_ratio_35}
\end{table}

% -------------------

% SHARP RATIO

\begin{table}[!htbp]
\centering
\footnotesize
%\resizebox{\textwidth}{!}{
{

\begin{tabular}{ccccccccccc}
\toprule
\multirow{2}{*}{} & \multirow{2}{*}{} & \multicolumn{4}{c}{\textbf{Louvain}} & \multicolumn{4}{c}{\textbf{Affinity propagation}} & \multirow{2}{*}{\begin{tabular}{@{}c@{}}\textbf{Row}\\\textbf{Average}\end{tabular}}  \\
\cmidrule{3-10}
& & \textit{C} & \textit{MI} & $\widetilde{\mathit{Cor}}$ & $\widetilde{\mathit{MI}}$ & \textit{C} & \textit{MI} & $\widetilde{\mathit{Cor}}$ & $\widetilde{\mathit{MI}}$ & \\
\midrule
& max & 1.40 & 1.32 & 1.46 & 1.38 & 1.39 & 1.36 & 1.45 & 1.51 & 1.41 \\  
& med & 1.30 & 1.47 & 1.57 & 1.17 & 1.46 & 1.42 & 1.53 & 1.30 & 1.40 \\  
\multirow{-3}{*}{\textbf{PCA}} & min & 1.16 & 0.92 & 1.50 & 1.21 & 1.33 & 1.37 & 1.32 & 1.10 & 1.24 \\  
\midrule
& max & 0.62 & 0.97 & 1.58 & 1.39 & 1.47 & 1.21 & 1.39 & 1.17 & 1.22 \\  
& med & 1.22 & 1.37 & 1.49 & 1.30 & 1.42 & 1.38 & 1.53 & 1.56 & 1.41 \\  
\multirow{-3}{*}{\textbf{Degree (FG)}} & min & 1.36 & 1.42 & 1.53 & 1.54 & 1.61 & 1.44 & 1.29 & 1.65 & 1.48 \\  
\midrule
& max & 1.33 & 1.42 & 1.60 & 1.30 & 1.50 & 1.43 & 1.49 & 1.48 & 1.44 \\  
& med & 1.44 & 1.29 & 1.43 & 1.43 & 1.48 & 1.29 & 1.43 & 1.74 & 1.44 \\  
\multirow{-3}{*}{\textbf{Closeness (FG)}} & min & 0.83 & 0.85 & 1.51 & 1.27 & 1.19 & 1.23 & 1.40 & 1.04 & 1.17 \\  
\midrule
& max & 1.41 & 1.45 & 1.59 & 1.38 & 1.57 & 1.43 & 1.38 & 1.49 & 1.46 \\  
& med & 1.34 & 1.46 & 1.76 & 1.23 & 1.36 & 1.47 & 1.49 & 1.55 & 1.46 \\  
\multirow{-3}{*}{\textbf{Degree (MST)}} & min & 1.49 & 1.65 & 1.84 & 1.91 & 1.41 & 1.51 & 1.53 & 1.38 & 1.59 \\  
\midrule
& max & 1.32 & 1.39 & 1.53 & 1.38 & 1.42 & 1.44 & 1.40 & 1.60 & 1.43 \\  
& med & 1.39 & 1.56 & 1.62 & 1.30 & 1.58 & 1.39 & 1.52 & 1.50 & 1.48 \\  
\multirow{-3}{*}{\textbf{Closeness (MST)}} & min & 1.05 & 0.86 & 1.58 & 1.33 & 1.46 & 1.28 & 1.54 & 1.20 & 1.29 \\  
\midrule
\textbf{Column Average} & & 1.24 & 1.29 & 1.57 & 1.37 & 1.44 & 1.37 & 1.45 & 1.42 & \textbf{1.39} \\  
\bottomrule
\end{tabular}
}
\caption{Sharpe ratio of portfolios of 15 stocks (1.2019 -- 8.2022).}
\label{tab:sharp_ratio_stock_15}
\end{table}

\begin{table}[!htbp]
\centering
%\resizebox{\textwidth}{!}{
{\fontsize{6pt}{7pt}\selectfont

\begin{tabular}{cccccccccc}
\toprule
\multirow{2}{*}{} & \multirow{2}{*}{} & \multicolumn{4}{c}{\textbf{Louvain}} & \multicolumn{4}{c}{\textbf{Affinity propagation}} \\
\cmidrule{3-10}
& & \textit{C} & \textit{MI} & $\widetilde{\mathit{Cor}}$ & $\widetilde{\mathit{MI}}$ & \textit{C} & \textit{MI} & $\widetilde{\mathit{Cor}}$ & $\widetilde{\mathit{MI}}$ \\
\midrule
&max & 11.09 & 5.48 & 14.37 & 3.6 & 10.16 & 11.53 & 12.5 & 10.65 \\ 
&med & -1.26 & 6.53 & 16.07 & 3.7 & 5.63 & 8.18 & 5.36 & 9.37 \\ 

\multirow{-3}{*}{\textbf{PCA}} & min & 18.57 & 19.33 & 27.23 & 16.38 & 4.67 & 6.72 & 18.59 & 16.06 \\ 
\midrule

&max & 13.47 & 0.87 & 10.2 & 12.42 & 20.16 & 10.41 & 39.12 & 10.73 \\ 
&med & 7.31 & 9.11 & 25.09 & 29.38 & 5.27 & 8.57 & 8.61 & 21.9 \\
\multirow{-3}{*}{\textbf{Degree (FG)}} & min & 23.4 & 18.52 & 13.67 & -1.9 & 7.82 & 6.37 & 3.42 & 9.47 \\ 
\midrule
&max & 15.37 & 10.06 & 18.35 & -0.96 & 7.86 & 9.88 & 7.8 & 10.67 \\  
&med & 8.08 & 43.62 & 19.61 & 11.72 & 6.7 & 3.66 & -2.85 & 9.13 \\  
\multirow{-3}{*}{\textbf{Closeness (FG)}} &min & 12.69 & 8.65 & 19.32 & 11.81 & 8.79 & 9.74 & 33.72 & 5.21 \\ 
\midrule
&max & 11 & 4.39 & 16.5 & 1.41 & 7.41 & 9.41 & 1.87 & 6.28 \\  
&med & 3.44 & 7.13 & 10.26 & 1.93 & 6.07 & 14.48 & 15.35 & 8.62 \\ 
\multirow{-3}{*}{\textbf{Degree (MST)}} & min & -3.29 & 9.81 & 22.52 & 2.74 & 17.9 & 6.19 & 4.74 & 8.84 \\ 
\midrule
&max & 14.53 & 15.11 & 20.37 & 11.34 & 2.3 & 12.51 & 16.2 & 11.47 \\ 
&med & 45.58 & 14.94 & 16.26 & 5.17 & 7.08 & 3.6 & 32.52 & 9.51 \\ 
\multirow{-3}{*}{\textbf{Closeness (MST)}} & min & 6.43 & -8.83 & 13.19 & -0.91 & 9.45 & 3.37 & 20.59 & 9.07 \\

\bottomrule
\end{tabular}
}
\caption{Returns (\%) of portfolios of 25 stocks during COVID-19 (1.2020 -- 12.2020).}
\label{tab:stock_covid_return}
\end{table}

% -------------------

\begin{table}[!htbp]
\centering
%\resizebox{\textwidth}{!}{
{\fontsize{6pt}{7pt}\selectfont

\begin{tabular}{cccccccccc}
\toprule
\multirow{2}{*}{} & \multirow{2}{*}{} & \multicolumn{4}{c}{\textbf{Louvain}} & \multicolumn{4}{c}{\textbf{Affinity propagation}} \\
\cmidrule{3-10}
& & \textit{C} & \textit{MI} & $\widetilde{\mathit{Cor}}$ & $\widetilde{\mathit{MI}}$ & \textit{C} & \textit{MI} & $\widetilde{\mathit{Cor}}$ & $\widetilde{\mathit{MI}}$ \\
\midrule
&max & 0.398 & 0.468 & 0.391 & 0.457 & 0.403 & 0.424 & 0.405 & 0.412 \\ 
&med & 0.391 & 0.453 & 0.358 & 0.429 & 0.403 & 0.371 & 0.398 & 0.410 \\ 
\multirow{-3}{*}{\textbf{PCA}} & min & 0.219 & 0.251 & 0.213 & 0.232 & 0.320 & 0.283 & 0.372 & 0.385 \\ 
\midrule
&max & 0.237 & 0.327 & 0.296 & 0.351 & 0.280 & 0.311 & 0.292 & 0.386 \\ 
&med & 0.398 & 0.366 & 0.382 & 0.437 & 0.342 & 0.354 & 0.393 & 0.409 \\ 
\multirow{-3}{*}{\textbf{Degree (FG)}} & min & 0.410 & 0.437 & 0.406 & 0.395 & 0.403 & 0.432 & 0.409 & 0.420 \\ 
\midrule
&max & 0.435 & 0.462 & 0.402 & 0.397 & 0.378 & 0.453 & 0.385 & 0.374 \\ 
&med & 0.401 & 0.384 & 0.398 & 0.318 & 0.374 & 0.353 & 0.409 & 0.416 \\  
\multirow{-3}{*}{\textbf{Closeness (FG)}} & min & 0.234 & 0.289 & 0.296 & 0.375 & 0.315 & 0.342 & 0.304 & 0.394 \\ 
\midrule
&max & 0.412 & 0.331 & 0.429 & 0.392 & 0.407 & 0.426 & 0.410 & 0.405 \\  
&med & 0.344 & 0.384 & 0.290 & 0.351 & 0.375 & 0.333 & 0.380 & 0.365 \\ 
\multirow{-3}{*}{\textbf{Degree (MST)}} & min & 0.359 & 0.356 & 0.339 & 0.333 & 0.329 & 0.353 & 0.407 & 0.385 \\ 
\midrule
&max & 0.449 & 0.404 & 0.407 & 0.377 & 0.351 & 0.381 & 0.403 & 0.405 \\ 
&med & 0.399 & 0.362 & 0.382 & 0.371 & 0.373 & 0.434 & 0.369 & 0.416 \\ 
\midrule
&max & 0.449 & 0.404 & 0.407 & 0.377 & 0.351 & 0.381 & 0.403 & 0.405 \\ 
&med & 0.399 & 0.362 & 0.382 & 0.371 & 0.373 & 0.434 & 0.369 & 0.416 \\ 
\multirow{-3}{*}{\textbf{Closeness (MST)}} & min & 0.262 & 0.382 & 0.291 & 0.380 & 0.344 & 0.333 & 0.361 & 0.391 \\

\bottomrule
\end{tabular}
}
\caption{Annualized log return volatility of portfolios of 25 stocks during COVID-19 (1.2020 -- 12.2020).}
\label{tab:stock_covid_volatility}
\end{table}

\begin{table}[!htbp]
\centering
{\fontsize{6pt}{7pt}\selectfont
\begin{tabular}{ccccccccccc}
\toprule
\multirow{2}{*}{} & \multirow{2}{*}{} & \multicolumn{4}{c}{\textbf{Louvain}} & \multicolumn{4}{c}{\textbf{Affinity propagation}} & \multirow{2}{*}{\begin{tabular}{@{}c@{}}\textbf{Row}\\\textbf{Average}\end{tabular}} \\
\cmidrule{3-10}
& & \textit{C} & \textit{MI} & $\widetilde{\mathit{Cor}}$ & $\widetilde{\mathit{MI}}$ & \textit{C} & \textit{MI} & $\widetilde{\mathit{Cor}}$ & $\widetilde{\mathit{MI}}$ & \\
\midrule
& max & 0.19 & 0.08 & 0.45 & 0.04 & 0.16 & 0.29 & 0.36 & -0.05 & 0.19 \\  
& med & 0.03 & -0.01 & 0.38 & -0.05 & 0.09 & 0.27 & 0.22 & -0.11 & 0.10 \\  
\multirow{-3}{*}{\textbf{PCA}} & min & 0.86 & 1.03 & 1.41 & 0.78 & 0.22 & 0.46 & 0.55 & 0.34 & 0.71 \\  
\midrule
& max & 0.78 & 0.12 & 0.15 & 0.31 & 0.43 & 0.07 & 0.61 & 0.13 & 0.32 \\  
& med & 0.21 & 0.07 & 0.54 & 1.10 & 0.32 & 0.29 & 0.02 & 0.48 & 0.38 \\  
\multirow{-3}{*}{\textbf{Degree (FG)}} & min & 0.59 & 0.46 & 0.32 & -0.22 & 0.27 & 0.30 & 0.20 & -0.03 & 0.23 \\  
\midrule
& max & 0.16 & 0.17 & 0.54 & -0.22 & 0.31 & 0.17 & 0.23 & 0.17 & 0.19 \\  
& med & 0.22 & 0.37 & 0.74 & 0.27 & 0.34 & 0.28 & -0.14 & -0.05 & 0.25 \\  
\multirow{-3}{*}{\textbf{Closeness (FG)}} & min & 0.85 & 0.40 & 0.23 & 0.38 & 0.30 & -0.15 & 0.33 & -0.10 & 0.28 \\  
\midrule
& max & 0.21 & 0.62 & -0.02 & 0.03 & 0.25 & 0.38 & 0.38 & -0.04 & 0.23 \\  
& med & 0.36 & 0.21 & 0.36 & 0.53 & 0.14 & 0.40 & 0.42 & 0.28 & 0.34 \\  
\multirow{-3}{*}{\textbf{Degree (MST)}} & min & 0.36 & 0.32 & 0.83 & 0.09 & 0.40 & 0.22 & 0.09 & -0.01 & 0.29 \\  
\midrule
& max & 0.16 & 0.38 & 0.43 & 0.47 & 0.16 & 0.52 & 0.42 & 0.17 & 0.34 \\  
& med & 1.92 & 0.57 & 0.45 & 0.17 & 0.21 & -0.05 & 0.03 & -0.09 & 0.40 \\  
\multirow{-3}{*}{\textbf{Closeness (MST)}} & min & 0.47 & -0.16 & 0.41 & -0.12 & 0.30 & -0.01 & 0.58 & 0.10 & 0.19 \\  
\midrule
\textbf{Column Average} & & 0.49 & 0.31 & 0.48 & 0.24 & 0.26 & 0.23 & 0.29 & 0.08 & \textbf{0.30} \\  
\bottomrule
\end{tabular}
}
\caption{Sharpe ratio of portfolios of 25 stocks during COVID-19 (1.2020 -- 12.2020).}
\label{tab:stock_covid_sharp_ratio}
\end{table}

% ----------------------

\begin{table}[!htbp]
\centering
%\resizebox{\textwidth}{!}{
{\fontsize{6pt}{7pt}\selectfont
\begin{tabular}{cccccccccc}
\toprule
\multirow{2}{*}{} & \multirow{2}{*}{} & \multicolumn{4}{c}{\textbf{Louvain}} & \multicolumn{4}{c}{\textbf{Affinity propagation}} \\
\cmidrule{3-10}
& & \textit{C} & \textit{MI} & $\widetilde{\mathit{Cor}}$ & $\widetilde{\mathit{MI}}$ & \textit{C} & \textit{MI} & $\widetilde{\mathit{Cor}}$ & $\widetilde{\mathit{MI}}$ \\
\midrule
&max & 9.16 & 3.6 & 3.88 & -3.01 & 0.56 & 2.11 & -4.13 & -2.03 \\ 
&med & -5.54 & -3.84 & -0.43 & -2.71 & -0.2 & 1.67 & -5.41 & -4.73 \\ 
\multirow{-3}{*}{\textbf{PCA}} & min & -5.25 & 0.61 & -6.89 & -3.95 & -2.28 & 4.12 & -9.25 & -5.35 \\ 
\midrule
&max & -8.67 & 2.72 & -7.59 & -3.82 & -8.98 & -9.59 & -7.07 & -3.8 \\ 
&med & -11.12 & 9.98 & -3.47 & -6.09 & -6.27 & 4.36 & -1.38 & -3.85 \\ 
\multirow{-3}{*}{\textbf{Degree (FG)}} & min & 5.06 & 2.22 & 0.12 & 3.46 & 2.54 & 0.51 & 1.32 & -0.97 \\ 
\midrule
&max & -1.81 & -3.29 & -1.68 & 0.26 & -0.71 & -3.94 & 6.41 & -0.97 \\ 
&med & 1.05 & -12.44 & -0.89 & -9.66 & -0.71 & -7.86 & 0.68 & 2.47 \\ 
\multirow{-3}{*}{\textbf{Closeness (FG)}} & min & -7.08 & -4.99 & -2.99 & 2.38 & -12.29 & -6.64 & -10.95 & -4.9 \\ 
\midrule
&max & 3.93 & 2.48 & -0.29 & 3.06 & 1.91 & 2.16 & -2.99 & 1.3 \\ 
&med & -14.16 & -4.05 & 5.66 & -1.76 & -1.74 & -4.39 & -2.6 & -0.14 \\ 
\multirow{-3}{*}{\textbf{Degree (MST)}} & min & -3.4 & -5.41 & -6.87 & -5.64 & -9.59 & -3.21 & -13.79 & 17.39 \\ 
\midrule
&max & -0.44 & -7.84 & -1.36 & 0.66 & -4.78 & -3.48 & -5.74 & 22.78 \\ 
&med & 3.47 & -4.91 & -11.36 & -4.18 & -4.47 & -3.06 & -3.64 & -7.96 \\ 
\multirow{-3}{*}{\textbf{Closeness (MST)}} & min & -2.26 & -1.3 & 7.41 & -8 & -3.04 & -3.36 & -7.08 & -10.69 \\

\bottomrule
\end{tabular}
}
\caption{Returns (\%) of portfolios of 25 stocks during the Russian invasion of Ukraine (8.2021 -- 7.2022).}
\label{tab:stock_war_return}
\end{table}

% --------------

\begin{table}[!htbp]
\centering
{\fontsize{6pt}{7pt}\selectfont
\begin{tabular}{cccccccccc}
\toprule
\multirow{2}{*}{} & \multirow{2}{*}{} & \multicolumn{4}{c}{\textbf{Louvain}} & \multicolumn{4}{c}{\textbf{Affinity propagation}} \\
\cmidrule{3-10}
& & \textit{C} & \textit{MI} & $\widetilde{\mathit{Cor}}$ & $\widetilde{\mathit{MI}}$ & \textit{C} & \textit{MI} & $\widetilde{\mathit{Cor}}$ & $\widetilde{\mathit{MI}}$ \\
\midrule
&max & 0.187 & 0.258 & 0.263 & 0.247 & 0.198 & 0.232 & 0.238 & 0.277 \\ 
&med & 0.183 & 0.181 & 0.178 & 0.191 & 0.214 & 0.192 & 0.217 & 0.227 \\ 
\multirow{-3}{*}{\textbf{PCA}} & min & 0.132 & 0.137 & 0.164 & 0.134 & 0.171 & 0.192 & 0.229 & 0.141 \\ 
\midrule
&max & 0.175 & 0.182 & 0.166 & 0.188 & 0.175 & 0.180 & 0.222 & 0.162 \\ 
&med & 0.192 & 0.204 & 0.194 & 0.175 & 0.209 & 0.198 & 0.246 & 0.214 \\ 
\multirow{-3}{*}{\textbf{Degree (FG)}} & min & 0.193 & 0.202 & 0.195 & 0.204 & 0.202 & 0.211 & 0.208 & 0.278 \\ 
\midrule
&max & 0.201 & 0.233 & 0.190 & 0.231 & 0.195 & 0.217 & 0.226 & 0.280 \\ 
&med & 0.208 & 0.223 & 0.199 & 0.188 & 0.188 & 0.224 & 0.249 & 0.204 \\  
\multirow{-3}{*}{\textbf{Closeness (FG)}} & min & 0.151 & 0.160 & 0.169 & 0.178 & 0.206 & 0.192 & 0.231 & 0.162 \\ 
\midrule
&max & 0.169 & 0.186 & 0.177 & 0.215 & 0.199 & 0.224 & 0.216 & 0.203 \\  
&med & 0.208 & 0.174 & 0.224 & 0.200 & 0.213 & 0.218 & 0.225 & 0.222 \\ 
\multirow{-3}{*}{\textbf{Degree (MST)}} & min & 0.182 & 0.229 & 0.190 & 0.189 & 0.189 & 0.210 & 0.219 & 0.236 \\ 
\midrule
&max & 0.215 & 0.261 & 0.183 & 0.164 & 0.175 & 0.209 & 0.241 & 0.295 \\ 
&med & 0.210 & 0.183 & 0.218 & 0.191 & 0.199 & 0.210 & 0.224 & 0.183 \\ 
\multirow{-3}{*}{\textbf{Closeness (MST)}} & min & 0.159 & 0.150 & 0.236 & 0.212 & 0.198 & 0.180 & 0.218 & 0.177 \\

\bottomrule
\end{tabular}
}
\caption{Annualized log return volatility of portfolios of 25 stocks during the Russian invasion of Ukraine (8.2021 -- 7.2022).}
\label{tab:stock_war_volatility}
\end{table}

% -----------------------------

% SHARP RATIO
\begin{table}[!htbp]
\centering
{\fontsize{6pt}{7pt}\selectfont
\begin{tabular}{ccccccccccc}
\toprule
\multirow{2}{*}{} & \multirow{2}{*}{} & \multicolumn{4}{c}{\textbf{Louvain}} & \multicolumn{4}{c}{\textbf{Affinity propagation}} & \multirow{2}{*}{\begin{tabular}{@{}c@{}}\textbf{Row}\\\textbf{Average}\end{tabular}} \\
\cmidrule{3-10}
& & \textit{C} & \textit{MI} & $\widetilde{\mathit{Cor}}$ & $\widetilde{\mathit{MI}}$ & \textit{C} & \textit{MI} & $\widetilde{\mathit{Cor}}$ & $\widetilde{\mathit{MI}}$ & \\
\midrule
& max & 1.21 & 0.30 & 0.13 & -0.11 & 0.30 & 0.23 & -0.02 & 0.10 & 0.27 \\  
& med & -0.39 & -0.25 & 0.27 & -0.15 & 0.16 & -0.14 & -0.27 & -0.26 & -0.13 \\  
\multirow{-3}{*}{\textbf{PCA}} & min & 0.49 & 0.11 & -0.91 & -0.36 & -0.16 & -0.19 & -0.45 & -0.42 & -0.24 \\  
\midrule
& max & -1.24 & 0.21 & -0.46 & -0.62 & -0.23 & -0.53 & -0.13 & 0.06 & -0.37 \\  
& med & -0.18 & 0.78 & -0.41 & -0.01 & -0.18 & 0.30 & 0.37 & -0.19 & 0.06 \\  
\multirow{-3}{*}{\textbf{Degree (FG)}} & min & 0.03 & 0.15 & 0.51 & -0.24 & 0.15 & 0.03 & 0.12 & 0.03 & 0.10 \\  
\midrule
& max & -0.00 & -0.03 & 0.16 & -0.19 & 0.11 & -0.42 & 0.79 & -0.02 & 0.05 \\  
& med & 0.06 & -0.54 & -0.26 & -0.78 & 0.06 & -0.59 & 0.37 & 0.09 & -0.20 \\  
\multirow{-3}{*}{\textbf{Closeness (FG)}} & min & -0.92 & -0.23 & -0.44 & 0.43 & -0.35 & -0.64 & -0.46 & 0.06 & -0.32 \\  
\midrule
& max & 0.37 & -0.08 & 0.09 & 0.31 & 0.13 & 0.10 & -0.01 & -0.30 & 0.08 \\  
& med & 0.17 & -0.05 & 0.39 & 0.00 & 0.05 & -0.17 & 0.25 & -0.15 & 0.06 \\  
\multirow{-3}{*}{\textbf{Degree (MST)}} & min & -0.13 & -0.30 & 0.15 & 0.18 & -0.40 & -0.56 & -0.59 & 0.30 & -0.17 \\  
\midrule
& max & -0.20 & -0.30 & 0.05 & 0.20 & -0.36 & -0.18 & 0.04 & 0.79 & 0.01 \\  
& med & 0.55 & -0.12 & -0.69 & -0.15 & -0.17 & -0.23 & 0.10 & -0.29 & -0.13 \\  
\multirow{-3}{*}{\textbf{Closeness (MST)}} & min & 0.04 & -0.15 & 0.40 & -0.45 & -0.31 & -0.18 & -0.28 & -0.84 & -0.22 \\  
\midrule
\textbf{Column Average} & & -0.01 & -0.03 & -0.07 & -0.13 & -0.08 & -0.21 & -0.01 & -0.07 & \textbf{-0.08} \\  
\bottomrule
\end{tabular}
}
\caption{Sharpe ratio of portfolios of 25 stocks during the Russian invasion of Ukraine (8.2021 -- 7.2022).}
\label{tab:stock_war_sharp_ratio}
\end{table}

\clearpage
% -------------------

\section{Cryptocurrency portfolios}

\begin{table}[!htbp]
\centering
\resizebox{\textwidth}{!}{
%{\footnotesize

}
\caption{Mean return (\%) $\pm$ standard error of portfolios of 20 cryptocurrencies (1.2019 -- 8.2022).}
\label{tab:crypto_three_return_sme}
\end{table}

% % ------------------

\begin{table}[!htbp]
\centering
\resizebox{\textwidth}{!}{
%
}
\caption{Mean annualized log return volatility $\pm$ standard error of portfolios of 20 cryptocurrencies (1.2019 -- 8.2022).}
\label{tab:crypto_three_volatility_sme}
\end{table}

\begin{table}[!htbp]
\centering
\resizebox{\textwidth}{!}{
%{\footnotesize
%
}
\caption{Mean return (\%) $\pm$ standard error of portfolios of 30 cryptocurrencies (1.2019 -- 8.2022).}
\label{tab:crypto_return_data_30_crypto}
\end{table}

% Volatility Test

\begin{table}[!htbp]
\centering
\resizebox{\textwidth}{!}{
%
}
\caption{Mean annualized log return volatility $\pm$ standard error of portfolios of 30 cryptocurrencies (1.2019 -- 8.2022).}
\label{tab:crypto_volatility_data_30_crypto}
\end{table}

\begin{table}[!htbp]
\centering
\resizebox{\textwidth}{!}{
%{\footnotesize
%
}
\caption{Mean return (\%) $\pm$ standard error of portfolios of 10 cryptocurrencies (1.2019 -- 8.2022).}
\label{tab:crypto_return_data_10_crypto}
\end{table}

% Return 10 Volatility

\begin{table}[!htbp]
\centering
\resizebox{\textwidth}{!}{
%
}
\caption{Mean annualized log return volatility $\pm$ standard error of portfolios of 10 cryptocurrencies (1.2019 -- 8.2022).}
\label{tab:crypto_volatility_data_10_crypto}
\end{table}

% % ------------------

\begin{table}[!htbp]
\centering
\footnotesize{
%
} \\
\cmidrule{3-10}
& & \textit{C} & \textit{MI} & $\widetilde{\mathit{Cor}}$ & $\widetilde{\mathit{MI}}$ & \textit{C} & \textit{MI} & $\widetilde{\mathit{Cor}}$ & $\widetilde{\mathit{MI}}$ & \\
\midrule
& max & 2.80 & 2.54 & 2.57 & 2.53 & 6.31 & 1.83 & 5.03 & 2.50 & 3.26 \\  
& med & 3.44 & 3.03 & 2.68 & 3.10 & 6.21 & 3.05 & 2.64 & 4.27 & 3.55 \\  
\multirow{-3}{*}{\textbf{PCA}} & min & 2.85 & 4.38 & 1.80 & 5.12 & 6.64 & 4.77 & 5.68 & 3.86 & 4.39 \\  
\midrule
& max & 6.05 & 4.77 & 3.70 & 2.28 & 6.68 & 2.59 & 6.50 & 2.81 & 4.42 \\  
& med & 2.62 & 2.86 & 3.35 & 2.83 & 6.07 & 1.81 & 2.32 & 5.05 & 3.36 \\  
\multirow{-3}{*}{\textbf{Degree (FG)}} & min & 2.88 & 2.91 & 2.14 & 2.40 & 6.32 & 1.84 & 2.65 & 2.59 & 2.96 \\  
\midrule
& max & 3.19 & 2.88 & 3.07 & 3.19 & 6.37 & 1.56 & 3.26 & 3.33 & 3.36 \\  
& med & 2.03 & 1.89 & 2.66 & 2.86 & 6.17 & 2.76 & 2.30 & 4.49 & 3.14 \\  
\multirow{-3}{*}{\textbf{Closeness (FG)}} & min & 5.88 & 4.86 & 2.47 & 3.10 & 6.70 & 2.58 & 3.02 & 3.33 & 3.99 \\  
\midrule
& max & 3.11 & 1.69 & 2.65 & 3.18 & 6.37 & 2.08 & 3.37 & 3.39 & 3.23 \\  
& med & 4.56 & 2.93 & 2.57 & 2.63 & 6.11 & 2.34 & 3.24 & 1.68 & 3.26 \\  
\multirow{-3}{*}{\textbf{Degree (MST)}} & min & 5.18 & 5.53 & 3.20 & 2.77 & 6.19 & 2.92 & 6.63 & 6.11 & 4.81 \\  
\midrule
& max & 3.43 & 4.16 & 2.67 & 2.89 & 6.35 & 2.12 & 5.53 & 3.55 & 3.84 \\  
& med & 2.46 & 2.23 & 2.92 & 2.97 & 6.22 & 2.28 & 2.67 & 3.06 & 3.10 \\  
\multirow{-3}{*}{\textbf{Closeness (MST)}} & min & 2.38 & 3.16 & 2.75 & 3.18 & 6.87 & 3.42 & 2.17 & 3.52 & 3.43 \\  
\midrule
\textbf{Column Average} & & 3.52 & 3.32 & 2.75 & 3.00 & 6.37 & 2.53 & 3.80 & 3.57 & \textbf{3.61} \\  
\bottomrule
\end{tabular}
}
\caption{Sharpe ratio of portfolios of 20 cryptocurrencies (1.2019 -- 8.2022).}
\label{tab:crypto_sharp_ratio_full_period}
\end{table}

\begin{table}[!htbp]
\centering
\footnotesize{
\begin{tabular}{ccccccccccc}
\toprule
\multirow{2}{*}{} & \multirow{2}{*}{} & \multicolumn{4}{c}{\textbf{Louvain}} & \multicolumn{4}{c}{\textbf{Affinity propagation}} & \multirow{2}{*}{\begin{tabular}{@{}c@{}}\textbf{Row}\\\textbf{Average}\end{tabular}} \\
\cmidrule{3-10}
& & \textit{C} & \textit{MI} & $\widetilde{\mathit{Cor}}$ & $\widetilde{\mathit{MI}}$ & \textit{C} & \textit{MI} & $\widetilde{\mathit{Cor}}$ & $\widetilde{\mathit{MI}}$ & \\
\midrule
& max & 3.21 & 2.28 & 2.64 & 2.80 & 3.24 & 1.67 & 2.51 & 2.88 & 2.65 \\  
& med & 2.16 & 2.13 & 3.13 & 3.15 & 2.60 & 2.68 & 2.47 & 2.65 & 2.62 \\  
\multirow{-3}{*}{\textbf{PCA}} & min & 1.92 & 2.10 & 1.57 & 4.44 & 2.70 & 2.28 & 2.56 & 2.19 & 2.47 \\  
\midrule
& max & 2.77 & 1.95 & 2.42 & 5.32 & 2.97 & 1.74 & 3.28 & 2.82 & 2.91 \\  
& med & 2.02 & 3.52 & 2.07 & 1.98 & 3.67 & 2.66 & 2.84 & 2.48 & 2.65 \\  
\multirow{-3}{*}{\textbf{Degree (FG)}} & min & 3.19 & 2.89 & 2.22 & 2.07 & 2.57 & 1.98 & 2.01 & 2.83 & 2.47 \\  
\midrule
& max & 3.36 & 2.73 & 2.87 & 1.58 & 3.45 & 2.89 & 2.70 & 2.79 & 2.80 \\  
& med & 1.42 & 3.28 & 2.34 & 2.51 & 2.96 & 2.15 & 2.03 & 2.42 & 2.39 \\  
\multirow{-3}{*}{\textbf{Closeness (FG)}} & min & 2.29 & 2.05 & 1.94 & 4.80 & 2.58 & 2.18 & 4.51 & 2.42 & 2.85 \\  
\midrule
& max & 1.77 & 2.52 & 2.55 & 3.77 & 2.93 & 2.65 & 2.33 & 2.91 & 2.68 \\  
& med & 2.74 & 2.77 & 2.52 & 2.18 & 2.87 & 1.55 & 1.81 & 1.97 & 2.30 \\  
\multirow{-3}{*}{\textbf{Degree (MST)}} & min & 1.34 & 1.75 & 1.71 & 3.38 & 3.14 & 2.02 & 2.13 & 1.72 & 2.15 \\  
\midrule
& max & 3.89 & 3.37 & 2.45 & 3.54 & 3.24 & 2.80 & 2.74 & 3.34 & 3.17 \\  
& med & 2.55 & 1.80 & 3.71 & 3.28 & 3.08 & 1.82 & 2.03 & 1.97 & 2.53 \\  
\multirow{-3}{*}{\textbf{Closeness (MST)}} & min & 1.94 & 2.66 & 1.75 & 2.10 & 2.32 & 2.24 & 1.93 & 2.51 & 2.18 \\  
\midrule
\textbf{Column Average} & & 2.44 & 2.52 & 2.39 & 3.13 & 2.95 & 2.22 & 2.53 & 2.53 & \textbf{2.59} \\  
\bottomrule
\end{tabular}
}
\caption{Sharpe ratio of portfolios of 30 cryptocurrencies (1.2019 -- 8.2022).}
\label{tab:crypto_sharp_ratio_30_coins}
\end{table}

\begin{table}[!htbp]
\centering
\footnotesize{
\begin{tabular}{ccccccccccc}
\toprule
\multirow{2}{*}{} & \multirow{2}{*}{} & \multicolumn{4}{c}{\textbf{Louvain}} & \multicolumn{4}{c}{\textbf{Affinity propagation}} & \multirow{2}{*}{\begin{tabular}{@{}c@{}}\textbf{Row}\\\textbf{Average}\end{tabular}} \\
\cmidrule{3-10}
& & \textit{C} & \textit{MI} & $\widetilde{\mathit{Cor}}$ & $\widetilde{\mathit{MI}}$ & \textit{C} & \textit{MI} & $\widetilde{\mathit{Cor}}$ & $\widetilde{\mathit{MI}}$ & \\
\midrule
& max & 2.31 & 2.73 & 3.15 & 1.96 & 1.14 & 2.98 & 2.68 & 2.57 & 2.44 \\  
& med & 4.10 & 2.54 & 2.37 & 2.65 & 1.00 & 4.98 & 1.93 & 4.00 & 2.95 \\  
\multirow{-3}{*}{\textbf{PCA}} & min & 2.70 & 5.79 & 1.57 & 2.45 & 2.38 & 3.54 & 6.31 & 3.54 & 3.54 \\  
\midrule
& max & 3.22 & 7.61 & 3.28 & 1.80 & 2.11 & 3.78 & 4.49 & 2.74 & 3.63 \\  
& med & 2.24 & 2.62 & 3.51 & 3.59 & 1.03 & 3.63 & 1.38 & 2.62 & 2.58 \\  
\multirow{-3}{*}{\textbf{Degree (FG)}} & min & 3.10 & 3.16 & 1.75 & 2.61 & 1.38 & 2.87 & 2.99 & 2.93 & 2.60 \\  
\midrule
& max & 2.76 & 3.47 & 3.01 & 2.83 & 1.36 & 2.50 & 2.59 & 2.86 & 2.67 \\  
& med & 1.74 & 1.68 & 3.02 & 1.87 & 1.30 & 3.22 & 1.74 & 6.25 & 2.60 \\  
\multirow{-3}{*}{\textbf{Closeness (FG)}} & min & 1.54 & 6.05 & 1.65 & 2.44 & 2.07 & 4.04 & 3.10 & 2.21 & 2.89 \\  
\midrule
& max & 3.09 & 3.24 & 2.26 & 3.03 & 1.24 & 3.33 & 3.68 & 3.00 & 2.86 \\  
& med & 1.51 & 4.06 & 2.85 & 2.38 & 0.87 & 2.68 & 3.41 & 5.95 & 2.97 \\  
\multirow{-3}{*}{\textbf{Degree (MST)}} & min & 5.68 & 5.15 & 3.36 & 3.71 & 1.63 & 3.18 & 2.04 & 3.53 & 3.53 \\  
\midrule
& max & 3.17 & 3.90 & 2.85 & 2.70 & 1.36 & 3.43 & 5.39 & 2.37 & 3.15 \\  
& med & 1.82 & 2.35 & 3.37 & 2.72 & 1.18 & 2.72 & 1.82 & 1.67 & 2.21 \\  
\multirow{-3}{*}{\textbf{Closeness (MST)}} & min & 3.42 & 3.39 & 2.27 & 2.88 & 2.41 & 4.46 & 1.93 & 3.32 & 3.01 \\  
\midrule
\textbf{Column Average} & & 2.83 & 3.85 & 2.68 & 2.64 & 1.50 & 3.42 & 3.03 & 3.30 & \textbf{2.91} \\  
\bottomrule
\end{tabular}
}
\caption{Sharpe ratio of portfolios of 10 cryptocurrencies (1.2019 -- 8.2022).}
\label{tab:crypto_10_sharp_ratio}
\end{table}

\begin{table}[!htbp]
\centering
%\resizebox{\textwidth}{!}{
{\fontsize{6pt}{7pt}\selectfont
\begin{tabular}{cccccccccc}
\toprule
\multirow{2}{*}{} & \multirow{2}{*}{} & \multicolumn{4}{c}{\textbf{Louvain}} & \multicolumn{4}{c}{\textbf{Affinity propagation}} \\
\cmidrule{3-10}
& & \textit{C} & \textit{MI} & $\widetilde{\mathit{Cor}}$ & $\widetilde{\mathit{MI}}$ & \textit{C} & \textit{MI} & $\widetilde{\mathit{Cor}}$ & $\widetilde{\mathit{MI}}$ \\
\midrule
& max & 153.77 & 69.93 & 104.04 & 118.15 & 331.11 & 376.74 & 2652.26 & 123.63 \\ 
& med & 2612.81 & 251.63 & 312.7 & 77.75 & 289.85 & 490.25 & 196.59 & 2772.85 \\ 
\multirow{-3}{*}{\textbf{PCA}} & min & 496.63 & 101.81 & 338.58 & 373.89 & 426.92 & 352.55 & 102.62 & 306.71 \\ 
\midrule
& max & 379.45 & 355.85 & 309.46 & 103.39 & 468.64 & 478.02 & 2700.43 & 66.67 \\ 
& med & 206.2 & 92.03 & 76.22 & 183.78 & 278.75 & 433.62 & 104.58 & 82.53 \\ 
\multirow{-3}{*}{\textbf{Degree (FG)}} & min & 120.94 & 101.03 & 184.21 & 66.81 & 329.92 & 353.93 & 215.46 & 215.4 \\ 
\midrule
& max & 97.49 & 166.36 & 213.96 & 121.7 & 322.58 & 355.59 & 193.27 & 132.41 \\ 
& med & 59.82 & 35.73 & 118.31 & 289.05 & 287.2 & 281.43 & 75.24 & 113.08 \\ 
\multirow{-3}{*}{\textbf{Closeness (FG)}} & min & 358.8 & 342.47 & 307.27 & 34.49 & 453.8 & 324.25 & 136.82 & 2599.67 \\ 
\midrule
& max & 177.67 & 83.94 & 90.95 & 206.41 & 325.65 & 364.92 & 100.37 & 211.96 \\ 
& med & 228.08 & 138.83 & 202.32 & 84.97 & 264.79 & 384.23 & 261.94 & 295.8 \\ 
\multirow{-3}{*}{\textbf{Degree (MST)}} & min & 306.42 & 33.78 & 137.39 & 81.22 & 283.65 & 377.8 & 2715.91 & 62.74 \\ 
\midrule
& max & 193.86 & 113.22 & 167.18 & 125.01 & 335.28 & 330.61 & 2701.89 & 125.93 \\ 
& med & 177.77 & 87.46 & 100.61 & 89.33 & 259.68 & 391.04 & 139.77 & 2906.11 \\ 
\multirow{-3}{*}{\textbf{Closeness (MST)}} & min & 244.81 & 142.87 & 240.64 & 166.66 & 424.33 & 413.76 & 128.63 & 53.94 \\

\bottomrule
\end{tabular}
}
\caption{Return (\%) of portfolios of 20 cryptocurrencies during COVID-19 (1.2020 -- 12.2020).}
\label{tab:crypto_covid_return}
\end{table}

% ---------------------------

\begin{table}[!htbp]
\centering
{\fontsize{6pt}{7pt}\selectfont
\begin{tabular}{cccccccccc}
\toprule
\multirow{2}{*}{} & \multirow{2}{*}{} & \multicolumn{4}{c}{\textbf{Louvain}} & \multicolumn{4}{c}{\textbf{Affinity propagation}} \\
\cmidrule{3-10}
& & \textit{C} & \textit{MI} & $\widetilde{\mathit{Cor}}$ & $\widetilde{\mathit{MI}}$ & \textit{C} & \textit{MI} & $\widetilde{\mathit{Cor}}$ & $\widetilde{\mathit{MI}}$ \\
\midrule
& max & 0.972 & 0.867 & 0.925 & 0.945 & 0.961 & 0.953 & 1.549 & 0.944 \\ 
& med & 1.523 & 0.945 & 2.119 & 0.899 & 0.989 & 0.929 & 1.441 & 1.541 \\ 
\multirow{-3}{*}{\textbf{PCA}} & min & 1.416 & 0.993 & 0.998 & 0.976 & 2.541 & 1.032 & 0.962 & 0.989 \\ 
\midrule
& max & 1.049 & 0.958 & 2.404 & 0.888 & 2.567 & 0.978 & 1.525 & 0.969 \\ 
& med & 0.959 & 0.948 & 0.910 & 1.540 & 0.983 & 0.994 & 0.918 & 0.981 \\ 
\multirow{-3}{*}{\textbf{Degree (FG)}} & min & 1.376 & 0.923 & 0.973 & 0.940 & 0.966 & 0.974 & 0.959 & 0.931 \\ 
\midrule
& max & 0.942 & 0.943 & 0.969 & 0.929 & 0.989 & 1.013 & 0.941 & 0.923 \\ 
& med & 0.862 & 0.924 & 0.952 & 2.062 & 1.009 & 1.031 & 1.610 & 1.573 \\ 
\multirow{-3}{*}{\textbf{Closeness (FG)}} & min & 1.009 & 0.965 & 2.311 & 0.883 & 2.622 & 1.023 & 1.065 & 1.570 \\ 
\midrule
& max & 1.012 & 0.925 & 0.932 & 1.000 & 0.981 & 0.981 & 0.961 & 0.944 \\ 
& med & 0.977 & 0.960 & 0.947 & 0.971 & 0.999 & 0.943 & 0.855 & 1.021 \\ 
\multirow{-3}{*}{\textbf{Degree (MST)}} & min & 1.483 & 1.574 & 1.525 & 1.679 & 1.046 & 1.042 & 1.552 & 0.979 \\ 
\midrule
& max & 0.932 & 0.923 & 0.972 & 0.962 & 0.946 & 0.993 & 1.500 & 0.923 \\ 
& med & 0.949 & 0.873 & 0.935 & 1.510 & 1.033 & 1.038 & 0.949 & 1.511 \\ 
\multirow{-3}{*}{\textbf{Closeness (MST)}} & min & 2.456 & 0.996 & 2.407 & 0.986 & 2.518 & 1.013 & 1.065 & 0.892 \\ 
\bottomrule
\end{tabular}
}
\caption{Annualized log return volatility of portfolios of 20 cryptocurrencies during COVID-19 (1.2020 -- 12.2020).}
\label{tab:crypto_covid_volatility}
\end{table}

% -------------------------

% SHARP RATIO

\begin{table}[!htbp]
\centering
{\fontsize{6pt}{7pt}\selectfont
\begin{tabular}{ccccccccccc}
\toprule
\multirow{2}{*}{} & \multirow{2}{*}{} & \multicolumn{4}{c}{\textbf{Louvain}} & \multicolumn{4}{c}{\textbf{Affinity propagation}} & \multirow{2}{*}{\begin{tabular}{@{}c@{}}\textbf{Row}\\\textbf{Average}\end{tabular}} \\
\cmidrule{3-10}
& & \textit{C} & \textit{MI} & $\widetilde{\mathit{Cor}}$ & $\widetilde{\mathit{MI}}$ & \textit{C} & \textit{MI} & $\widetilde{\mathit{Cor}}$ & $\widetilde{\mathit{MI}}$ & \\
\midrule
& max & 1.58 & 0.80 & 1.12 & 1.25 & 3.44 & 3.95 & 17.12 & 1.31 & 3.82 \\  
& med & 17.16 & 2.66 & 1.47 & 0.86 & 2.93 & 5.27 & 1.36 & 17.99 & 6.21 \\  
\multirow{-3}{*}{\textbf{PCA}} & min & 3.51 & 1.02 & 3.39 & 3.83 & 1.68 & 3.41 & 1.06 & 3.10 & 2.62 \\  
\midrule
& max & 3.61 & 3.71 & 1.29 & 1.16 & 1.82 & 4.89 & 17.71 & 0.69 & 4.36 \\  
& med & 2.15 & 0.97 & 0.83 & 1.19 & 2.83 & 4.36 & 1.14 & 0.84 & 1.79 \\  
\multirow{-3}{*}{\textbf{Degree (FG)}} & min & 0.88 & 1.09 & 1.89 & 0.71 & 3.41 & 3.63 & 2.24 & 2.31 & 2.02 \\  
\midrule
& max & 1.03 & 1.76 & 2.21 & 1.31 & 3.26 & 3.51 & 2.05 & 1.43 & 2.07 \\  
& med & 0.69 & 0.38 & 1.24 & 1.40 & 2.84 & 2.73 & 0.47 & 0.72 & 1.31 \\  
\multirow{-3}{*}{\textbf{Closeness (FG)}} & min & 3.55 & 3.55 & 1.33 & 0.39 & 1.73 & 3.17 & 1.28 & 16.56 & 3.94 \\  
\midrule
& max & 1.75 & 0.90 & 0.97 & 2.06 & 3.32 & 3.72 & 1.04 & 2.24 & 2.00 \\  
& med & 2.33 & 1.44 & 2.13 & 0.87 & 2.65 & 4.07 & 3.06 & 2.89 & 2.43 \\  
\multirow{-3}{*}{\textbf{Degree (MST)}} & min & 2.06 & 0.21 & 0.90 & 0.48 & 2.71 & 3.62 & 17.49 & 0.64 & 3.51 \\  
\midrule
& max & 2.08 & 1.22 & 1.72 & 1.30 & 3.54 & 3.33 & 18.01 & 1.36 & 4.07 \\  
& med & 1.87 & 1.00 & 1.07 & 0.59 & 2.51 & 3.76 & 1.47 & 19.23 & 3.94 \\  
\multirow{-3}{*}{\textbf{Closeness (MST)}} & min & 1.00 & 1.43 & 1.00 & 1.69 & 1.68 & 4.08 & 1.21 & 0.60 & 1.59 \\  
\midrule
\textbf{Column Average} & & 3.02 & 1.48 & 1.50 & 1.27 & 2.69 & 3.83 & 5.78 & 4.79 & \textbf{3.05} \\  
\bottomrule
\end{tabular}
}
\caption{Sharpe ratio of portfolios of 20 cryptocurrencies during COVID-19 (1.2020 -- 12.2020).}
\label{tab:crypto_sharp_ratio_covid_period}
\end{table}

\begin{table}[!htbp]
\centering
%\resizebox{\textwidth}{!}{
{\fontsize{6pt}{7pt}\selectfont
\begin{tabular}{cccccccccc}
\toprule
\multirow{2}{*}{} & \multirow{2}{*}{} & \multicolumn{4}{c}{\textbf{Louvain}} & \multicolumn{4}{c}{\textbf{Affinity propagation}} \\
\cmidrule{3-10}
& & \textit{C} & \textit{MI} & $\widetilde{\mathit{Cor}}$ & $\widetilde{\mathit{MI}}$ & \textit{C} & \textit{MI} & $\widetilde{\mathit{Cor}}$ & $\widetilde{\mathit{MI}}$ \\
\midrule
&max & -74.95 & -75.13 & -72.54 & -74.38 & -50.93 & -67.47 & -72.02 & -69.25 \\ 
&med & -69.87 & -70.01 & -51.66 & -72.71 & -51.18 & -55.37 & -64.07 & -65.08 \\ 
\multirow{-3}{*}{\textbf{PCA}} & min & -45.95 & -25.4 & -55.69 & -55.02 & -53.35 & -69.84 & -63.06 & -60.27 \\ 
\midrule
&max & -55.44 & -45.82 & -58.03 & -81.71 & -42.91 & -34.28 & -69.52 & -62.62 \\ 
&med & -66.85 & -79.68 & -69.72 & -57.7 & -57.33 & -72.08 & -71.82 & -70.88 \\ 
\multirow{-3}{*}{\textbf{Degree (FG)}} & min & -66.19 & -70.97 & -60.2 & -67.88 & -51.7 & -66.12 & -68.34 & -68.38 \\ 
\midrule
&max & -71.52 & -69.12 & -69.88 & -46.64 & -49.83 & -70.11 & -70.77 & -65.64 \\ 
&med & -63 & -67.28 & -59.71 & -79.22 & -53.04 & -68.96 & -70.25 & -76.03 \\ 
\multirow{-3}{*}{\textbf{Closeness (FG)}} & min & -57.29 & -61.44 & -52.2 & -55.37 & -43.65 & -40.55 & -59.78 & -68.55 \\ 
\midrule
&max & -70.34 & -70.67 & -75.31 & -61.14 & -52.66 & -67.35 & -66.57 & -62.46 \\ 
&med & -71.18 & -69.34 & -57.58 & -72.81 & -54.6 & -71.61 & -77.14 & -71.6 \\ 
\multirow{-3}{*}{\textbf{Degree (MST)}} & min & -62.89 & -64.54 & -60.07 & -74.47 & -44.8 & -67.09 & -37.87 & -68.82 \\ 
\midrule
&max & -76.07 & -56.86 & -59.72 & -65.66 & -51.62 & -66.46 & -71.78 & -69.9 \\ 
&med & -57.8 & -67.1 & -73.19 & -76.56 & -50.34 & -38.69 & -73.93 & -72.46 \\ 
\multirow{-3}{*}{\textbf{Closeness (MST)}} & min & -70 & -72.5 & -59.8 & -77.49 & -53.15 & -70.04 & -68.66 & -61.45 \\

\bottomrule
\end{tabular}
}
\caption{Return (\%) of portfolios of 20 cryptocurrencies during the Russian invasion of Ukraine (8.2021 -- 7.2022).}
\label{tab:crypto_war_return}
\end{table}

% ------------------------

\begin{table}[!htbp]
\centering
{\fontsize{6pt}{7pt}\selectfont
\begin{tabular}{cccccccccc}
\toprule
\multirow{2}{*}{} & \multirow{2}{*}{} & \multicolumn{4}{c}{\textbf{Louvain}} & \multicolumn{4}{c}{\textbf{Affinity propagation}} \\
\cmidrule{3-10}
& & \textit{C} & \textit{MI} & $\widetilde{\mathit{Cor}}$ & $\widetilde{\mathit{MI}}$ & \textit{C} & \textit{MI} & $\widetilde{\mathit{Cor}}$ & $\widetilde{\mathit{MI}}$ \\
\midrule
& max & 1.963 & 0.791 & 0.835 & 0.798 & 1.066 & 1.799 & 2.099 & 1.902 \\  
& med & 0.956 & 0.783 & 0.969 & 1.887 & 1.059 & 1.868 & 0.801 & 2.236 \\  
\multirow{-3}{*}{\textbf{PCA}} & min & 1.903 & 0.985 & 1.840 & 2.002 & 1.071 & 2.674 & 0.999 & 1.413 \\  
\midrule
& max & 2.131 & 2.488 & 1.860 & 0.873 & 1.065 & 2.314 & 0.900 & 1.666 \\  
& med & 0.778 & 0.863 & 0.816 & 1.674 & 1.126 & 1.856 & 2.166 & 2.259 \\  
\multirow{-3}{*}{\textbf{Degree (FG)}} & min & 1.694 & 1.573 & 0.980 & 1.702 & 1.072 & 1.973 & 0.877 & 0.751 \\  
\midrule
& max & 0.828 & 1.529 & 0.871 & 2.029 & 1.064 & 2.029 & 0.839 & 1.377 \\  
& med & 0.980 & 0.783 & 0.898 & 1.924 & 1.098 & 2.039 & 0.875 & 0.843 \\  
\multirow{-3}{*}{\textbf{Closeness (FG)}} & min & 2.275 & 3.101 & 1.100 & 0.875 & 1.076 & 2.344 & 0.807 & 1.722 \\  
\midrule
& max & 1.435 & 1.484 & 2.558 & 0.897 & 1.077 & 1.996 & 0.896 & 0.786 \\  
& med & 0.840 & 1.880 & 1.028 & 0.794 & 1.107 & 2.092 & 1.229 & 2.198 \\  
\multirow{-3}{*}{\textbf{Degree (MST)}} & min & 0.896 & 2.411 & 1.423 & 0.836 & 1.061 & 2.891 & 1.954 & 1.811 \\  
\midrule
& max & 0.939 & 1.343 & 0.955 & 0.729 & 1.072 & 1.949 & 0.751 & 2.113 \\  
& med & 0.800 & 1.949 & 0.755 & 1.487 & 1.053 & 1.759 & 0.867 & 0.986 \\  
\multirow{-3}{*}{\textbf{Closeness (MST)}} & min & 2.744 & 2.529 & 0.903 & 2.286 & 1.148 & 2.956 & 1.711 & 1.535 \\  
\bottomrule
\end{tabular}
}
\caption{Annualized log return volatility of portfolios of 20 cryptocurrencies during the Russian invasion of Ukraine (8.2021 -- 7.2022).}
\label{tab:crypto_war_volatility}
\end{table}

% -----------------------------

% SHARP RATIO

\begin{table}[!htbp]
\centering
{\fontsize{6pt}{7pt}\selectfont
\begin{tabular}{ccccccccccc}
\toprule
\multirow{2}{*}{} & \multirow{2}{*}{} & \multicolumn{4}{c}{\textbf{Louvain}} & \multicolumn{4}{c}{\textbf{Affinity propagation}} & \multirow{2}{*}{\begin{tabular}{@{}c@{}}\textbf{Row}\\\textbf{Average}\end{tabular}} \\
\cmidrule{3-10}
& & \textit{C} & \textit{MI} & $\widetilde{\mathit{Cor}}$ & $\widetilde{\mathit{MI}}$ & \textit{C} & \textit{MI} & $\widetilde{\mathit{Cor}}$ & $\widetilde{\mathit{MI}}$ & \\
\midrule
& max & -0.39 & -0.96 & -0.88 & -0.94 & -0.49 & -0.38 & -0.35 & -0.37 & -0.59 \\  
& med & -0.74 & -0.91 & -0.54 & -0.39 & -0.49 & -0.30 & -0.81 & -0.30 & -0.56 \\  
\multirow{-3}{*}{\textbf{PCA}} & min & -0.25 & -0.27 & -0.31 & -0.28 & -0.51 & -0.27 & -0.64 & -0.43 & -0.37 \\  
\midrule
& max & -0.27 & -0.19 & -0.32 & -0.95 & -0.41 & -0.15 & -0.78 & -0.38 & -0.43 \\  
& med & -0.87 & -0.94 & -0.87 & -0.35 & -0.52 & -0.39 & -0.34 & -0.32 & -0.57 \\  
\multirow{-3}{*}{\textbf{Degree (FG)}} & min & -0.40 & -0.46 & -0.62 & -0.41 & -0.49 & -0.34 & -0.79 & -0.92 & -0.55 \\  
\midrule
& max & -0.88 & -0.46 & -0.81 & -0.24 & -0.48 & -0.35 & -0.86 & -0.48 & -0.57 \\  
& med & -0.65 & -0.87 & -0.68 & -0.42 & -0.49 & -0.34 & -0.81 & -0.91 & -0.65 \\  
\multirow{-3}{*}{\textbf{Closeness (FG)}} & min & -0.26 & -0.20 & -0.48 & -0.64 & -0.42 & -0.18 & -0.75 & -0.40 & -0.42 \\  
\midrule
& max & -0.50 & -0.48 & -0.30 & -0.69 & -0.50 & -0.34 & -0.75 & -0.81 & -0.55 \\  
& med & -0.86 & -0.37 & -0.57 & -0.93 & -0.50 & -0.35 & -0.64 & -0.33 & -0.57 \\  
\multirow{-3}{*}{\textbf{Degree (MST)}} & min & -0.71 & -0.27 & -0.43 & -0.90 & -0.43 & -0.24 & -0.20 & -0.39 & -0.45 \\  
\midrule
& max & -0.82 & -0.43 & -0.64 & -0.91 & -0.49 & -0.35 & -0.97 & -0.34 & -0.62 \\  
& med & -0.73 & -0.35 & -0.98 & -0.52 & -0.49 & -0.23 & -0.86 & -0.75 & -0.61 \\  
\multirow{-3}{*}{\textbf{Closeness (MST)}} & min & -0.26 & -0.29 & -0.67 & -0.34 & -0.47 & -0.24 & -0.41 & -0.41 & -0.39 \\  
\midrule
\textbf{Column Average} & & -0.57 & -0.50 & -0.61 & -0.59 & -0.48 & -0.30 & -0.66 & -0.50 & \textbf{-0.53} \\  
\bottomrule
\end{tabular}
}
\caption{Sharpe ratio of portfolios of 20 cryptocurrencies during the Russian invasion of Ukraine (8.2021 -- 7.2022).}
\label{tab:crypto_sharp_ratio_war_period}
\end{table}

\end{appendices}

% \bibliographystyle{bmc-mathphys} % Style BST file (bmc-mathphys, vancouver, spbasic).
%\afterpage{\clearpage}
\clearpage
\setlength{\bibsep}{0pt plus 0.3ex} % minimal vertical space

\bibliography{sn-bibliography}% common bib file


\begin{thebibliography}{72}
\ifx \bisbn   \undefined \def \bisbn  #1{ISBN #1}\fi
\ifx \binits  \undefined \def \binits#1{#1}\fi
\ifx \bauthor  \undefined \def \bauthor#1{#1}\fi
\ifx \batitle  \undefined \def \batitle#1{#1}\fi
\ifx \bjtitle  \undefined \def \bjtitle#1{#1}\fi
\ifx \bvolume  \undefined \def \bvolume#1{\textbf{#1}}\fi
\ifx \byear  \undefined \def \byear#1{#1}\fi
\ifx \bissue  \undefined \def \bissue#1{#1}\fi
\ifx \bfpage  \undefined \def \bfpage#1{#1}\fi
\ifx \blpage  \undefined \def \blpage #1{#1}\fi
\ifx \burl  \undefined \def \burl#1{\textsf{#1}}\fi
\ifx \doiurl  \undefined \def \doiurl#1{\url{https://doi.org/#1}}\fi
\ifx \betal  \undefined \def \betal{\textit{et al.}}\fi
\ifx \binstitute  \undefined \def \binstitute#1{#1}\fi
\ifx \binstitutionaled  \undefined \def \binstitutionaled#1{#1}\fi
\ifx \bctitle  \undefined \def \bctitle#1{#1}\fi
\ifx \beditor  \undefined \def \beditor#1{#1}\fi
\ifx \bpublisher  \undefined \def \bpublisher#1{#1}\fi
\ifx \bbtitle  \undefined \def \bbtitle#1{#1}\fi
\ifx \bedition  \undefined \def \bedition#1{#1}\fi
\ifx \bseriesno  \undefined \def \bseriesno#1{#1}\fi
\ifx \blocation  \undefined \def \blocation#1{#1}\fi
\ifx \bsertitle  \undefined \def \bsertitle#1{#1}\fi
\ifx \bsnm \undefined \def \bsnm#1{#1}\fi
\ifx \bsuffix \undefined \def \bsuffix#1{#1}\fi
\ifx \bparticle \undefined \def \bparticle#1{#1}\fi
\ifx \barticle \undefined \def \barticle#1{#1}\fi
\bibcommenthead
\ifx \bconfdate \undefined \def \bconfdate #1{#1}\fi
\ifx \botherref \undefined \def \botherref #1{#1}\fi
\ifx \url \undefined \def \url#1{\textsf{#1}}\fi
\ifx \bchapter \undefined \def \bchapter#1{#1}\fi
\ifx \bbook \undefined \def \bbook#1{#1}\fi
\ifx \bcomment \undefined \def \bcomment#1{#1}\fi
\ifx \oauthor \undefined \def \oauthor#1{#1}\fi
\ifx \citeauthoryear \undefined \def \citeauthoryear#1{#1}\fi
\ifx \endbibitem  \undefined \def \endbibitem {}\fi
\ifx \bconflocation  \undefined \def \bconflocation#1{#1}\fi
\ifx \arxivurl  \undefined \def \arxivurl#1{\textsf{#1}}\fi
\csname PreBibitemsHook\endcsname

%%% 1
\bibitem[\protect\citeauthoryear{Markowitz}{1952}]{markowitz1952utility}
\begin{barticle}
\bauthor{\bsnm{Markowitz}, \binits{H.}}:
\batitle{The utility of wealth}.
\bjtitle{Journal of political Economy}
\bvolume{60}(\bissue{2}),
\bfpage{151}--\blpage{158}
(\byear{1952})
\end{barticle}
\endbibitem

%%% 2
\bibitem[\protect\citeauthoryear{Sharpe}{1964}]{sharpe1964capital}
\begin{barticle}
\bauthor{\bsnm{Sharpe}, \binits{W.F.}}:
\batitle{Capital asset prices: A theory of market equilibrium under conditions of risk}.
\bjtitle{The journal of finance}
\bvolume{19}(\bissue{3}),
\bfpage{425}--\blpage{442}
(\byear{1964})
\end{barticle}
\endbibitem

%%% 3
\bibitem[\protect\citeauthoryear{Lintner}{1965}]{lintner1965security}
\begin{barticle}
\bauthor{\bsnm{Lintner}, \binits{J.}}:
\batitle{Security prices, risk, and maximal gains from diversification}.
\bjtitle{The journal of finance}
\bvolume{20}(\bissue{4}),
\bfpage{587}--\blpage{615}
(\byear{1965})
\end{barticle}
\endbibitem

%%% 4
\bibitem[\protect\citeauthoryear{Mossin}{1966}]{mossin1966equilibrium}
\begin{botherref}
\oauthor{\bsnm{Mossin}, \binits{J.}}:
Equilibrium in a capital asset market.
Econometrica: Journal of the econometric society,
768--783
(1966)
\end{botherref}
\endbibitem

%%% 5
\bibitem[\protect\citeauthoryear{Fama and French}{1993}]{fama1993common}
\begin{barticle}
\bauthor{\bsnm{Fama}, \binits{E.F.}},
\bauthor{\bsnm{French}, \binits{K.R.}}:
\batitle{Common risk factors in the returns on stocks and bonds}.
\bjtitle{Journal of financial economics}
\bvolume{33}(\bissue{1}),
\bfpage{3}--\blpage{56}
(\byear{1993})
\end{barticle}
\endbibitem

%%% 6
\bibitem[\protect\citeauthoryear{Hinich and Patterson}{1985}]{hinich1985evidence}
\begin{barticle}
\bauthor{\bsnm{Hinich}, \binits{M.J.}},
\bauthor{\bsnm{Patterson}, \binits{D.M.}}:
\batitle{Evidence of nonlinearity in daily stock returns}.
\bjtitle{Journal of Business \& Economic Statistics}
\bvolume{3}(\bissue{1}),
\bfpage{69}--\blpage{77}
(\byear{1985})
\end{barticle}
\endbibitem

%%% 7
\bibitem[\protect\citeauthoryear{Franses and Van~Dijk}{1996}]{franses1996forecasting}
\begin{barticle}
\bauthor{\bsnm{Franses}, \binits{P.H.}},
\bauthor{\bsnm{Van~Dijk}, \binits{D.}}:
\batitle{Forecasting stock market volatility using (non-linear) garch models}.
\bjtitle{Journal of forecasting}
\bvolume{15}(\bissue{3}),
\bfpage{229}--\blpage{235}
(\byear{1996})
\end{barticle}
\endbibitem

%%% 8
\bibitem[\protect\citeauthoryear{Caginalp and DeSantis}{2011}]{caginalp2011nonlinearity}
\begin{barticle}
\bauthor{\bsnm{Caginalp}, \binits{G.}},
\bauthor{\bsnm{DeSantis}, \binits{M.}}:
\batitle{Nonlinearity in the dynamics of financial markets}.
\bjtitle{Nonlinear Analysis: Real World Applications}
\bvolume{12}(\bissue{2}),
\bfpage{1140}--\blpage{1151}
(\byear{2011})
\end{barticle}
\endbibitem

%%% 9
\bibitem[\protect\citeauthoryear{Hartman and Hlinka}{2018}]{hartman2018nonlinearity}
\begin{botherref}
\oauthor{\bsnm{Hartman}, \binits{D.}},
\oauthor{\bsnm{Hlinka}, \binits{J.}}:
Nonlinearity in stock networks.
Chaos: An Interdisciplinary Journal of Nonlinear Science
\textbf{28}(8)
(2018)
\end{botherref}
\endbibitem

%%% 10
\bibitem[\protect\citeauthoryear{Inglada-Perez}{2020}]{inglada2020comprehensive}
\begin{barticle}
\bauthor{\bsnm{Inglada-Perez}, \binits{L.}}:
\batitle{A comprehensive framework for uncovering non-linearity and chaos in financial markets: Empirical evidence for four major stock market indices}.
\bjtitle{Entropy}
\bvolume{22}(\bissue{12}),
\bfpage{1435}
(\byear{2020})
\end{barticle}
\endbibitem

%%% 11
\bibitem[\protect\citeauthoryear{Allen and Babus}{2009}]{allen2009networks}
\begin{botherref}
\oauthor{\bsnm{Allen}, \binits{F.}},
\oauthor{\bsnm{Babus}, \binits{A.}}:
Networks in finance.
The network challenge: strategy, profit, and risk in an interlinked world
\textbf{367}
(2009)
\end{botherref}
\endbibitem

%%% 12
\bibitem[\protect\citeauthoryear{Huang et~al.}{2009}]{huang2009network}
\begin{barticle}
\bauthor{\bsnm{Huang}, \binits{W.-Q.}},
\bauthor{\bsnm{Zhuang}, \binits{X.-T.}},
\bauthor{\bsnm{Yao}, \binits{S.}}:
\batitle{A network analysis of the chinese stock market}.
\bjtitle{Physica A: Statistical Mechanics and its Applications}
\bvolume{388}(\bissue{14}),
\bfpage{2956}--\blpage{2964}
(\byear{2009})
\end{barticle}
\endbibitem

%%% 13
\bibitem[\protect\citeauthoryear{Chi et~al.}{2010}]{chi2010network}
\begin{barticle}
\bauthor{\bsnm{Chi}, \binits{K.T.}},
\bauthor{\bsnm{Liu}, \binits{J.}},
\bauthor{\bsnm{Lau}, \binits{F.C.}}:
\batitle{A network perspective of the stock market}.
\bjtitle{Journal of Empirical Finance}
\bvolume{17}(\bissue{4}),
\bfpage{659}--\blpage{667}
(\byear{2010})
\end{barticle}
\endbibitem

%%% 14
\bibitem[\protect\citeauthoryear{Brida and Risso}{2009}]{BridaRisso2010}
\begin{barticle}
\bauthor{\bsnm{Brida}, \binits{J.}},
\bauthor{\bsnm{Risso}, \binits{W.}}:
\batitle{Dynamics and structure of the 30 largest north american companies}.
\bjtitle{Society for Computational Economics}
\bvolume{35}(\bissue{1}),
\bfpage{85}--\blpage{99}
(\byear{2009})
\end{barticle}
\endbibitem

%%% 15
\bibitem[\protect\citeauthoryear{Silva et~al.}{2016}]{silva2016structure}
\begin{barticle}
\bauthor{\bsnm{Silva}, \binits{T.C.}},
\bauthor{\bsnm{Souza}, \binits{S.R.S.}},
\bauthor{\bsnm{Tabak}, \binits{B.M.}}:
\batitle{Structure and dynamics of the global financial network}.
\bjtitle{Chaos, Solitons \& Fractals}
\bvolume{88},
\bfpage{218}--\blpage{234}
(\byear{2016})
\end{barticle}
\endbibitem

%%% 16
\bibitem[\protect\citeauthoryear{Battiston et~al.}{2016}]{battiston2016price}
\begin{barticle}
\bauthor{\bsnm{Battiston}, \binits{S.}},
\bauthor{\bsnm{Caldarelli}, \binits{G.}},
\bauthor{\bsnm{May}, \binits{R.M.}},
\bauthor{\bsnm{Roukny}, \binits{T.}},
\bauthor{\bsnm{Stiglitz}, \binits{J.E.}}:
\batitle{The price of complexity in financial networks}.
\bjtitle{Proceedings of the National Academy of Sciences}
\bvolume{113}(\bissue{36}),
\bfpage{10031}--\blpage{10036}
(\byear{2016})
\end{barticle}
\endbibitem

%%% 17
\bibitem[\protect\citeauthoryear{Kim and Sayama}{2017}]{kim2017predicting}
\begin{barticle}
\bauthor{\bsnm{Kim}, \binits{M.}},
\bauthor{\bsnm{Sayama}, \binits{H.}}:
\batitle{Predicting stock market movements using network science: An information theoretic approach}.
\bjtitle{Applied network science}
\bvolume{2}(\bissue{1}),
\bfpage{1}--\blpage{14}
(\byear{2017})
\end{barticle}
\endbibitem

%%% 18
\bibitem[\protect\citeauthoryear{Iori et~al.}{2018}]{iori2018empirical}
\begin{bchapter}
\bauthor{\bsnm{Iori}, \binits{G.}},
\bauthor{\bsnm{Mantegna}, \binits{R.}}, \betal:
\bctitle{Empirical analyses of networks in finance}.
In: \bbtitle{Handbook of Computational Economics-Volume 4-Heterogeneous Agent Modeling}
vol. \bseriesno{4},
pp. \bfpage{637}--\blpage{685}
(\byear{2018})
\end{bchapter}
\endbibitem

%%% 19
\bibitem[\protect\citeauthoryear{Mantegna and Stanley}{1999}]{Mantegna_Stanley_1999}
\begin{bbook}
\bauthor{\bsnm{Mantegna}, \binits{R.N.}},
\bauthor{\bsnm{Stanley}, \binits{H.E.}}:
\bbtitle{Introduction to Econophysics: Correlations and Complexity in Finance},
(\byear{1999})
\end{bbook}
\endbibitem

%%% 20
\bibitem[\protect\citeauthoryear{Bouchaud and Potters}{2003}]{Bouchaud_Potters_2003}
\begin{bbook}
\bauthor{\bsnm{Bouchaud}, \binits{J.P.}},
\bauthor{\bsnm{Potters}, \binits{M.}}:
\bbtitle{Theory of Financial Risk and Derivative Pricing: From Statistical Physics to Risk Management},
\bedition{2}nd edn.
(\byear{2003})
\end{bbook}
\endbibitem

%%% 21
\bibitem[\protect\citeauthoryear{Bun et~al.}{2017}]{Bun_2017}
\begin{barticle}
\bauthor{\bsnm{Bun}, \binits{J.}},
\bauthor{\bsnm{Bouchaud}, \binits{J.P.}},
\bauthor{\bsnm{Potters}, \binits{M.}}:
\batitle{Cleaning large correlation matrices: Tools from random matrix theory}.
\bjtitle{Physics Reports}
\bvolume{666},
\bfpage{1}--\blpage{109}
(\byear{2017})
\end{barticle}
\endbibitem

%%% 22
\bibitem[\protect\citeauthoryear{Dose and Cincotti}{2005}]{dose2005clustering}
\begin{barticle}
\bauthor{\bsnm{Dose}, \binits{C.}},
\bauthor{\bsnm{Cincotti}, \binits{S.}}:
\batitle{Clustering of financial time series with application to index and enhanced index tracking portfolio}.
\bjtitle{Physica A: Statistical Mechanics and its Applications}
\bvolume{355}(\bissue{1}),
\bfpage{145}--\blpage{151}
(\byear{2005})
\end{barticle}
\endbibitem

%%% 23
\bibitem[\protect\citeauthoryear{Tola et~al.}{2008}]{tola2008cluster}
\begin{barticle}
\bauthor{\bsnm{Tola}, \binits{V.}},
\bauthor{\bsnm{Lillo}, \binits{F.}},
\bauthor{\bsnm{Gallegati}, \binits{M.}},
\bauthor{\bsnm{Mantegna}, \binits{R.N.}}:
\batitle{Cluster analysis for portfolio optimization}.
\bjtitle{Journal of Economic Dynamics and Control}
\bvolume{32}(\bissue{1}),
\bfpage{235}--\blpage{258}
(\byear{2008})
\end{barticle}
\endbibitem

%%% 24
\bibitem[\protect\citeauthoryear{Nanda et~al.}{2010}]{Nanda2010ClusteringIS}
\begin{barticle}
\bauthor{\bsnm{Nanda}, \binits{S.R.}},
\bauthor{\bsnm{Mahanty}, \binits{B.}},
\bauthor{\bsnm{Tiwari}, \binits{M.K.}}:
\batitle{Clustering {Indian} stock market data for portfolio management}.
\bjtitle{Expert System with Applications}
\bvolume{37},
\bfpage{8793}--\blpage{8798}
(\byear{2010})
\end{barticle}
\endbibitem

%%% 25
\bibitem[\protect\citeauthoryear{Duarte and De~Castro}{2020}]{Duarte2020}
\begin{barticle}
\bauthor{\bsnm{Duarte}, \binits{F.G.}},
\bauthor{\bsnm{De~Castro}, \binits{L.N.}}:
\batitle{A framework to perform asset allocation based on partitional clustering}.
\bjtitle{IEEE Access}
\bvolume{8},
\bfpage{110775}--\blpage{110788}
(\byear{2020})
\end{barticle}
\endbibitem

%%% 26
\bibitem[\protect\citeauthoryear{Mizuno et~al.}{2006}]{Mizuno_2006}
\begin{barticle}
\bauthor{\bsnm{Mizuno}, \binits{T.}},
\bauthor{\bsnm{Takayasu}, \binits{H.}},
\bauthor{\bsnm{Takayasu}, \binits{M.}}:
\batitle{Correlation networks among currencies}.
\bjtitle{Physica A: Statistical Mechanics and its Applications}
\bvolume{364},
\bfpage{336}--\blpage{342}
(\byear{2006})
\end{barticle}
\endbibitem

%%% 27
\bibitem[\protect\citeauthoryear{Basnarkov et~al.}{2019}]{basnarkov2019correlation}
\begin{barticle}
\bauthor{\bsnm{Basnarkov}, \binits{L.}},
\bauthor{\bsnm{Stojkoski}, \binits{V.}},
\bauthor{\bsnm{Utkovski}, \binits{Z.}},
\bauthor{\bsnm{Kocarev}, \binits{L.}}:
\batitle{Correlation patterns in foreign exchange markets}.
\bjtitle{Physica A: Statistical Mechanics and its Applications}
\bvolume{525},
\bfpage{1026}--\blpage{1037}
(\byear{2019})
\end{barticle}
\endbibitem

%%% 28
\bibitem[\protect\citeauthoryear{Basnarkov et~al.}{2020}]{basnarkov2020lead}
\begin{barticle}
\bauthor{\bsnm{Basnarkov}, \binits{L.}},
\bauthor{\bsnm{Stojkoski}, \binits{V.}},
\bauthor{\bsnm{Utkovski}, \binits{Z.}},
\bauthor{\bsnm{Kocarev}, \binits{L.}}:
\batitle{Lead--lag relationships in foreign exchange markets}.
\bjtitle{Physica A: Statistical Mechanics and its Applications}
\bvolume{539},
\bfpage{122986}
(\byear{2020})
\end{barticle}
\endbibitem

%%% 29
\bibitem[\protect\citeauthoryear{Stosic et~al.}{2018}]{Stosic2018CollectiveBO}
\begin{barticle}
\bauthor{\bsnm{Stosic}, \binits{D.}},
\bauthor{\bsnm{Stosic}, \binits{D.}},
\bauthor{\bsnm{Ludermir}, \binits{T.B.}},
\bauthor{\bsnm{Stosic}, \binits{T.}}:
\batitle{Collective behavior of cryptocurrency price changes}.
\bjtitle{Physica A: Statistical Mechanics and its Applications}
\bvolume{507},
\bfpage{499}--\blpage{509}
(\byear{2018})
\end{barticle}
\endbibitem

%%% 30
\bibitem[\protect\citeauthoryear{Papadimitriou et~al.}{2020}]{papadimitriou2020evolution}
\begin{barticle}
\bauthor{\bsnm{Papadimitriou}, \binits{T.}},
\bauthor{\bsnm{Gogas}, \binits{P.}},
\bauthor{\bsnm{Gkatzoglou}, \binits{F.}}:
\batitle{The evolution of the cryptocurrencies market: A complex networks approach}.
\bjtitle{Journal of Computational and Applied Mathematics}
\bvolume{376},
\bfpage{112831}
(\byear{2020})
\end{barticle}
\endbibitem

%%% 31
\bibitem[\protect\citeauthoryear{Newman}{2003}]{newman2003structure}
\begin{barticle}
\bauthor{\bsnm{Newman}, \binits{M.E.}}:
\batitle{The structure and function of complex networks}.
\bjtitle{SIAM review}
\bvolume{45}(\bissue{2}),
\bfpage{167}--\blpage{256}
(\byear{2003})
\end{barticle}
\endbibitem

%%% 32
\bibitem[\protect\citeauthoryear{Battiston et~al.}{2016}]{battiston2016complexity}
\begin{barticle}
\bauthor{\bsnm{Battiston}, \binits{S.}},
\bauthor{\bsnm{Farmer}, \binits{J.D.}},
\bauthor{\bsnm{Flache}, \binits{A.}},
\bauthor{\bsnm{Garlaschelli}, \binits{D.}},
\bauthor{\bsnm{Haldane}, \binits{A.G.}},
\bauthor{\bsnm{Heesterbeek}, \binits{H.}},
\bauthor{\bsnm{Hommes}, \binits{C.}},
\bauthor{\bsnm{Jaeger}, \binits{C.}},
\bauthor{\bsnm{May}, \binits{R.}},
\bauthor{\bsnm{Scheffer}, \binits{M.}}:
\batitle{Complexity theory and financial regulation}.
\bjtitle{Science}
\bvolume{351}(\bissue{6275}),
\bfpage{818}--\blpage{819}
(\byear{2016})
\end{barticle}
\endbibitem

%%% 33
\bibitem[\protect\citeauthoryear{Niu et~al.}{2021}]{niu2021implicit}
\begin{barticle}
\bauthor{\bsnm{Niu}, \binits{X.}},
\bauthor{\bsnm{Niu}, \binits{X.}},
\bauthor{\bsnm{Wu}, \binits{K.}}:
\batitle{Implicit government guarantees and the externality of portfolio diversification: A complex network approach}.
\bjtitle{Physica A: Statistical Mechanics and its Applications}
\bvolume{572},
\bfpage{125908}
(\byear{2021})
\end{barticle}
\endbibitem

%%% 34
\bibitem[\protect\citeauthoryear{Mantegna}{1999}]{mantegna1999hierarchical}
\begin{barticle}
\bauthor{\bsnm{Mantegna}, \binits{R.N.}}:
\batitle{Hierarchical structure in financial markets}.
\bjtitle{The European Physical Journal B - Condensed Matter and Complex Systems}
\bvolume{11}(\bissue{1}),
\bfpage{193}--\blpage{197}
(\byear{1999})
\end{barticle}
\endbibitem

%%% 35
\bibitem[\protect\citeauthoryear{Bonanno et~al.}{2004}]{bonanno2004networks}
\begin{barticle}
\bauthor{\bsnm{Bonanno}, \binits{G.}},
\bauthor{\bsnm{Caldarelli}, \binits{G.}},
\bauthor{\bsnm{Lillo}, \binits{F.}},
\bauthor{\bsnm{Micciche}, \binits{S.}},
\bauthor{\bsnm{Vandewalle}, \binits{N.}},
\bauthor{\bsnm{Mantegna}, \binits{R.N.}}:
\batitle{Networks of equities in financial markets}.
\bjtitle{The European Physical Journal B}
\bvolume{38}(\bissue{2}),
\bfpage{363}--\blpage{371}
(\byear{2004})
\end{barticle}
\endbibitem

%%% 36
\bibitem[\protect\citeauthoryear{Onnela et~al.}{2002}]{onnela2002dynamic}
\begin{barticle}
\bauthor{\bsnm{Onnela}, \binits{J.-P.}},
\bauthor{\bsnm{Chakraborti}, \binits{A.}},
\bauthor{\bsnm{Kaski}, \binits{K.}},
\bauthor{\bsnm{Kerti{\'e}sz}, \binits{J.}}:
\batitle{Dynamic asset trees and portfolio analysis}.
\bjtitle{The European Physical Journal B-Condensed Matter and Complex Systems}
\bvolume{30}(\bissue{3}),
\bfpage{285}--\blpage{288}
(\byear{2002})
\end{barticle}
\endbibitem

%%% 37
\bibitem[\protect\citeauthoryear{Onnela et~al.}{2003}]{onnela2003dynamics}
\begin{barticle}
\bauthor{\bsnm{Onnela}, \binits{J.-P.}},
\bauthor{\bsnm{Chakraborti}, \binits{A.}},
\bauthor{\bsnm{Kaski}, \binits{K.}},
\bauthor{\bsnm{Kertesz}, \binits{J.}},
\bauthor{\bsnm{Kanto}, \binits{A.}}:
\batitle{Dynamics of market correlations: Taxonomy and portfolio analysis}.
\bjtitle{Physical Review E}
\bvolume{68}(\bissue{5}),
\bfpage{056110}
(\byear{2003})
\end{barticle}
\endbibitem

%%% 38
\bibitem[\protect\citeauthoryear{Tumminello et~al.}{2005}]{tumminello2005tool}
\begin{barticle}
\bauthor{\bsnm{Tumminello}, \binits{M.}},
\bauthor{\bsnm{Aste}, \binits{T.}},
\bauthor{\bsnm{Di~Matteo}, \binits{T.}},
\bauthor{\bsnm{Mantegna}, \binits{R.N.}}:
\batitle{A tool for filtering information in complex systems}.
\bjtitle{Proceedings of the National Academy of Sciences}
\bvolume{102}(\bissue{30}),
\bfpage{10421}--\blpage{10426}
(\byear{2005})
\end{barticle}
\endbibitem

%%% 39
\bibitem[\protect\citeauthoryear{Tumminello et~al.}{2006}]{Tumminello_2006}
\begin{barticle}
\bauthor{\bsnm{Tumminello}, \binits{M.}},
\bauthor{\bsnm{Di~Matteo}, \binits{T.}},
\bauthor{\bsnm{Aste}, \binits{T.}},
\bauthor{\bsnm{Mantegna}, \binits{R.N.}}:
\batitle{Correlation based networks of equity returns sampled at different time horizons}.
\bjtitle{The European Physical Journal B}
\bvolume{55}(\bissue{2}),
\bfpage{209}--\blpage{217}
(\byear{2006})
\end{barticle}
\endbibitem

%%% 40
\bibitem[\protect\citeauthoryear{Pozzi et~al.}{2013}]{pozzi2013spread}
\begin{barticle}
\bauthor{\bsnm{Pozzi}, \binits{F.}},
\bauthor{\bsnm{Di~Matteo}, \binits{T.}},
\bauthor{\bsnm{Aste}, \binits{T.}}:
\batitle{Spread of risk across financial markets: better to invest in the peripheries}.
\bjtitle{Scientific reports}
\bvolume{3}(\bissue{1}),
\bfpage{1}--\blpage{7}
(\byear{2013})
\end{barticle}
\endbibitem

%%% 41
\bibitem[\protect\citeauthoryear{Peralta and Zareei}{2016}]{peralta2016network}
\begin{barticle}
\bauthor{\bsnm{Peralta}, \binits{G.}},
\bauthor{\bsnm{Zareei}, \binits{A.}}:
\batitle{A network approach to portfolio selection}.
\bjtitle{Journal of Empirical Finance}
\bvolume{38},
\bfpage{157}--\blpage{180}
(\byear{2016})
\end{barticle}
\endbibitem

%%% 42
\bibitem[\protect\citeauthoryear{Nicolo~Musmeci}{2014}]{Musmeci2014ClusteringAH}
\begin{botherref}
\oauthor{\bsnm{Nicolo~Musmeci}, \binits{T.d.M.} \bsuffix{Tomaso~Aste}}:
Clustering and hierarchy of financial markets data: advantages of the {DBHT}
(2014)
\end{botherref}
\endbibitem

%%% 43
\bibitem[\protect\citeauthoryear{Ren et~al.}{2017}]{ren2017dynamic}
\begin{barticle}
\bauthor{\bsnm{Ren}, \binits{F.}},
\bauthor{\bsnm{Lu}, \binits{Y.-N.}},
\bauthor{\bsnm{Li}, \binits{S.-P.}},
\bauthor{\bsnm{Jiang}, \binits{X.-F.}},
\bauthor{\bsnm{Zhong}, \binits{L.-X.}},
\bauthor{\bsnm{Qiu}, \binits{T.}}:
\batitle{Dynamic portfolio strategy using clustering approach}.
\bjtitle{PloS one}
\bvolume{12}(\bissue{1}),
\bfpage{0169299}
(\byear{2017})
\end{barticle}
\endbibitem

%%% 44
\bibitem[\protect\citeauthoryear{L{\'o}pez~de Prado}{2016}]{dePrado59}
\begin{barticle}
\bauthor{\bsnm{Prado}, \binits{M.}}:
\batitle{Building diversified portfolios that outperform out of sample}.
\bjtitle{The Journal of Portfolio Management}
\bvolume{42}(\bissue{4}),
\bfpage{59}--\blpage{69}
(\byear{2016})
\end{barticle}
\endbibitem

%%% 45
\bibitem[\protect\citeauthoryear{Raffinot}{2017}]{raffinot2017hierarchical}
\begin{barticle}
\bauthor{\bsnm{Raffinot}, \binits{T.}}:
\batitle{Hierarchical clustering-based asset allocation}.
\bjtitle{The Journal of Portfolio Management}
\bvolume{44}(\bissue{2}),
\bfpage{89}--\blpage{99}
(\byear{2017})
\end{barticle}
\endbibitem

%%% 46
\bibitem[\protect\citeauthoryear{Li et~al.}{2019}]{li2019portfolio}
\begin{barticle}
\bauthor{\bsnm{Li}, \binits{Y.}},
\bauthor{\bsnm{Jiang}, \binits{X.-F.}},
\bauthor{\bsnm{Tian}, \binits{Y.}},
\bauthor{\bsnm{Li}, \binits{S.-P.}},
\bauthor{\bsnm{Zheng}, \binits{B.}}:
\batitle{Portfolio optimization based on network topology}.
\bjtitle{Physica A: Statistical Mechanics and its Applications}
\bvolume{515},
\bfpage{671}--\blpage{681}
(\byear{2019})
\end{barticle}
\endbibitem

%%% 47
\bibitem[\protect\citeauthoryear{Ricca and Scozzari}{2024}]{ricca2024portfolio}
\begin{barticle}
\bauthor{\bsnm{Ricca}, \binits{F.}},
\bauthor{\bsnm{Scozzari}, \binits{A.}}:
\batitle{Portfolio optimization through a network approach: Network assortative mixing and portfolio diversification}.
\bjtitle{European Journal of Operational Research}
\bvolume{312}(\bissue{2}),
\bfpage{700}--\blpage{717}
(\byear{2024})
\end{barticle}
\endbibitem

%%% 48
\bibitem[\protect\citeauthoryear{Giudici et~al.}{2022}]{giudici2022network}
\begin{barticle}
\bauthor{\bsnm{Giudici}, \binits{P.}},
\bauthor{\bsnm{Polinesi}, \binits{G.}},
\bauthor{\bsnm{Spelta}, \binits{A.}}:
\batitle{Network models to improve robot advisory portfolios}.
\bjtitle{Annals of Operations Research}
\bvolume{313}(\bissue{2}),
\bfpage{965}--\blpage{989}
(\byear{2022})
\end{barticle}
\endbibitem

%%% 49
\bibitem[\protect\citeauthoryear{Ioannidis et~al.}{2023}]{ioannidis2023portfolio}
\begin{barticle}
\bauthor{\bsnm{Ioannidis}, \binits{E.}},
\bauthor{\bsnm{Sarikeisoglou}, \binits{I.}},
\bauthor{\bsnm{Angelidis}, \binits{G.}}:
\batitle{Portfolio construction: A network approach}.
\bjtitle{Mathematics}
\bvolume{11}(\bissue{22}),
\bfpage{4670}
(\byear{2023})
\end{barticle}
\endbibitem

%%% 50
\bibitem[\protect\citeauthoryear{Chaudhari and Crane}{2020}]{chaudhari2020cross}
\begin{barticle}
\bauthor{\bsnm{Chaudhari}, \binits{H.}},
\bauthor{\bsnm{Crane}, \binits{M.}}:
\batitle{Cross-correlation dynamics and community structures of cryptocurrencies}.
\bjtitle{Journal of Computational Science}
\bvolume{44},
\bfpage{101130}
(\byear{2020})
\end{barticle}
\endbibitem

%%% 51
\bibitem[\protect\citeauthoryear{Gavin and Crane}{2021}]{gavin2021community}
\begin{botherref}
\oauthor{\bsnm{Gavin}, \binits{J.}},
\oauthor{\bsnm{Crane}, \binits{M.}}:
Community detection in cryptocurrencies with potential applications to portfolio diversification.
arXiv preprint arXiv:2108.09763
(2021)
\end{botherref}
\endbibitem

%%% 52
\bibitem[\protect\citeauthoryear{Kitanovski et~al.}{2022}]{kitanovski2022cryptocurrency}
\begin{bchapter}
\bauthor{\bsnm{Kitanovski}, \binits{D.}},
\bauthor{\bsnm{Mirchev}, \binits{M.}},
\bauthor{\bsnm{Chorbev}, \binits{I.}},
\bauthor{\bsnm{Mishkovski}, \binits{I.}}:
\bctitle{Cryptocurrency portfolio diversification using network community detection}.
In: \bbtitle{2022 30th Telecommunications Forum (TELFOR)},
pp. \bfpage{1}--\blpage{4}
(\byear{2022}).
\bcomment{IEEE}
\end{bchapter}
\endbibitem

%%% 53
\bibitem[\protect\citeauthoryear{Das et~al.}{2023}]{das2023portfolio}
\begin{bchapter}
\bauthor{\bsnm{Das}, \binits{J.D.}},
\bauthor{\bsnm{Bowala}, \binits{S.}},
\bauthor{\bsnm{Thulasiram}, \binits{R.K.}},
\bauthor{\bsnm{Thavaneswaran}, \binits{A.}}:
\bctitle{Portfolio diversification with clustering techniques}.
In: \bbtitle{2023 IEEE Symposium Series on Computational Intelligence (SSCI)},
pp. \bfpage{97}--\blpage{102}
(\byear{2023}).
\bcomment{IEEE}
\end{bchapter}
\endbibitem

%%% 54
\bibitem[\protect\citeauthoryear{V{\`y}rost et~al.}{2019}]{vyrost2019network}
\begin{barticle}
\bauthor{\bsnm{V{\`y}rost}, \binits{T.}},
\bauthor{\bsnm{Ly{\'o}csa}, \binits{{\v{S}}.}},
\bauthor{\bsnm{Baum{\"o}hl}, \binits{E.}}:
\batitle{Network-based asset allocation strategies}.
\bjtitle{The North American Journal of Economics and Finance}
\bvolume{47},
\bfpage{516}--\blpage{536}
(\byear{2019})
\end{barticle}
\endbibitem

%%% 55
\bibitem[\protect\citeauthoryear{Wang et~al.}{2024}]{wang2024portfolio}
\begin{barticle}
\bauthor{\bsnm{Wang}, \binits{G.-J.}},
\bauthor{\bsnm{Huai}, \binits{H.}},
\bauthor{\bsnm{Zhu}, \binits{Y.}},
\bauthor{\bsnm{Xie}, \binits{C.}},
\bauthor{\bsnm{Uddin}, \binits{G.S.}}:
\batitle{Portfolio optimization based on network centralities: Which centrality is better for asset selection during global crises?}
\bjtitle{Journal of Management Science and Engineering}
\bvolume{9}(\bissue{3}),
\bfpage{348}--\blpage{375}
(\byear{2024})
\end{barticle}
\endbibitem

%%% 56
\bibitem[\protect\citeauthoryear{Marti et~al.}{2021}]{marti2021review}
\begin{botherref}
\oauthor{\bsnm{Marti}, \binits{G.}},
\oauthor{\bsnm{Nielsen}, \binits{F.}},
\oauthor{\bsnm{Bi{\'n}kowski}, \binits{M.}},
\oauthor{\bsnm{Donnat}, \binits{P.}}:
A review of two decades of correlations, hierarchies, networks and clustering in financial markets.
Progress in Information Geometry,
245--274
(2021)
\end{botherref}
\endbibitem

%%% 57
\bibitem[\protect\citeauthoryear{Sutiene et~al.}{2024}]{sutiene2024enhancing}
\begin{barticle}
\bauthor{\bsnm{Sutiene}, \binits{K.}},
\bauthor{\bsnm{Schwendner}, \binits{P.}},
\bauthor{\bsnm{Sipos}, \binits{C.}},
\bauthor{\bsnm{Lorenzo}, \binits{L.}},
\bauthor{\bsnm{Mirchev}, \binits{M.}},
\bauthor{\bsnm{Lameski}, \binits{P.}},
\bauthor{\bsnm{Kabasinskas}, \binits{A.}},
\bauthor{\bsnm{Tidjani}, \binits{C.}},
\bauthor{\bsnm{Ozturkkal}, \binits{B.}},
\bauthor{\bsnm{Cerneviciene}, \binits{J.}}:
\batitle{Enhancing portfolio management using artificial intelligence: literature review}.
\bjtitle{Frontiers in Artificial Intelligence}
\bvolume{7},
\bfpage{1371502}
(\byear{2024})
\end{barticle}
\endbibitem

%%% 58
\bibitem[\protect\citeauthoryear{Pacreau et~al.}{2021}]{pacreau2021graph}
\begin{botherref}
\oauthor{\bsnm{Pacreau}, \binits{G.}},
\oauthor{\bsnm{Lezmi}, \binits{E.}},
\oauthor{\bsnm{Xu}, \binits{J.}}:
Graph neural networks for asset management.
Available at SSRN 3976168
(2021)
\end{botherref}
\endbibitem

%%% 59
\bibitem[\protect\citeauthoryear{Soleymani and Paquet}{2021}]{soleymani2021deep}
\begin{barticle}
\bauthor{\bsnm{Soleymani}, \binits{F.}},
\bauthor{\bsnm{Paquet}, \binits{E.}}:
\batitle{Deep graph convolutional reinforcement learning for financial portfolio management--deeppocket}.
\bjtitle{Expert Systems with Applications}
\bvolume{182},
\bfpage{115127}
(\byear{2021})
\end{barticle}
\endbibitem

%%% 60
\bibitem[\protect\citeauthoryear{Babaei et~al.}{2022}]{babaei2022explainable}
\begin{barticle}
\bauthor{\bsnm{Babaei}, \binits{G.}},
\bauthor{\bsnm{Giudici}, \binits{P.}},
\bauthor{\bsnm{Raffinetti}, \binits{E.}}:
\batitle{Explainable artificial intelligence for crypto asset allocation}.
\bjtitle{Finance Research Letters}
\bvolume{47},
\bfpage{102941}
(\byear{2022})
\end{barticle}
\endbibitem

%%% 61
\bibitem[\protect\citeauthoryear{Giudici}{2024}]{giudici2024safe}
\begin{barticle}
\bauthor{\bsnm{Giudici}, \binits{P.}}:
\batitle{Safe machine learning}.
\bjtitle{Statistics}
\bvolume{58}(\bissue{3}),
\bfpage{473}--\blpage{477}
(\byear{2024})
\end{barticle}
\endbibitem

%%% 62
\bibitem[\protect\citeauthoryear{Dionisio et~al.}{2004}]{dionisio2004mutual}
\begin{barticle}
\bauthor{\bsnm{Dionisio}, \binits{A.}},
\bauthor{\bsnm{Menezes}, \binits{R.}},
\bauthor{\bsnm{Mendes}, \binits{D.A.}}:
\batitle{Mutual information: a measure of dependency for nonlinear time series}.
\bjtitle{Physica A: Statistical Mechanics and its Applications}
\bvolume{344}(\bissue{1-2}),
\bfpage{326}--\blpage{329}
(\byear{2004})
\end{barticle}
\endbibitem

%%% 63
\bibitem[\protect\citeauthoryear{Fiedor}{2014}]{fiedor2014networks}
\begin{barticle}
\bauthor{\bsnm{Fiedor}, \binits{P.}}:
\batitle{Networks in financial markets based on the mutual information rate}.
\bjtitle{Physical Review E}
\bvolume{89}(\bissue{5}),
\bfpage{052801}
(\byear{2014})
\end{barticle}
\endbibitem

%%% 64
\bibitem[\protect\citeauthoryear{Guo et~al.}{2018}]{guo2018development}
\begin{barticle}
\bauthor{\bsnm{Guo}, \binits{X.}},
\bauthor{\bsnm{Zhang}, \binits{H.}},
\bauthor{\bsnm{Tian}, \binits{T.}}:
\batitle{Development of stock correlation networks using mutual information and financial big data}.
\bjtitle{PloS one}
\bvolume{13}(\bissue{4}),
\bfpage{0195941}
(\byear{2018})
\end{barticle}
\endbibitem

%%% 65
\bibitem[\protect\citeauthoryear{Sharpe}{1994}]{sharpe1994sharpe}
\begin{barticle}
\bauthor{\bsnm{Sharpe}, \binits{W.F.}}:
\batitle{The sharpe ratio}.
\bjtitle{Journal of portfolio management}
\bvolume{21}(\bissue{1}),
\bfpage{49}--\blpage{58}
(\byear{1994})
\end{barticle}
\endbibitem

%%% 66
\bibitem[\protect\citeauthoryear{Khan and Niazi}{2017}]{khan2017network}
\begin{botherref}
\oauthor{\bsnm{Khan}, \binits{B.S.}},
\oauthor{\bsnm{Niazi}, \binits{M.A.}}:
Network community detection: A review and visual survey.
arXiv preprint arXiv:1708.00977
(2017)
\end{botherref}
\endbibitem

%%% 67
\bibitem[\protect\citeauthoryear{Delbert}{2009}]{dueck2009affinity}
\begin{botherref}
\oauthor{\bsnm{Delbert}, \binits{D.}}:
Affinity propagation: clustering data by passing messages.
Doctor of Philosophy, Graduate Department of Electrical \& Computer Engineering, University of Toronto
(2009)
\end{botherref}
\endbibitem

%%% 68
\bibitem[\protect\citeauthoryear{McLeod and van Vuuren}{2004}]{mcleod2004interpreting}
\begin{barticle}
\bauthor{\bsnm{McLeod}, \binits{W.}},
\bauthor{\bsnm{Vuuren}, \binits{G.}}:
\batitle{Interpreting the sharpe ratio when excess returns are negative}.
\bjtitle{Investment Analysts Journal}
\bvolume{33}(\bissue{59}),
\bfpage{15}--\blpage{20}
(\byear{2004})
\end{barticle}
\endbibitem

%%% 69
\bibitem[\protect\citeauthoryear{He et~al.}{2020}]{he2020impact}
\begin{barticle}
\bauthor{\bsnm{He}, \binits{Q.}},
\bauthor{\bsnm{Liu}, \binits{J.}},
\bauthor{\bsnm{Wang}, \binits{S.}},
\bauthor{\bsnm{Yu}, \binits{J.}}:
\batitle{The impact of covid-19 on stock markets}.
\bjtitle{Economic and Political Studies}
\bvolume{8}(\bissue{3}),
\bfpage{275}--\blpage{288}
(\byear{2020})
\end{barticle}
\endbibitem

%%% 70
\bibitem[\protect\citeauthoryear{Belhassine and Karamti}{2021}]{belhassine2021contagion}
\begin{barticle}
\bauthor{\bsnm{Belhassine}, \binits{O.}},
\bauthor{\bsnm{Karamti}, \binits{C.}}:
\batitle{Contagion and portfolio management in times of covid-19}.
\bjtitle{Economic Analysis and Policy}
\bvolume{72},
\bfpage{73}--\blpage{86}
(\byear{2021})
\end{barticle}
\endbibitem

%%% 71
\bibitem[\protect\citeauthoryear{Yousfi et~al.}{2024}]{yousfi2024pandemic}
\begin{botherref}
\oauthor{\bsnm{Yousfi}, \binits{M.}},
\oauthor{\bsnm{Farhani}, \binits{R.}},
\oauthor{\bsnm{Bouzgarrou}, \binits{H.}}:
From the pandemic to the russia--ukraine crisis: Dynamic behavior of connectedness between financial markets and implications for portfolio management.
Economic Analysis and Policy
(2024)
\end{botherref}
\endbibitem

%%% 72
\bibitem[\protect\citeauthoryear{Ahelegbey and Giudici}{2022}]{ahelegbey2022netvix}
\begin{barticle}
\bauthor{\bsnm{Ahelegbey}, \binits{D.F.}},
\bauthor{\bsnm{Giudici}, \binits{P.}}:
\batitle{Netvix—a network volatility index of financial markets}.
\bjtitle{Physica A: statistical mechanics and its applications}
\bvolume{594},
\bfpage{127017}
(\byear{2022})
\end{barticle}
\endbibitem

\end{thebibliography}
%\listoffigures
%\listoftables

\end{document}